\documentclass[onecolumn]{aa}
\usepackage{graphicx}
\usepackage{txfonts}

\usepackage{xcolor}
\definecolor{mBlue}{RGB}{51, 77, 167}

\usepackage{url}
\usepackage[bookmarks=false, pdfnewwindow=true, 
            colorlinks=true, linkcolor=magenta, citecolor=mBlue, filecolor=cyan, urlcolor=magenta,  
            final=true]{hyperref}

\graphicspath{{./}}

\begin{document}

   \title{Exploring the radial evolution of Interplanetary \\
   Coronal Mass Ejections using EUHFORIA}

   \author{C. Scolini\inst{1,2},
          S. Dasso\inst{3,4},
          L. Rodriguez\inst{2},
          A. N. Zhukov\inst{2,5},
          \and
          S. Poedts\inst{1,6}
          }

   \institute{Centre for mathematical Plasma Astrophysics, KU Leuven, Leuven, Belgium\\
             \email{camilla.scolini@gmail.com}
             \and
             Solar--Terrestrial Centre of Excellence --- SIDC, Royal Observatory of Belgium, Brussels, Belgium
             \and 
             CONICET, Universidad de Buenos Aires, Instituto de Astronomía y Física del Espacio, Grupo LAMP, Buenos Aires, Argentina
             \and
             Universidad de Buenos Aires, Facultad de Ciencias Exactas y Naturales, Departamento de Ciencias de la Atmósfera y los Océanos, Grupo LAMP, Buenos Aires, Argentina 
             \and
             Skobeltsyn Institute of Nuclear Physics, Moscow State University, Moscow, Russia
             \and 
             Institute of Physics, University of Maria Curie-Sk{\l}odowska, Lublin, Poland
             }

   \date{Submitted: December 23, 2020 | Revised: February 6, 2021 | Accepted: --}

 
  \abstract{
  Coronal Mass Ejections (CMEs) are large-scale eruptions from the Sun into interplanetary space. 
  Despite being major space weather drivers, our knowledge of the CME properties in the inner heliosphere remains constrained by the scarcity of observations at heliocentric distances other than 1~au. Furthermore, most CMEs are observed in situ by single spacecraft, requiring numerical models to complement the sparse observations available. 
  } 
  {
  We aim to assess the ability of the linear force-free spheromak CME model in EUropean Heliospheric FORecasting Information Asset (EUHFORIA) to describe the radial evolution of interplanetary CMEs, yielding new context for observational studies.
  }
  {
  We model one well-studied CME with EUHFORIA, investigating its radial evolution by placing virtual spacecraft along the Sun--Earth line in the simulation domain. To directly compare observational and modelling results, we characterise the interplanetary CME signatures between $0.2$ and $1.9$~au from modelled time series, exploiting techniques traditionally employed to analyse real in situ data. 
  }
  {
  Results show that the modelled radial evolution of the mean solar wind and CME values is consistent with observational and theoretical expectations.
  The CME expands as a consequence of the decaying pressure in the surrounding solar wind: 
  the expansion is rapid within $0.4$~au, and moderate at larger distances.
  The early rapid expansion could not explain the overestimated CME radial size in our simulation, 
  suggesting this is an intrinsic limitation of the spheromak geometry used.
  The magnetic field profile indicates a relaxation of the CME structure during propagation, 
  while CME ageing is most probably not a substantial source of magnetic asymmetry beyond 0.4~au.
  Finally, we report a CME wake that is significantly shorter than suggested by observations.
  }
  {
  Overall, EUHFORIA provides a consistent description of the radial evolution of solar wind and CMEs, at least close to their centres; nevertheless, improvements are required to better reproduce the CME radial extension.
  }
  

   \titlerunning{On the radial evolution of CMEs in EUHFORIA}
   \authorrunning{C. Scolini et al.}
   
   \maketitle

\section{Introduction}

The Sun continuously emits charged and neutral particles in the form of solar wind \citep{parker:1958,parker:1965,cranmer:2017}.
In addition, it also ejects gigantic structures of plasma and magnetic field called Coronal Mass Ejections \citep[CMEs; see e.g.][]{webb:2012}. 
These structures are observed as they propagate away from the Sun by coronagraphs \citep{illing:1985, vourlidas:2013}, and are eventually probed in situ by spacecraft monitoring the conditions of the interplanetary medium \citep[where they are named ICMEs, i.e.\ Interplanetary CMEs\footnote{In the following, we use the term ``CME'' rather than ``ICME'' to indicate a CME structure both in the solar corona and in interplanetary space.};][]{zurbuchen:2006, kilpua:2017}. 
In situ, CMEs can be distinguished from the ambient solar wind through which they propagate via a combination of plasma and magnetic field parameters \citep{burlaga:1981, bothmer:1996, wimmer:2006, zurbuchen:2006}.
Typically, when a fast-enough CME is detected, the first signature observed is a CME-driven interplanetary shock, 
followed by a sheath region composed of solar wind material accumulated between the shock and the actual CME structure. 
This region typically presents increased density and magnetic field, and it exhibits turbulent variations of the magnetic field magnitude and direction \citep{kilpua:2017}. 
If the configuration of the spacecraft crossing through the CME structure is appropriate \citep[see e.g.][for examples of different crossing configurations]{gopalswamy:2009, janvier:2014}, 
the sheath is followed by the actual CME body, referred to as a ``magnetic ejecta'' because of the magnetically-dominated plasma and low level of magnetic field fluctuations typically observed in this region.

The evolution of CMEs in the inner heliosphere is the result of the delicate interplay between the internal plasma and magnetic field properties, 
and the conditions of the surrounding solar wind, and it is primarily shaped by two effects: 
the expansion of the magnetic ejecta, which controls its internal magnetic field magnitude and size \citep{demoulin:2009, demoulin:2010, gulisano:2010},
and the interaction with the surrounding solar wind, which controls the CME kinematics and is often described in terms of a drag force \citep{cargill:2004, vrsnak:2010}. Furthermore, the erosion of the CME magnetic field (and its opposite, flux injection) as a consequence of magnetic reconnection processes with the surrounding solar wind \citep{dasso:2006, dasso:2007, ruffenach:2012, ruffenach:2015, lavraud:2014, manchester:2014, pal:2020} can also alter the CME magnetic configuration and size. The interaction of CMEs with other transients in the solar wind, such as high-speed streams \citep[e.g.][]{gopalswamy:2009, wood:2012, rodriguez:2016, winslow:2021}, the heliospheric current sheet \citep{winslow:2016}, and other CMEs \citep[e.g.][]{dasso:2009, lugaz:2017, shen:2017}, also majorly contribute to alter the CME kinematics and internal properties during propagation.
The geometry, morphology, and kinematics of CME structure in the solar corona and interplanetary space are all closely related to the complexity of the propagation scenario, which can result in major evolutionary deformations of various nature \citep[e.g.][]{isavnin:2016}.

Over the past decades, the radial evolution of CMEs in the inner heliosphere has been primarily investigated by means of statistical studies using observations of a large number of CMEs to derive an average scenario of CME propagation \citep{liu:2005, gulisano:2010, winslow:2015, janvier:2019, salman:2020}, and by means of numerical simulations \citep[see e.g. the review by][]{manchester:2017}.
More recently, the increasing number of spacecraft probing the solar wind plasma and magnetic field conditions in different regions of the inner heliosphere also allowed for more direct observations of individual CMEs at different heliocentric distances, thanks to the higher number of longitudinal conjunction events among different spacecraft \citep{rodriguez:2008, nakwacki:2011, winslow:2016, winslow:2018, good:2015, good:2018, lugaz:2020, salman:2020}. 
Despite the increased capabilities to observe the same CME event at multiple spacecraft located at different heliocentric distances, at the time of writing, for most CMEs in situ observations remain limited to a single spacecraft. 
Furthermore, in situ CME observations within 0.3~au from the Sun are still extremely rare \citep[e.g.][]{korreck:2020, moestl:2020, nieves:2020, weiss:2021}, with NASA's Parker Solar Probe \citep[PSP;][]{fox:2016} having so far operated during a low solar activity phase when only few CMEs were observed.
%
Because of these limitations, numerical simulations still represent an invaluable tool to fill in observational gaps and to provide context to the sparse observations available. Despite their potential, however, the performances of numerical solar wind and CME propagation models for the inner heliosphere developed for space weather research and forecasting purposes remain largely untested for heliocentric distances other than that of the Earth \citep[for a recent exception, see][]{alhaddad:2019}. For instance, the effect of the presence of the model inner boundary on the early CME propagation and evolution modelled is a critical -- but so far uninvestigated -- issue, shall such models be used for predictions of the CME properties close the Sun in the near future \citep[][]{odstrcil:2020}.

In this work, we aim to bridge the gap between observational and modelling studies 
by borrowing techniques traditionally employed to analyse in situ CME observations \citep[e.g.][]{gulisano:2010, janvier:2019}, and applying them to modelled in situ CME signatures. 
The double purpose of the study is: 
(1) to validate the numerical modelling of interplanetary CMEs performed by the EUropean Heliospheric FORecasting Information Asset \citep[EUHFORIA;][]{pomoell:2018} integrated with the linear force-free spheromak CME model \citep{verbeke:2019b}, over a broad range of heliocentric distances; 
and (2) to perform a comprehensive analysis of the radial evolution of interplanetary CMEs exploiting the potential of available observational studies and techniques in combination with numerical models.
We focus on addressing a number of fundamental questions on the interplanetary evolution of CME structures 
that have not been targeted by past studies using the EUHFORIA model, in particular: 
{how realistically is the radial evolution of the ambient solar wind modelled in EUHFORIA, 
and how does it affect the radial propagation of magnetised CME structures in numerical simulations?
How are the average CME properties and the size of the various CME structures evolving with radial distance?
How is expansion affecting the magnetic field and kinematics of CMEs at various heliocentric distances? 
How long does it take to the solar wind to recover to pre-event values after the passage of a CME?
}
To serve the purpose, we take as a case study the Earth-directed fast halo CME observed at the Sun on 12 July 2012. 
This event was previously analysed in detail using EUHFORIA simulations by \citet{scolini:2019}, hereafter referred to as ``Paper 1''.
In this work, we further analyse the simulations presented in Paper~1, focusing in particular on Run~03 therein, corresponding to an initial CME configuration which have been optimised based on physical arguments and a thorough comparison with observations.

This paper is structured as follows.
In Section~\ref{sec:methods}, we provide an overview of the observational CME characteristics of the event and introduce the modelling tool used to investigate its evolution in interplanetary space. 
In Sections~\ref{sec:results_sw} and \ref{sec:results_cme}, we present the results obtained from the analysis of the radial evolution of the ambient solar wind and CME structure. In Section~\ref{sec:results_interpretation} we discuss the main findings 
and provide a physical interpretation of the results. Finally, in Section~\ref{sec:conclusions}, we present our conclusions.


\section{Observations and modelling tools}
\label{sec:methods}

\subsection{Remote-sensing and in situ observations of the 12 July 2012 CME}
\label{subsec:data}

In this work, we investigate the radial evolution of an Earth-directed fast halo CME observed on 12 July 2012.
On the day of the eruption, its source region (NOAA~AR~11520) was located close to the solar disk centre,
and the CME was first observed in LASCO coronagraphs \citep{brueckner:1995} on-board the \textit{Solar and Heliospheric Observatory} \citep[SOHO;][]{domingo:1995} at 16:48~UT, appearing as a fast halo propagating with an average projected speed 
in the plane of the sky of 885~km~s$^{-1}$ (from the CDAW CME catalog\footnote{\url{https://cdaw.gsfc.nasa.gov/CME_list/}}). 
An intense X1.4-class flare was also detected by the GOES satellites starting from 15:37~UT.

\begin{figure}[ht]
\centering
{ \includegraphics[width=0.8\hsize]{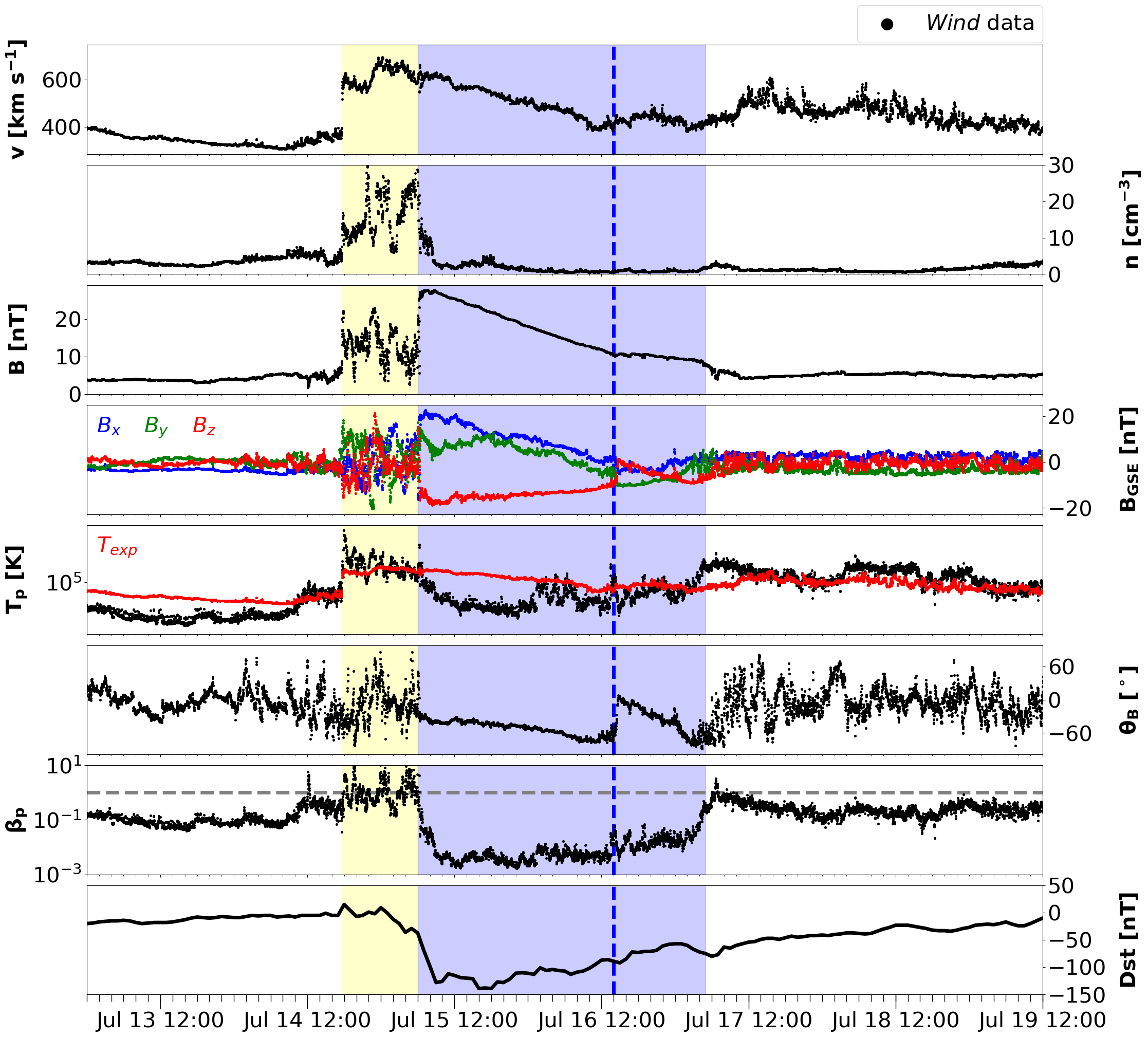}} 
\caption{
1-min averaged solar wind magnetic field and plasma parameters as detected by the \textit{Wind} spacecraft at L1. 
From top to bottom: 
bulk speed ($v$), 
proton number density ($n_p$), 
magnetic field magnitude ($B$), 
magnetic field components in GSE coordinates ($B_x$, $B_y$, $B_z$), 
proton temperature ($T_p$), 
magnetic field elevation ($\theta_B$), 
and proton $\beta$ ($\beta_p$).
The bottom panel shows the 1-hour $Dst$ index. 
The duration of the sheath is indicated by the yellow area, 
while the duration of the magnetic ejecta is bounded by the blue area. 
The dashed blue line marks the start of the perturbation in the speed profile due to an approaching fast solar wind stream at the back of the CME.
The temperature panel shows the comparison of the observed proton temperature $T_p$ (in black)
with the expected temperature $T_{exp}$ calculated using the speed-temperature relation by \citet{lopez:1986} (in red). 
In the $\beta_p$ panel, the value $\beta_p=1$ is indicated by the dashed grey line.
}
\label{fig:20120712_wind} 
\end{figure}

After propagating through the inner heliosphere, the CME was observed to arrive at the Sun--Earth Lagrangian point 1 (L1) a few days later, 
where its preceding shock was detected by the $Wind$ spacecraft \citep{ogilvie:1995} on 14 July 2012 at 17:39~UT 
\citep[from the Heliospheric Shock Database\footnote{\url{www.ipshocks.fi}};][]{kilpua:2015}.
According to the Richardson \& Cane list \citep[hereafter R\&C list\footnote{\url{http://www.srl.caltech.edu/ACE/ASC/DATA/level3/icmetable2.htm}};][]{cane:2003, richardson:2010}, 
a magnetic ejecta presenting the characteristics of a magnetic cloud \citep[i.e.\ an enhanced magnetic field, a smooth rotation of the magnetic field vector, and low proton temperature; e.g.][]{burlaga:1981, klein:1982}
was detected starting around 06:00~UT on 15 July, 
and ending at about 05:00~UT on 17 July.
The orientation of the magnetic field within the magnetic ejecta was characterised by a strongly-negative $B_z$ component in Geocentric Solar Ecliptic (GSE) coordinates, and therefore it was mostly anti-parallel with respect to the magnetic field of the Earth, triggering an intense geomagnetic storm as indicated by the Disturbance Storm Time (Dst) index, which reached a minimum of $-139$~nT on July 15.
Figure~\ref{fig:20120712_wind} shows the 1-min averaged measurements from the $Wind$ spacecraft taken during the passage of the CME. 
The region between the CME-driven shock and the starting of the magnetic ejecta (i.e. the CME sheath) is indicated by the yellow shaded area, 
while the duration of the magnetic ejecta is bounded by the blue shaded area.

The remote-sensing and in situ observational signatures of this event were discussed in detail in Paper~1.
Here we highlight two characteristics of the in situ speed profile $v(t)$ that are more relevant for the following discussion: 
(1) $v(t)$ is higher at the front than at the back of the magnetic ejecta, indicating that this was an expanding magnetic structure in the solar wind at 1~au; 
(2) $v(t)$ is decreasing with a nearly-linear behaviour only in the front part of the ejecta, while a distortion is present in the rear part of the ejecta starting around 14:00~UT on July 16 (marked by the dashed blue line in Figure~\ref{fig:20120712_wind}). This was most probably caused by the presence of an overtaking fast solar wind stream visible at the back of the magnetic ejecta. The CME under study therefore belongs to the class of ``perturbed'' magnetic ejectas as defined by \citet{gulisano:2010}, since it shows a strongly-distorted velocity profile compared to non-perturbed cases, which show a linear velocity profile for almost the full duration of the magnetic ejecta events. We will further discuss how this characteristics was reproduced by numerical simulations and how it affected the CME propagation in the following Sections.
Furthermore, we point out that the identification of the CME boundaries from in situ data, particularly the rear one, can be a difficult task. 
In fact, different authors frequently consider different boundaries for the same CME, typically with differences of few hours, due to the application of different identification criteria \citep[e.g.][]{dasso:2006, alhaddad:2013}, which can in turn significantly affect the reconstruction of the parameters describing the CME internal characteristics.

\subsection{The EUHFORIA model}
\label{subsec:euhforia}

EUHFORIA is a recent coronal and heliospheric model designed for space weather research and prediction purposes \citep{pomoell:2018}. 
The model is composed of a semi-empirical coronal model and a three-dimensional (3D) magnetohydrodynamics (MHD) model of the inner heliosphere. 
The solar wind conditions at the inner boundary of the heliospheric domain, located at 0.1~au, are derived through a semi-empirical approach based on the  Wang-Sheeley-Arge \citep[WSA; see e.g.][]{arge:2004} coronal model, which takes as input synoptic maps of the photospheric magnetic field from the Global Oscillation Network Group (GONG) network. 
A detailed discussion of the coronal model currently used in EUHFORIA is provided in the recent paper by \cite{pomoell:2018}.
The solar wind map generated by the coronal model is then used as inner boundary condition for the heliospheric model, which solves the time-dependent ideal MHD equations in 3D space in the Heliocentric Earth EQuatorial (HEEQ) coordinate system. 
The heliospheric model in EUHFORIA treats the solar wind plasma as a single fluid, 
solving the time-dependent MHD equations and providing as output the following eight MHD variables: 
the plasma number density $n$, 
the plasma velocity components $v_x$, $v_y$, $v_z$,
the plasma magnetic field components $B_x$, $B_y$, $B_z$, 
and the plasma thermal pressure $P_{th}$.
The standard computational domain for the heliospheric model extends from 0.1~au to 2~au in the radial direction ($D$), spanning $\pm 60^\circ$ in latitude ($\theta$), and $\pm 180^\circ$ in longitude ($\phi$). 

In addition to the modelling of the ambient solar wind, 
EUHFORIA admits the possibility to model CMEs either using a simplified cone model \citep{scolini:2018b},
or via a more realistic linear force-free spheromak (flux-rope) model \citep{verbeke:2019b}.
In both models, CMEs are inserted in to the heliosphere as time-dependent boundary conditions at the inner radial boundary of the domain, i.e. at the heliocentric distance of $D=0.1$~au.
The initial geometry and kinematics of cone and spheromak CMEs are defined via the following input parameters:
the CME insertion start time ($t_\mathrm{CME, i}$), 
speed ($v_\mathrm{CME, i}$), 
latitude ($\theta_\mathrm{CME, i}$), 
longitude ($\phi_\mathrm{CME, i}$), 
and half width ($\omega_\mathrm{CME, i}/2$).
These parameters are normally derived from remote-sensing white-light observations of CMEs in the corona.
In addition, due to more limited observational inputs, two additional parameters defining the CME mass density and temperature are set to be homogeneous and equal to the following default values: 
$\rho_\mathrm{CME, i} = 1 \times 10^{-18}$~kg~m$^{-3}$
and $T_\mathrm{CME, i} = 0.8 \times 10^{6}$~K \citep{pomoell:2018}.
A recent analysis on the remote-sensing estimation of the CME density and its comparison with observations at 1~au has been provided by \citet{temmer:2021}. In the future, this work will likely become an extremely valuable contribution to help constraining the CME initial density for heliospheric models such as EUHFORIA.
The magnetic configuration of spheromak CMEs is then specified via three additional parameters:
the chirality, or helicity sign ($H_\mathrm{CME, i}$), the tilt ($\gamma_\mathrm{CME, i}$), and the toroidal magnetic flux ($\varphi_{t, \mathrm{CME, i}}$), 
which can often be constrained using several EUV (low-coronal) and magnetic (photospheric) proxies
(see \citet{palmerio:2017} and references therein).

The comparison of EUHFORIA results with actual in situ measurements of the solar wind properties requires to relate the above single-fluid MHD variables to the macroscopic quantities measured by in situ spacecraft.
Employing a single-fluid approach, the EUHFORIA heliospheric model makes no distinction between different particle populations (e.g.\ protons, electrons, $\alpha$ particles, etc.), and it assumes all populations to be characterised by the same macroscopic parameters.
Considering the proton and electron populations as the two primary contributors to solar wind plasma 
\citep{vonsteiger:2000, zurbuchen:2002, reisenfeld:2007, heber:2009} we can treat the two species as characterised by the same (isotropic) temperature $T = T_p = T_e$. To further ensure the quasi-neutrality of the plasma at all times and spatial locations in the heliosphere, we also assume the two species to be characterised by a uniform number density $n_p = n_e$, with the plasma number density of all particle populations being $n = n_p + n_e$.
As a result, the solar wind thermal pressure $P_{th}$ provided by EUHFORIA can be described as
$P_{th} = P_{p,th} + P_{e, th} = n_p k_B T_p + n_e k_B T_e = 2 P_{p,th}$, i.e.\ twice the thermal pressure of the proton population alone.
In this picture, the plasma $\beta$ corresponds to $\beta = P_{th} / P_{mag} = 2 P_{p,th}/ P_{mag} = 2 \beta_p$, i.e.\ two times the proton $\beta$.
In the following, we therefore derive the proton thermal pressure and proton $\beta$ parameter from EUHFORIA outputs by dividing the plasma thermal pressure and plasma $\beta$ by a factor 2, i.e. $P_{p,th} = P_{th}/2$ and $\beta_p = \beta /2$. 
The plasma total pressure $P_{tot}$ from both EUHFORIA results and $Wind$ data is calculated as  
$P_{tot} = P_{mag} + P_{th} = P_{mag} + P_{p,th} + P_{e,th} = P_{mag} + 2 P_{p, th}$, 
while the plasma (and proton) temperature is calculated as $T = T_p = P / n k_B = P_p / n_p k_B$.
The speed $v$ of the solar wind plasma corresponds to the speed of the centre of mass of the system and can be expressed in terms of the proton and electron properties as $v = (v_p m_p + v_e m_e)/(m_p + m_e) \sim v_p$.

\subsection{Modelling the 12 July 2012 CME with EUHFORIA}
\label{subsubsec:euhforia_cme}

In this work, we further analyse the simulations of the 12 July 2012 CME presented in Paper~1. 
We focus in particular on the investigation of the results obtained from Run~03, as this was found to be the most realistic simulation of the CME performed using a magnetised flux-rope model.
The ambient solar wind in the heliosphere was obtained using as input the GONG standard synoptic map on 12 July 2012 at 11:54~UT. 
The spheromak CME structure was initialised using kinetic and geometric parameters derived from the 3D reconstruction based on the Graduated Cylindrical Shell \citep[GCS; ][]{thernisien:2009, thernisien:2011} model applied to multi-viewpoint observations of the CME in the corona. The GCS reconstruction provided the following set of CME initial parameters (extrapolated to 0.1~au assuming a self-similar propagation): 
passage time on July 12, 2012 at 19:24~UT;
translation speed of 763~km~s$^{-1}$;
longitude of $-4^\circ$;
latitude of $-8^\circ$;
half width of $38^\circ$.
We refer the reader to Paper~1 for further information on the analysis of the CME coronal signatures.
Here we only further note that, because of their large angular separations, the two STEREO spacecraft were in an optimal configuration to observe the radial propagation of the CME and provide accurate measurements of the CME kinematics in the direction most needed for the simulation (i.e.\ towards Earth).
The magnetic parameters of the spheromak CME were initialised based on EUV and photospheric magnetic observations of the source region, which indicated the eruption of a right-handed flux rope magnetic structure with inclination of about $45^\circ$ with respect to the solar equatorial plane and with the axial magnetic field pointing towards the south-east direction. Again, we redirect the reader to Paper~1 for further information on the analysis of the CME source region signatures.
The simulation was performed using EUHFORIA version~1.0.4 with the following resolution in the heliospheric domain: 
512 grid cells in the radial direction between $0.1$~au and $2.0$~au (corresponding to a radial resolution of $\Delta D = 0.00371 $~au), and a $2^\circ$ angular resolution in longitude and latitude.
To ensure the CME and its wake had time to propagate beyond 2~au, corresponding to the radial outer boundary in our simulation domain, we ran the model for 19~days after the time of the magnetogram, i.e. until 31 July 2012.

\begin{figure}
\centering
{\includegraphics[width=\hsize]{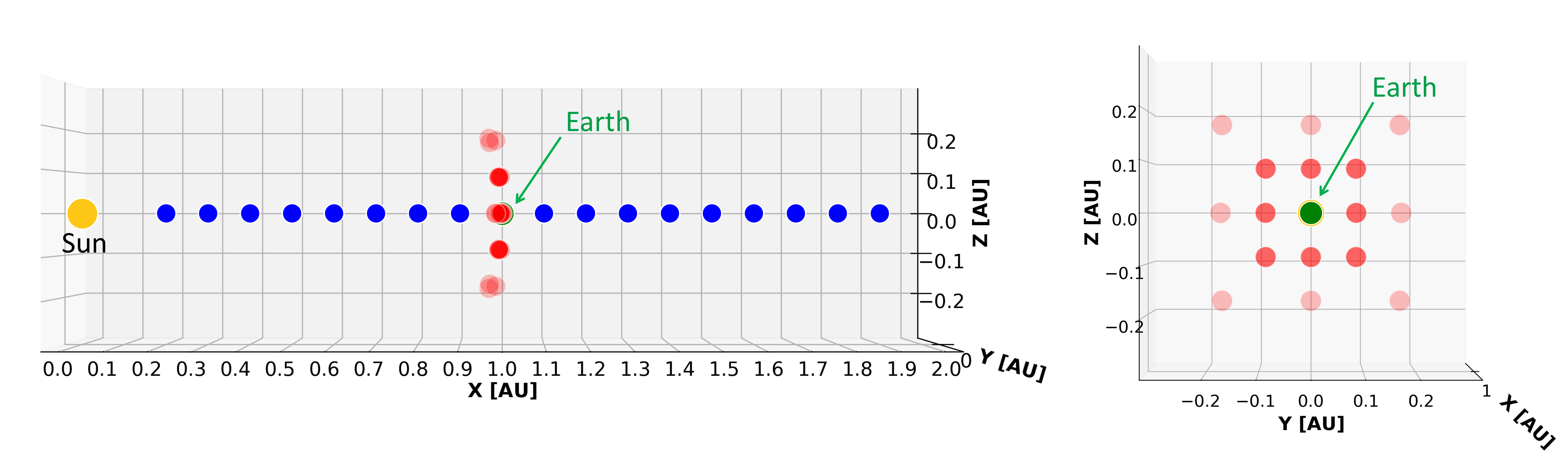}} 
\caption{Location of the virtual spacecraft placed in the EUHFORIA heliospheric simulation domain, as seen from two different views.
Blue dots: virtual spacecraft along the Sun--Earth line.
Red dots: virtual spacecraft near the location of the Earth.
Yellow dot: location of the Sun.
Green dot: location of the Earth.
We will focus on the virtual spacecraft in blue to study the radial evolution of the Earth-directed CME event considered.
}
\label{fig:virtual_spacecraft} 
\end{figure}

\begin{figure}
\centering
\includegraphics[width=0.8\hsize]{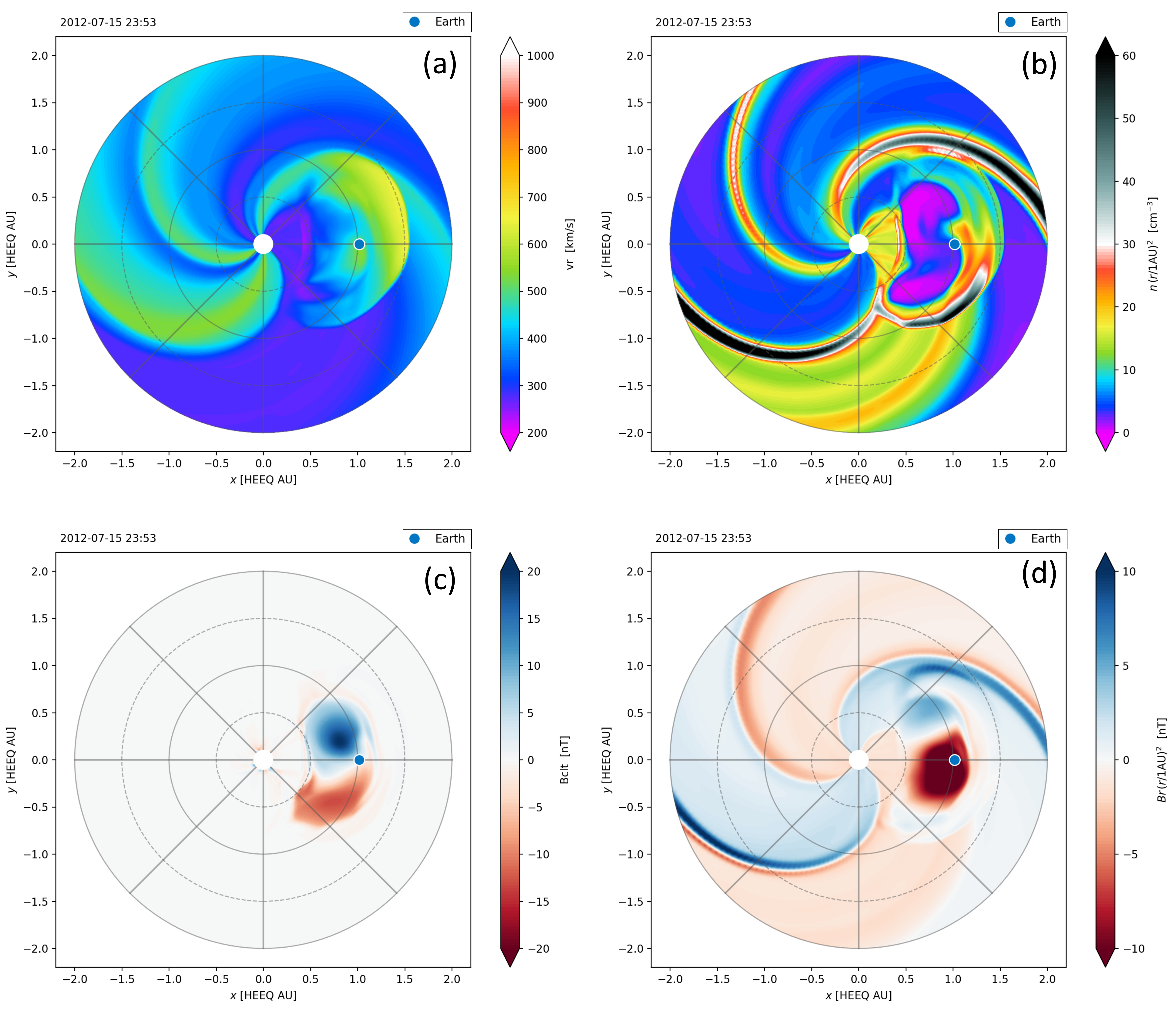}
\caption{Overview of EUHFORIA results in the heliographic equatorial plane.  
Snapshots of the radial speed $v_r$ (a), scaled number density $n \, (D/\mathrm{1 \,\, au})^2$ (b),
co-latitudinal magnetic field $B_{clt} = - B_\theta$ (c), and scaled radial magnetic field $B_{r} \, (D/\mathrm{1 \,\, au})^2$ (d)
on 15 July 2012 at 23:53~UT, when the CME was crossing through Earth.
}
\label{fig:euhforia_ecliptic} 
\end{figure}

\begin{figure}
\centering
\includegraphics[width=0.8\hsize]{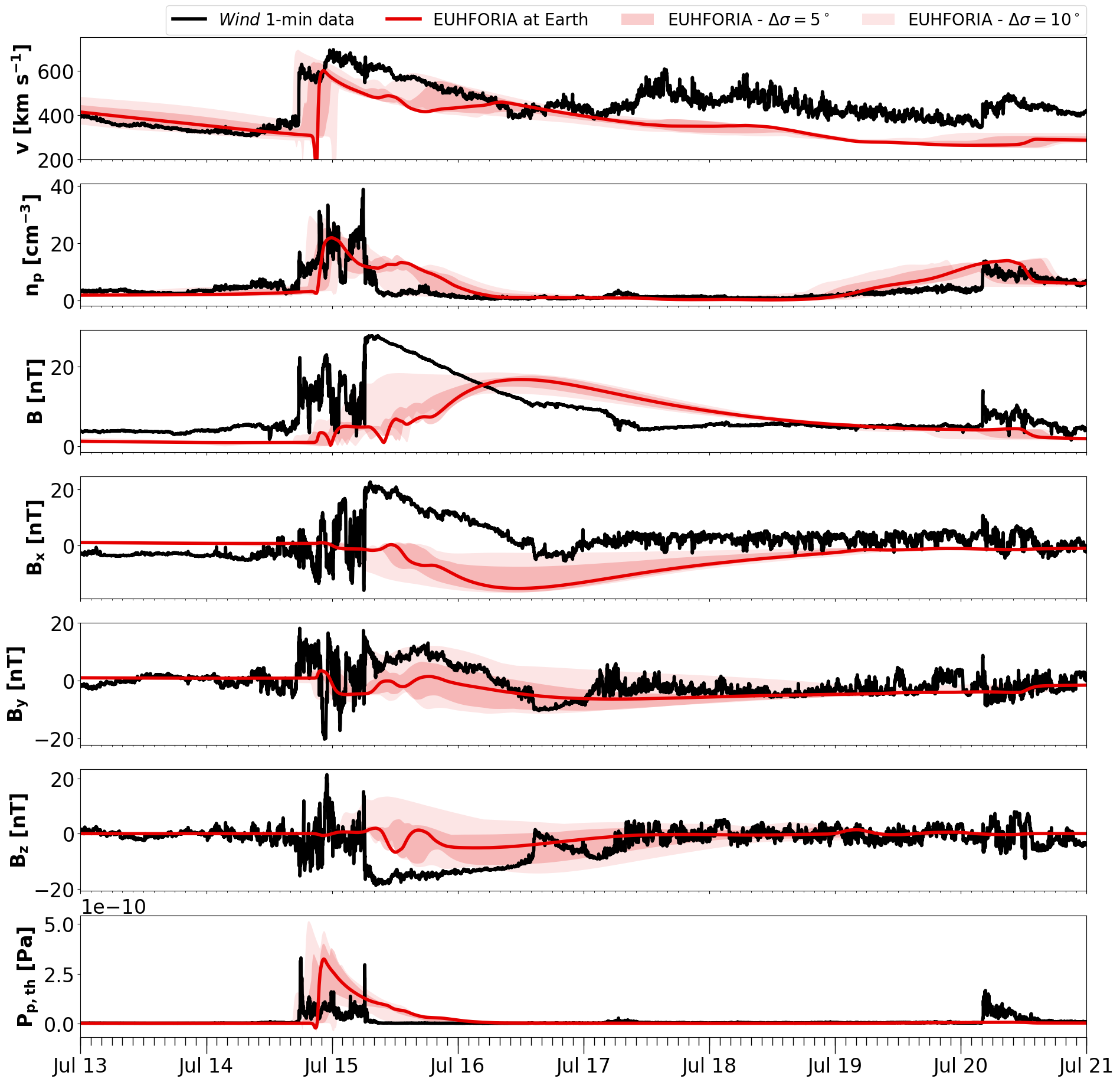}
\caption{Comparison of 10-min cadence EUHFORIA predictions at Earth (red solid line) and 1-min averaged \textit{Wind} proton data (black solid line).
The dark red and light red shaded areas show the maximum variation of EUHFORIA predictions 
at virtual spacecraft separated by $\Delta \sigma = 5^\circ$ and $\Delta \sigma = 10^\circ$ in longitude and/or latitude from Earth. 
From top to bottom: 
speed ($v$), 
proton number density ($n_p$), 
magnetic field magnitude ($B$), 
magnetic field components in GSE coordinates ($B_x$, $B_y$, $B_z$), 
and proton thermal pressure ($P_{p,th}$).
}
\label{fig:euhforia_earth} 
\end{figure}

To track the heliospheric propagation of the CME in the simulation, we placed a first set of virtual spacecraft along the Sun--Earth line.
Such virtual spacecraft were uniformly distributed in the radial direction between 0.2 and 1.9~au, with a separation of 0.1~au among each other. 
A second set of virtual spacecraft was placed around Earth position (i.e. at 1~au), 
at $\Delta \sigma = 5^\circ$ and $\Delta \sigma = 10^\circ$ longitudinal and/or latitudinal separation from Earth, in order to assess the spatial variability of the results in the vicinity of Earth. Figure~\ref{fig:virtual_spacecraft} shows a schematic representation of the location of the virtual spacecraft in the 3D heliospheric domain of EUHFORIA.
An overview of the CME propagation in the EUHFORIA heliospheric domain is provided in Figure~\ref{fig:euhforia_ecliptic}.
A comparison between EUHFORIA results at (and in the vicinity of) Earth and in situ measurements obtained by the $Wind$ spacecraft is provided in Figure~\ref{fig:euhforia_earth}. 

\section{Results: radial evolution of the ambient solar wind}
\label{sec:results_sw}

As a first step towards the understanding of the heliospheric propagation of the CME under study, we start by characterising the conditions of the ambient solar wind through which the CME propagates.  

\subsection{Theoretical and observational expectations}
To date, the most extensive database of solar wind measurements at different heliocentric distances on the ecliptic plane dates back to in situ observations carried out in the 1970s and 1980s by the \textit{Helios~1} and \textit{Helios~2} missions \citep{schwenn:1990}, which covered heliocentric distances between 0.29 and 0.98~au. 
Assuming a radial solar wind outflow, the radial dependence of the solar wind parameters extracted from \textit{Helios} data can be typically described through a power law scaling. Their radial behaviour can be therefore written in the form of a power law function,
\begin{equation}
    X(D) = (D/D_0)^{a_x} \cdot X_0,
    \label{eqn:power_law_lin} 
\end{equation}
where $X$ is the given solar wind parameter considered, 
$D$ is a generic heliocentric distance in astronomical units (au),
$D_0$ is a reference heliocentric distance in au,
$a_x$ is the power law exponent for the solar wind parameter $X$,
and $X_0$ is the magnitude of the solar wind parameter at the reference distance $D_0$ (in au).
Typically, the reference heliocentric distance is taken at 1~au, where continuous measurements of the solar wind parameters are available.
Numerous existing studies constrained the exponents of the various parameters based on \textit{Helios} observations.

\textbf{Speed.}
The Parker's solar wind model predicts the solar wind speed $v$ in interplanetary space to be approximately constant with respect to the heliocentric distance \citep{parker:1958, parker:1965}. Observational studies agree in considering the solar wind speed as essentially radial and independent on the heliocentric distance (i.e.\ $v \propto D^0$). \citet{schwenn:1983, schwenn:1990b} derived radial dependencies of $v \propto D^{0.083}$ and $v \propto D^{0.036}$ for \textit{Helios}~1 and \textit{Helios}~2, respectively, while \citet{venzmer:2018} reported a velocity exponent $v \propto D^{0.049}$ for the combined data sets. 
It is important to mention that the observed variability of the solar wind speed with $D$ can be expected to depend, among other factors, on the particular initial speed of individual plasma bubbles, and on their interaction with other structures in interplanetary space.

\textbf{Density.}
Assuming conservation of mass and a constant solar wind speed as predicted by the Parker's solar wind model, 
the proton number density is expected to fall off as $n_p \propto \, D^{-2}$.
\citet{bougeret:1984} derived the exponent for the solar wind electron number density, as $n_e \propto \, D^{-2.10}$ based on \textit{Helios}~1 and \textit{Helios}~2 observations.
More recently, \citet{venzmer:2018} reported a radial dependence for the solar wind proton number density of $n_p \propto \, D^{-2.01}$ by performing an analysis of the full \textit{Helios} database. 
Furthermore, \citet{hellinger:2011} and \citet{perrone:2019a} reported radial dependencies 
of $n_p \propto \, D^{-1.8}$ and $n_p \propto \, D^{-1.96 \pm 0.07}$, respectively, based on \textit{Helios} observations of solar wind high-speed streams ($v \ge 600$~km~s$^{-1}$).
In the slow solar wind ($v < 400$~km~s$^{-1}$), \citet{hellinger:2013} reported radial dependence
of $n_p \propto \, D^{-1.9}$ based on \textit{Helios} observations.
Under the assumption of quasi-neutrality, and treating the solar wind as composed by protons and electrons only \citep[neglecting the presence of $\alpha$ particles, which have typical abundances between 1 and 5\%; see e.g.][]{kasper:2012}, the number density of protons and electrons are identical, i.e.\ $n_p = n_e$, meaning that the above dependencies hold for both particle populations.

\textbf{Magnetic field magnitude.}
For an adiabatic solar wind expansion and assuming conservation of mass and magnetic flux, 
the three components (radial, latitudinal, and longitudinal) of the interplanetary magnetic field on the ecliptic plane are expected to fall off as $B_r \propto \, D^{-2}$, $B_\theta \propto \, D^{0}$, and $B_\phi \propto \, D^{-1}$, respectively.
Therefore, the total magnetic field magnitude can be expected to decay with a fall off index between $a_B = 0$ and $a_B = -2$.
Observationally, the radial dependence of the interplanetary magnetic field magnitude in the inner heliosphere was first characterised using early measurements carried out by the $Helios$ missions. 
\citet{mariani:1990} provides a summary of these early findings.
In particular, \textit{Helios}~1 data were best fitted by a radial dependence of $B \propto D^{-1.56 \pm 0.09}$,
while \textit{Helios}~2 data were best fitted via the relation $B  \propto  D^{-1.84 \pm 0.08}$.
More recently, \citet{venzmer:2018} reported a radial dependence of the total magnetic field as $B \propto \, D^{-1.55}$ by performing an analysis of the full \textit{Helios} database. 
Earlier analysis also reported a dependence of the scaling of the interplanetary magnetic field magnitude with the solar wind speed, e.g.\ 
\citet{mariani:1990} fitted \textit{Helios}~2 data with $B \propto D^{-1.86 \pm 0.05}$ for the fast solar wind ($v > 550$~km~s$^{-1}$),
and $B \propto  D^{-1.64 \pm 0.24}$ for the slow solar wind ($v < 450$~km~s$^{-1}$).
More recently, \citet{hellinger:2011} and \citet{perrone:2019a} obtained radial dependencies in fast solar wind streams of 
$B \propto \, D^{-1.6}$ and $B \propto \, D^{-1.59 \pm 0.06}$, respectively. For the slow solar wind, \citet{hellinger:2013} obtained a radial dependence of $B \propto \, D^{-1.6}$. 

\textbf{Temperature.}
Assuming an ideal gas in adiabatic expansion, the solar wind temperature (considering all populations in thermal equilibrium) is expected to fall off as $T \propto \, D^{-4/3}$.
However, the description of the actual solar wind temperature is much more complex, as the various species composing the solar wind plasma (i.e.\ protons and electrons, but also $\alpha$ particles and ionised heavy elements) exhibit distribution functions often characterised by different bulk temperatures and deviating from Maxwellian and/or isotropic conditions \citep[see e.g.][and references therein]{marsch:2006, marsch:2012, verscharen:2019}. 
Therefore, the temperature properties of the various particle populations have to be considered separately.
Furthermore, the existence of a speed--proton temperature relation in the solar wind has been also recognised \citep[e.g.][]{lopez:1986, lopez:1987, perrone:2019b}, requiring to treat the radial dependence of the solar wind temperature in slow and fast solar wind streams on a separate basis.
Limiting our attention to proton populations, \citet{schwenn:1981} obtained the following radial dependencies: 
$T_{p} \propto D^{-1.21}$ for the slow solar wind, 
and $T_{p} \propto D^{-0.69}$ for the fast solar wind, 
requiring the existence of an additional heating mechanism to explain the non-adiabatic expansion of fast solar wind streams in interplanetary space \citep[e.g.][]{totten:1995}.
More recently, for the fast solar wind \citet{hellinger:2011} reported a dependence of $T_{p} \propto D^{-0.74}$, 
while \citet{perrone:2019a} reported a dependence of $T_{p} \propto D^{-0.9 \pm 0.1}$.
\citet{hellinger:2013} reported a dependence of $T_{p} \propto D^{-0.58}$ in the slow solar wind.
An average proton temperature exponent of $T_p \propto D^{-0.79}$ was also reported by \citet{venzmer:2018}.
A detailed investigation of the physical origin of the temperature--speed relation in the solar wind, including a description of its relation to heating mechanisms and adiabatic expansion, can be found in \citet{demoulin:2009c}.

\textbf{Proton pressure and proton $\beta$.}
Based on the above expected theoretical scalings for the solar wind speed, density, magnetic field, and temperature, we can estimate the scaling of relevant derived quantities such the as the proton thermal pressure, magnetic pressure, and proton $\beta$. In particular, 
$P_{p, th} \propto n_p T_p \propto D^{a_{n_p} + a_{T_p}}$ 
(i.e.\ $a_{P_{p, th}} = a_{n_p} + a_{T_p} \sim -10/3 $),  
$P_{mag} \propto B^2 \propto D^{2 a_B}$ 
(i.e.\ $a_{P_{mag}} = 2 a_B \sim -4 \leftrightarrow 0 $), 
and hence $\beta_p = P_{p, th}/P_{mag} \propto D^{a_{n_p} + a_{T_p} - 2 a_B} $ 
(i.e.\ $a_{\beta_{p}} = a_{P_{p, th}} -  a_{P_{mag}} \sim -10/3 \leftrightarrow 2/3 $).
As a reference, \citet{perrone:2019a} reported a dependence of the proton thermal pressure $P_{p, th} \propto D^{-2.9 \pm 0.1}$, 
and a dependence of the magnetic pressure $P_{mag} \propto D^{-3.2 \pm 0.1}$ in high-speed solar wind streams.
The resulting radial dependence of the proton $\beta$, defined as the ratio between the proton thermal and magnetic pressures, 
i.e.\ $\beta_{p} = P_{p, th}/P_{mag} $, was $\beta_{p} \propto D^{0.4 \pm 0.1}$.

\subsection{Modelling results and comparison with expectations}

\begin{figure}
\centering
\includegraphics[width=0.95\hsize]{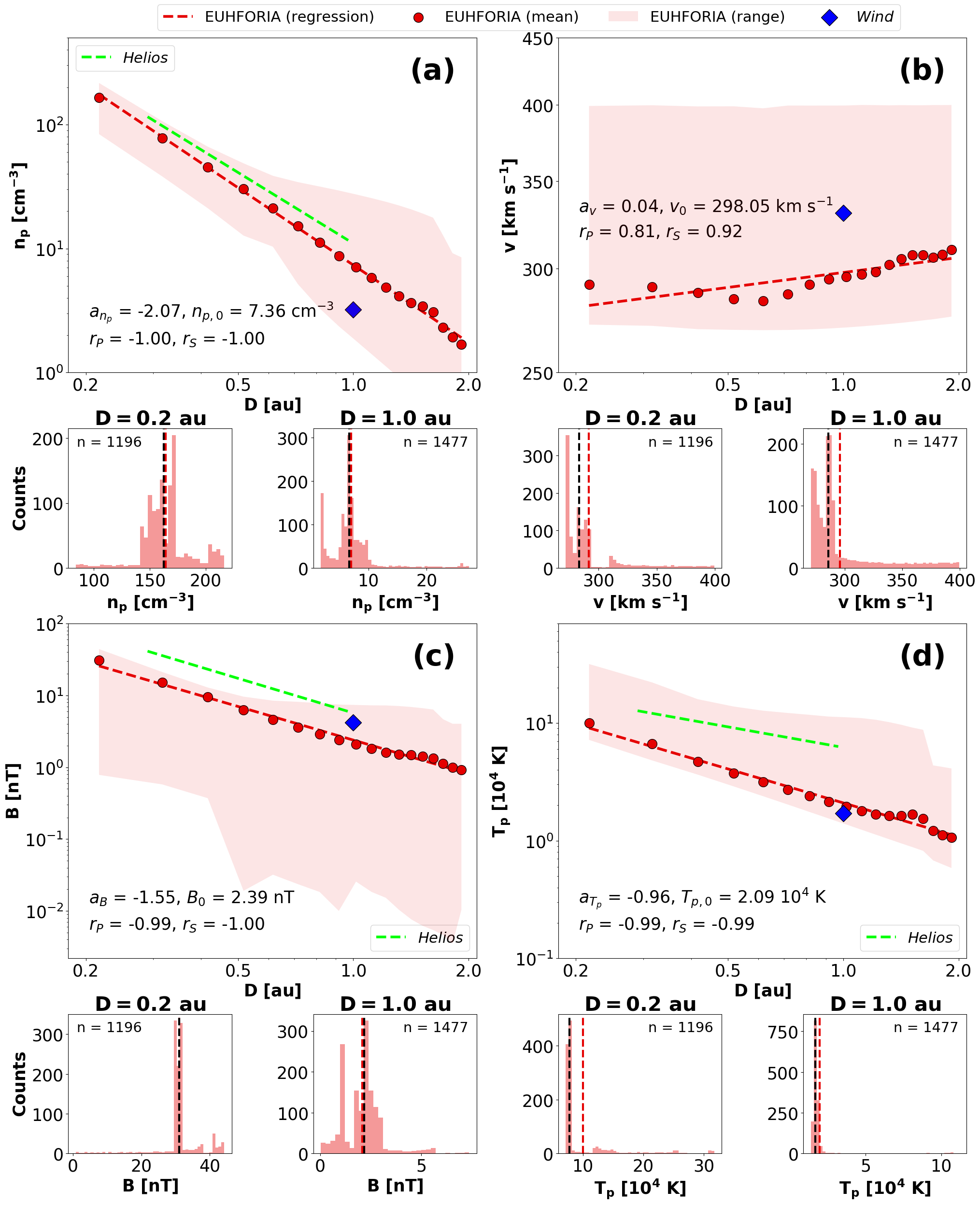}
\caption{Ambient solar wind from EUHFORIA: 
radial dependence of the mean solar wind parameters extracted at virtual spacecraft located along the Sun--Earth line (red dots).
(a): proton number density $\langle n_p \rangle_{sw}$.
(b): speed $\langle v \rangle_{sw}$. 
(c): magnetic field $\langle B \rangle_{sw}$.
(d): proton temperature $\langle T_p \rangle_{sw}$.
The red shaded areas show the maximum variation of EUHFORIA predictions as a function of the heliocentric distance.
The results from the fitting of the mean values are indicated as dashed red lines.
Observation-based relations from \citet{hellinger:2013} are indicated as dashed green lines.
Observed mean values from $Wind$ in the 24 hours prior to the CME arrival are shown as blue diamonds.
Histograms showing the frequency of occurrence of each parameter are provided for $D=0.2$~au and $D=1.0$~au radial distances.
The mean and median are indicated by the red and black vertical dashed lines, respectively.}
\label{fig:sw_radial_01} 
\end{figure}

\begin{figure}
\centering
\includegraphics[width=0.95\hsize]{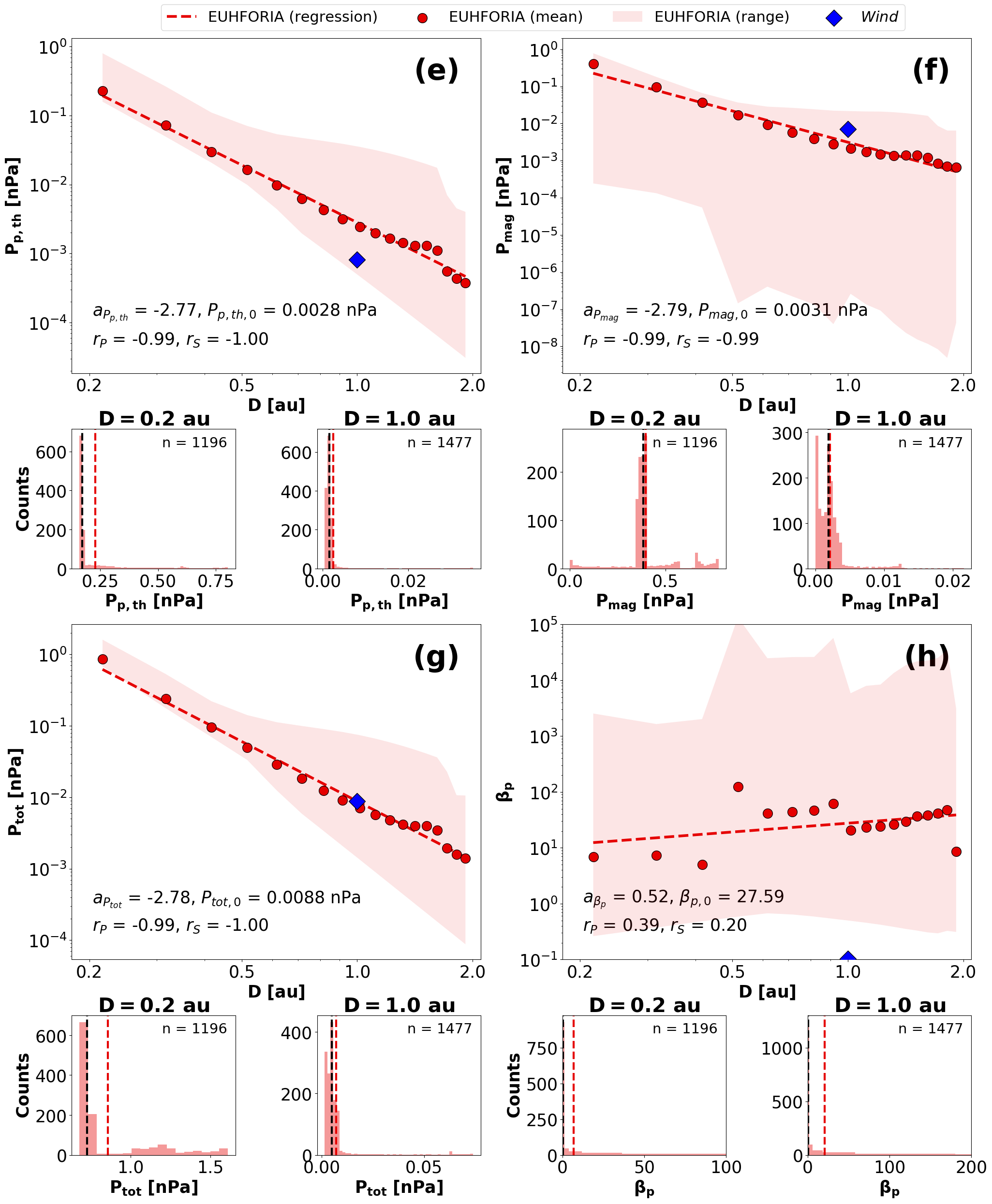}
\caption{Ambient solar wind from EUHFORIA: radial dependence of the mean solar wind parameters extracted at virtual spacecraft located along the Sun--Earth line (red dots).
(e): proton thermal pressure $\langle P_{p,th} \rangle_{sw}$. 
(f): magnetic pressure $\langle P_{mag} \rangle_{sw}$.
(g): total pressure $\langle P_{tot} \rangle_{sw}$.
(h): proton $\beta$ $\langle \beta_p \rangle_{sw}$.
The red shaded areas show the maximum variation of EUHFORIA predictions as a function of the heliocentric distance.
The results from the fitting of the mean values are indicated as dashed red lines.
Mean observed values from $Wind$ in the 24 hours prior to the CME arrival are shown as blue diamonds.
Histograms showing the frequency of occurrence of each parameter are provided for $D=0.2$~au and $D=1.0$~au radial distances.
The mean and median are indicated by the red and black vertical dashed lines, respectively.}
\label{fig:sw_radial_02} 
\end{figure}

To characterise the radial dependence of the solar wind parameters in EUHFORIA,
we consider modelling results at the virtual spacecraft distributed along the Sun--Earth line (Figure~\ref{fig:virtual_spacecraft}).
In order to make sure to consider similar solar wind conditions at each spacecraft (i.e.\ evolution with heliocentric distance), 
and because according to EUHFORIA simulations, the CME propagated through a slow solar wind background along the Sun--Earth line (Figure~\ref{fig:euhforia_ecliptic}),
we identify periods of slow solar wind using the criteria $v \le 400$~km~s$^{-1}$ \citep[similar to][]{hellinger:2013}.
For each of these time periods at different virtual spacecraft, 
we calculate the maximum, minimum, mean and median values of the solar wind parameters observed at each location, 
and we then perform a linear fit of the mean conditions in log--log space. 
By re-writing Equation~\ref{eqn:power_law_lin} in log--log space for a reference heliocentric distance $D_0=1$~au, 
we obtain a linear fitting relation equal to
\begin{equation}
    \log{X(D)} = a_x \cdot \log(D) + \log(X_0),
    \label{eqn:power_law_log} 
\end{equation}
where $\log{X(D)}$ is the logarithm of the given solar wind parameter $X$ considered, 
$\log(D)$ is the logarithm of the generic heliocentric distance in au,
$a_x$ is the power law exponent for the solar wind parameter $X$ (same as in Equation~\ref{eqn:power_law_lin}),
and $\log(X_0)$ is the logarithm of the typical value of the particular solar wind parameter $X$ at 1~au.
The quality of the fitting is evaluated in terms of the Pearson correlation coefficient ($r_P$), 
and of the Spearman's rank correlation coefficient ($r_S$).
The results from this procedure are shown in Figures~\ref{fig:sw_radial_01}~and~\ref{fig:sw_radial_02}.
Table~\ref{tab:solar_wind} reports the radial dependencies ($a_x$ parameter in Equations~\ref{eqn:power_law_lin} and~\ref{eqn:power_law_log}) of the mean slow solar wind parameters in EUHFORIA, derived from the fitting of the individual values at different heliocentric distances using Equation~\ref{eqn:power_law_lin} in log--log space. 
To assess the consistency between the mean- and median-based values (with the former typically being more sensitive to outliers than the latter), in the rightmost column in Table~\ref{tab:solar_wind} we also provide the dependencies estimated from the median values, and for both we indicate the uncertainties derived from the linear fitting based on $95 \%$ confidence intervals. 
The two methods give similar results, particularly for the magnetic field strength and for the proton density, speed, temperature, and $\beta$. Such a result is also suggested by the close mean and median values reported in the small panels (for 0.2 and 1.0~au) in Figures~\ref{fig:sw_radial_01} and \ref{fig:sw_radial_02}. 
Because of the larger number of observational studies investigating mean values rather than median values, in the following we focus on mean values only. A comparison with mean-based observational values (when available) is provided in Table~\ref{tab:solar_wind}.

Overall, the fall off (i.e.\ the slope) of the mean of each solar wind parameter with heliocentric distance is consistent with observational values 
(see summary of the exponents $a_x$ for the various parameters and solar wind components in Table~\ref{tab:solar_wind}), and most parameters considered also exhibit a strong correlation ($\ge 0.9$) with the heliocentric distance $D$ in log--log scale. The two parameters that exhibit the weakest dependencies on the heliocentric distance are the solar wind speed ($a_v = 0.04$, correlations between 0.81 and 0.92) and proton $\beta$ ($a_{\beta_p} = 0.52$, correlations between 0.20 and 0.39), consistent with theoretical and observational arguments.
Regarding $v$, it is important to note that a modelled non-zero dependence with heliocentric distance might in turn affect the modelled decay of all the other parameters, whose radial devolution is strictly dependent on the radial evolution of the speed.  

Since our characterisation of the solar wind radial evolution has the ultimate scope of assessing how the solar wind modelling in EUHFORIA affects the ultimate radial evolution of the CME, we focus here on the discussion of the observational--modelling discrepancies in the light of their potential influence on the CME propagation.
First of all, we note that the drag force is primarily dependent on the CME geometry, i.e.\ its cross-section and width, as well as on its density and speed relative to the ambient solar wind \citep[e.g.][]{vrsnak:2010}. The solar wind modelled in EUHFORIA shows that $\langle n_p \rangle_{sw}$ ($\langle v \rangle_{sw}$) is slightly overestimated (underestimated) compared to the mean values measured by $Wind$ at 1~au, suggesting the drag acted by the solar wind on the CME is probably slightly overestimated in our simulation than it was in reality.
Secondly, the CME expansion is primarily controlled by the relative pressure of the CME and the surrounding solar wind (contributing to the momentum equation primarily via the thermal and magnetic pressure gradients; see Equation~2 in Paper~1, 
and \citet{demoulin:2009}, \citet{gulisano:2010}), i.e.\ the most relevant solar wind parameters are $\langle P_{p,th} \rangle_{sw}$, $\langle P_{mag} \rangle_{sw}$, and $\langle P_{tot} \rangle_{sw}$.
We find that $\langle P_{p,th} \rangle_{sw}$ is slightly overestimated in EUHFORIA, while $\langle P_{mag} \rangle_{sw}$ is slightly underestimated (as a consequence of the slight underestimation of $\langle B \rangle_{sw}$). However, since $\langle \beta_{p} \rangle_{sw}$ is higher than 1 at all heliocentric distances, i.e.\ the total pressure in the solar wind is primarily determined by the thermal pressure rather than by the magnetic pressure, we do not expect this modelling discrepancy to significantly affect the expansion of the CME. Most importantly, we see from Figure~\ref{fig:sw_radial_02}~(g) that $\langle P_{tot} \rangle_{sw}$ well matches $Wind$ observations at L1, and therefore the CME expansion in EUHFORIA is most probably not (or only weakly) affected by any discrepancy with in situ observations.

Overall, we can therefore conclude that the EUHFORIA modelling of the slow solar wind in the range $0.2-1.9$~au reproduces the observational expectations sufficiently well to ensure a realistic modelling of the CME radial evolution for what concerns the CME--solar wind interactions resulting in drag and expansion.

\begin{table*}
\centering
\begin{tabular}{l|ccccc} 
 \hline
 \textbf{Parameter}         & \textbf{Theory}
                            & \multicolumn{3}{c}{\textbf{Observations}}  & \textbf{Simulations} \\ 
 \cline{3-5}
                            &        & All   & Fast   & Slow  & Slow  \\ 
 \hline
 $a_{ v }$           & $0$          & $[0.036 , 0.083]$     & --                & --                & $0.04 \,\, \pm \,\, 0.01  \,\, ( 0.02 \,\, \pm \,\, 0.01)$ \\ 
 $a_{ n_p }$         & $-2$         & $[-2.10 , -2.01]$     & $[-2.03 , -1.89]$ & $-1.9$            & $-2.07 \,\, \pm \,\, 0.05  \,\, ( -2.14 \,\, \pm \,\, 0.05)$ \\ 
 $a_{ B }$           & $[-2 , 0]$   & $[-1.92 , -1.47]$     & $[-1.91 , -1.53]$ & $[-1.88 , -1.40]$ & $-1.55 \,\, \pm \,\, 0.09  \,\, ( -1.65 \,\, \pm \,\, 0.05)$ \\
 $a_{ T_p }$         & $-4/3$       & $-0.79$               & $[-0.9 , -0.69]$  & $[-1.21 , -0.58]$ & $-0.96 \,\, \pm \,\, 0.06  \,\, ( -1.03 \,\, \pm \,\, 0.02)$\\ 
 $a_{ P_{p,th} }$    & $-10/3$      & --                    & $[-3.0 , -2.8]$   & --                & $-2.77 \,\, \pm \,\, 0.16  \,\, ( -3.16 \,\, \pm \,\, 0.07)$ \\ 
 $a_{ P_{mag} }$     & $[-4 , 0]$   & --                    & $[-3.3 , -3.1]$   & --                & $-2.79 \,\, \pm \,\, 0.17  \,\, ( -3.31 \,\, \pm \,\, 0.11)$ \\ 
 $a_{ \beta_{p} }$   & $[-10/3 , 2/3]$ & --                 & $[0.3 , 0.5]$     & --                & $0.52 \,\, \pm \,\, 0.67  \,\, ( 0.15 \,\, \pm \,\, 0.10)$\\ 
 \hline
\end{tabular}
\caption{Summary of the various exponents $a_x$ describing the radial scaling of the mean solar wind parameters $\langle X \rangle_{sw}$ in the inner heliosphere based on theoretical arguments, \textit{Helios} observations, and EUHFORIA modelling results. The $[...,...]$ notation is used to indicate the lower and upper boundaries of each value range, as derived from theoretical and observational arguments.
In the rightmost column, we have indicated the uncertainties ($\pm$) on the simulated values based on $95 \%$ confidence intervals.
In this column, we have also indicated in brackets the values based on the median, in order to provide a comparison with the mean-based parameters used in the analysis.}
\label{tab:solar_wind}
\end{table*}

\section{Results: radial evolution of the CME}
\label{sec:results_cme}

\subsection{Identification of the sheath and ejecta boundaries in EUHFORIA}
\label{subsec:results_boundaries}

Characterising the radial evolution of the CME properties in EUHFORIA simulations
requires to first distinguish among the various CME substructures in EUHFORIA time series, {i.e.} the CME-driven shock, the CME sheath, and the magnetic ejecta. This passes through the determination of the shock time ($t_{shock}$), and of the start ($t_{start}$) and end time ($t_{end}$) of the magnetic ejecta from the EUHFORIA time series extracted at different heliocentric distances. 
To identify the arrival time of the CME-driven shock at various radial distances, we apply the following threshold conditions to EUHFORIA time series:
\begin{equation}
\left (v(t_i)-v_{sw}(t_i) \ge 20~\mathrm{km~s}^{-1} \right ) 
\land \left (\frac{n_p(t_i)}{n_{p,sw}(t_i)} \ge 1.2 \right ) 
\land \left (\frac{B(t_i)}{B_{sw}(t_i)} \ge 1.2 \right ),
\label{eqn:shock_identification}   
\end{equation}
where $t_i$ is a generic time in the time series,
$v(t)$, $n_p(t)$, $B(t)$ are the speed, proton number density and magnetic field magnitude time series obtained from the simulation of the CME with EUHFORIA, and $v_{sw}(t)$, $n_{p,sw}(t)$, $B_{sw}(t)$ are the speed, proton number density and magnetic field magnitude time series of the ambient solar wind as determined by performing one simulation of the ambient solar wind alone. We note that in our simulations, time series are characterised by a cadence of 
$\Delta t_{out} = t_i - t_{i-1} = 10$~min.
The arrival time of the CME shock $t_{shock}$ at each spacecraft is marked based on the earliest time $t_i$ in the EUHFORIA time series at which Equation~\ref{eqn:shock_identification} is satisfied. This set of conditions is an adapted version of the conditions used to detect interplanetary shocks from in situ solar wind measurements employed by the Heliospheric Shock Database, and it was chosen due to its reliability in detecting the arrival of CME fronts in EUHFORIA in situ time series at various heliocentric distances, as verified by their visual inspection (see also Figure~\ref{fig:spheromak_boundaries}).

In the inner heliosphere, several proxies have been used to identify magnetic ejectas from in situ solar wind plasma and magnetic field time series, including proton temperature \citep{lopez:1986, lopez:1987, richardson:1995}, proton and plasma $\beta$ \citep{richardson:1995, lepping:2005}, and the characteristics of the magnetic field \citep[e.g.][]{burlaga:1981, gulisano:2012}.
For completeness, we note that the identification of magnetic ejectas in the outer heliosphere (beyond $\sim 5$~au) is even more complex, both because of the blurring of some of their characteristics through interaction with the ambient solar wind, and because of the limited instrumentation precluding the observations of some of the key parameters on board of past and current missions \citep{richardson:j:2006}. As a result, the most reliable characteristics for tracing magnetic ejectas in the outer heliosphere are those less affected by the interaction with solar wind streams, e.g.\ the elemental abundance ratios and the relative charge state abundance for given elements \citep[e.g.][]{goldstein:1998, paularena:2001}.

At 1~au, a common proton $\beta$ threshold used to automatically identify magnetic ejectas 
from in situ solar wind measurements is $\beta_{p,obs} \le 0.3$ \citep[e.g.][]{lepping:2005}.
To be as consistent as possible with this observational approach, in this work we identify the start time $t_{start}$ and end time $t_{end}$ of CME/magnetic ejecta based on an similar analysis of the proton $\beta$ EUHFORIA time series.
Because of the fundamental differences between the actual solar wind characteristics and its single-fluid description performed by EUHFORIA, 
we test a wide range of thresholds, spanning between $\beta_p \le 0.1$ and $\beta_p \le 1.0$.
We find that a $\beta_p \le 0.5$ threshold provides the best compromise to allow a robust identification of the CME boundaries over a wide range of latitudes, 
i.e.\ between 0.2~au and 1.9~au (see Appendix~\ref{app:appendix} for the comparison of different thresholds). In order to remove possible spurious results before and after the CME passage, we limit the application of the $\beta_p$ threshold condition to times following the detection of the CME shock. 
We define the start time of the CME/magnetic ejecta $t_{start}$ as the first time when $\beta_p(t_i) \le 0.5$, with $t_i \ge t_{shock}$. 
Furthermore, the passage of the CME rear edge is marked by the first time at which $\beta_p$ passes from $\beta_p \le 0.5$ to $\beta_p > 0.5$, with $t_{end} \ge t_{start}$. We further impose that the $\beta_p > 0.5$ condition is met for a period of at least 1~hour in order to make sure to identify the same feature at various heliocentric distances (the results of this approach are shown in Figure~\ref{fig:spheromak_boundaries}).
We note that the use of a fixed $\beta_p$ threshold condition to identify the magnetic ejecta at different heliocentric distances is also justified by the almost constant solar wind $\beta_p$ with respect to the heliocentric distance (Figure~\ref{fig:sw_radial_02}~(h)), which provides a stable reference value to distinguish between the magnetic ejecta and the solar wind regardless of the heliocentric distance.

From the identification of the time of the shock and of the start and end times of the ejecta, we identify the periods associated with the sheath and magnetic ejecta at each virtual spacecraft. Figure~\ref{fig:spheromak_boundaries} in Appendix~\ref{app:appendix} shows the results obtained applying the shock identification conditions (Equation~\ref{eqn:shock_identification}) and chosen plasma $\beta_p$ threshold to EUHFORIA time series at various radial distances. 

\subsection{Radial evolution of mean CME values}

\begin{figure}
\centering
\includegraphics[width=0.95\hsize]{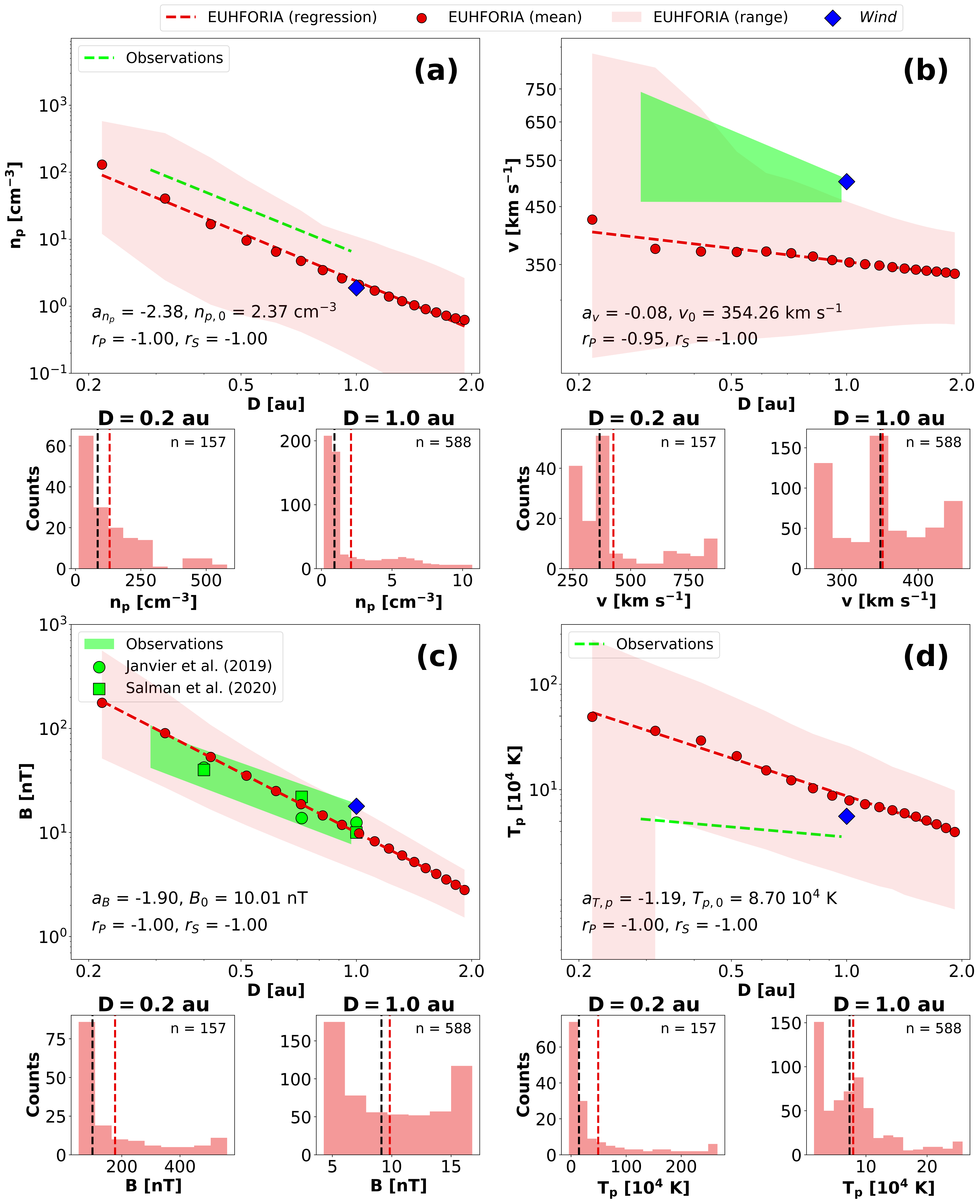}
\caption{Radial dependence of the mean CME parameters in EUHFORIA, extracted at virtual spacecraft located along the Sun--Earth line (red dots).
(a): speed $\langle v \rangle_\mathrm{CME}$. 
(b): proton number density $\langle n_p \rangle_\mathrm{CME}$.
(c): magnetic field $\langle B \rangle_\mathrm{CME}$.
(d): proton temperature $\langle T_p \rangle_\mathrm{CME}$.
The red shaded areas show the maximum variation of EUHFORIA predictions as a function of the heliocentric distance.
The results from the fitting of the mean values are indicated as dashed red lines.
Observation-based relations are indicated as dashed green lines or green shaded areas.
Observation-based values at selected heliocentric distances are indicated as green dots.
Mean values from $Wind$ are shown as blue diamonds.
Histograms showing the frequency of occurrence of each parameter are provided for $D=0.2$~au and $D=1.0$~au radial distances.
The mean and median are indicated by the red and black vertical dashed lines, respectively.
}
\label{fig:spheromak_radial_boundaries_01} 
\end{figure}

\begin{figure}
\centering
\includegraphics[width=0.95\hsize]{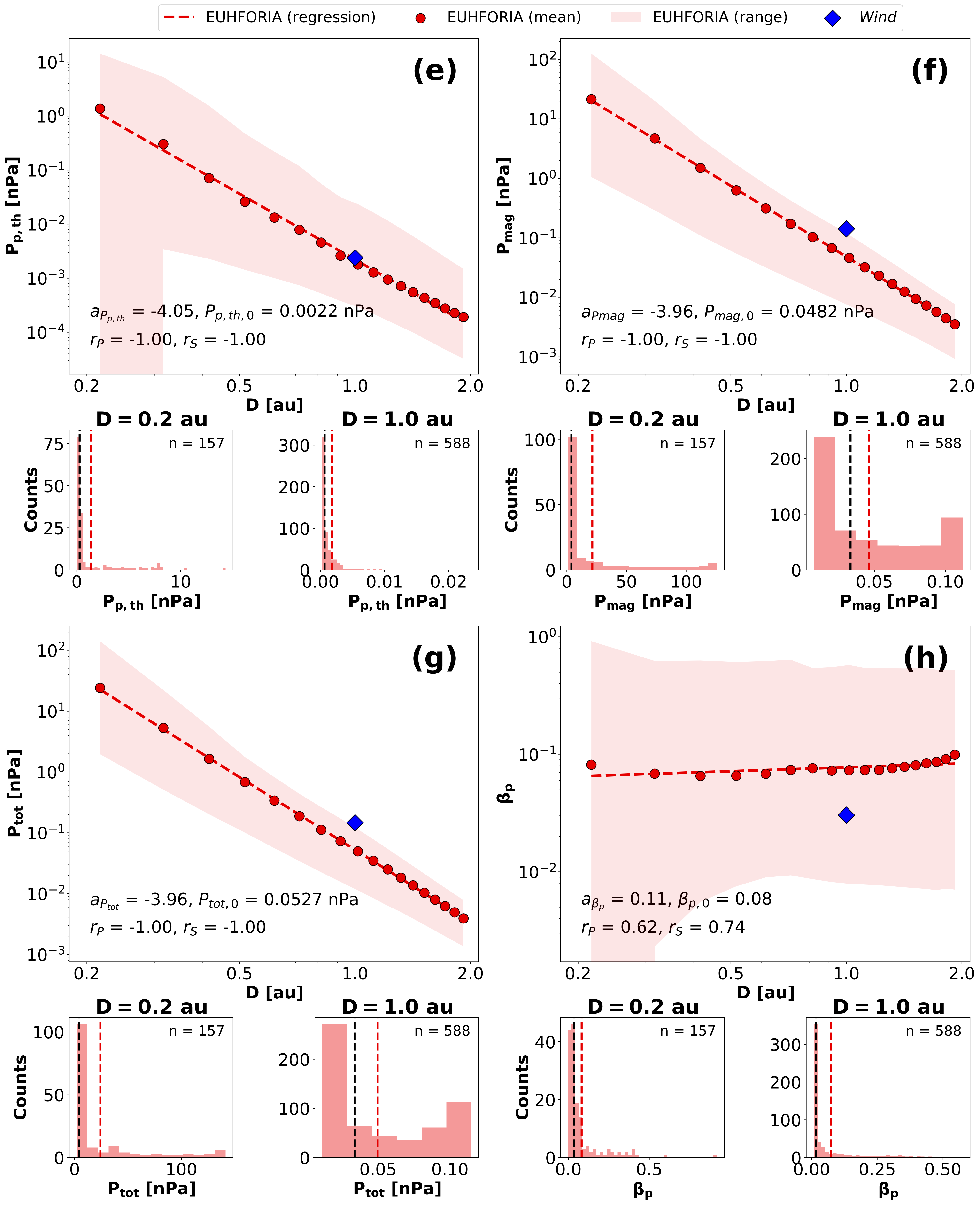} 
\caption{Radial dependence of the mean CME parameters in EUHFORIA, extracted at virtual spacecraft located along the Sun--Earth line (red dots).
(e): thermal pressure $\langle P_{th} \rangle_\mathrm{CME}$. 
(f): magnetic pressure $\langle P_{mag} \rangle_\mathrm{CME}$.
(g): total pressure $\langle P_{tot} \rangle_\mathrm{CME}$.
(h): proton $\beta$, $\langle \beta_p \rangle_\mathrm{CME}$.
The red shaded areas show the maximum variation of EUHFORIA predictions as a function of the heliocentric distance.
The results from the fitting of the mean values are indicated as dashed red lines.
Observation-based relations are indicated in green.
Mean values from $Wind$ are shown as blue diamonds.
Histograms showing the frequency of occurrence of each parameter are provided for $D=0.2$~au and $D=1.0$~au radial distances.
The mean and median are indicated by the red and black vertical dashed lines, respectively.
}
\label{fig:spheromak_radial_boundaries_02} 
\end{figure}

Based on the determination of the sheath and CME boundaries discussed above, we calculate the mean values of the parameters characterising the CME structure at various heliocentric distances, and compare their radial evolution with results reported in previous studies. As discussed below, we limit our attention to mean values because of the larger number of observational studies investigating mean values rather than median values \citep[despite the latter being more robust and resistant to outliers; see][]{janvier:2019}.
In particular, we focus on the determination of the CME 
mean magnetic field $\langle B \rangle_\mathrm{CME}$, 
mean proton number density $\langle n_p \rangle_\mathrm{CME}$,
mean speed $\langle v \rangle_\mathrm{CME}$,
and mean proton temperature $\langle T_p \rangle_\mathrm{CME}$.
Figures~\ref{fig:spheromak_radial_boundaries_01}~and~\ref{fig:spheromak_radial_boundaries_02} show the radial dependence of the mean CME properties in a similar manner as done in Figures~\ref{fig:sw_radial_01}~and~~\ref{fig:sw_radial_02} for the ambient solar wind. Table~\ref{tab:cme} reports the heliocentric distance dependencies ($a_x$ parameter in Equations~\ref{eqn:power_law_lin}~and~\ref{eqn:power_law_log}), derived from the fitting of the mean individual values at different heliocentric distances using Equation~\ref{eqn:power_law_lin} in log--log space. 
To assess the consistency between mean and median values, in Table~\ref{tab:cme} we also include the dependencies estimated from the median-based values. For both, we indicate the uncertainties derived from the linear fitting based on 95\% confidence intervals. In this case, the two methods give more different results than in the solar wind case, as also suggested by the different mean and median values reported in the small panels (for 0.2 and 1.0 au) in Figures~\ref{fig:spheromak_radial_boundaries_01} and \ref{fig:spheromak_radial_boundaries_02}. Because of the larger number of observational studies investigating mean values rather than median values, in the following we focus on mean values only. A comparison with mean-based observational values (when available) is also provided in Table~\ref{tab:cme}.

\begin{table*}
\centering
\begin{tabular}{l|cc} 
 \hline
 \textbf{Parameter}         & \textbf{Observations}  & \textbf{Simulations} \\ 
 \hline
 $a_{ v }$                  & $[-0.305 , -0.002]$  & $-0.08 \,\, \pm \,\, 0.01\,\, ( 0.01 \,\, \pm \,\, 0.02)$ \\ 
 $a_{ n_p }$                & $-2.32$              & $-2.38 \,\, \pm \,\, 0.11 \,\, ( -2.78 \,\, \pm \,\, 0.12)$ \\ 
 $a_{ B }$              & $[-1.95 , -1.30]$        & $-1.90 \,\, \pm \,\, 0.01 \,\, ( -1.72 \,\, \pm \,\, 0.06)$ \\ 
 $a_{ T_{p} }$              & $-0.32$              & $-1.19 \,\, \pm \,\, 0.05\,\, ( -0.79 \,\, \pm \,\, 0.15)$\\ 
 $a_{ P_{p,th} }$           & --                   & $-4.05 \,\, \pm \,\, 0.11\,\, ( -3.86 \,\, \pm \,\, 0.03)$ \\ 
 $a_{ P_{mag} }$            & --                   & $-3.96 \,\, \pm \,\, 0.03\,\, ( -3.44 \,\, \pm \,\, 0.12)$ \\ 
 $a_{ P_{tot} }$            & --                   & $-3.96 \,\, \pm \,\, 0.03\,\, ( -3.43 \,\, \pm \,\, 0.13)$\\ 
 $a_{ \beta_{p} }$          & --                   & $0.11 \,\, \pm \,\, 0.07\,\, ( -0.16 \,\, \pm \,\, 0.13)$\\ 
 \hline
 $a_{\Delta t_\mathrm{CME}}$        & $1.08$                    & $0.77 \,\, \pm \,\, 0.03$ \\ 
 $a_{S_\mathrm{CME}}$               & $[0.61 , 0.92]$           & $0.69 \,\, \pm \,\, 0.03$ \\ 
 $a_{\zeta_\mathrm{CME}}$           & $[0.16 , 1.7]$            & $-0.17 \,\, \pm \,\, 0.07$ \\ 
 \hline
\end{tabular}
\caption{Summary of the various exponents $a_x$ describing the radial scaling of the mean CME parameters $\langle X \rangle_\mathrm{CME}$ and of the CME duration/size in the inner heliosphere, based on observational results and EUHFORIA simulations.
The $[...,...]$ notation is used to indicate the lower and upper boundaries of each value range, as derived from observational studies. 
Only for the exponents provided in this table, in the rightmost column we have also indicated the uncertainties ($\pm$) on the simulated values based on $95 \%$ confidence intervals.
In this column, we have also indicated in brackets the values based on the median, in order to provide a comparison with the mean-based parameters used in the analysis.
}
\label{tab:cme}
\end{table*}

\textbf{Speed.}
The CME mean speed in EUHFORIA simulations is found to be mildly decreasing with heliocentric distance, with a slope of $a_{v} = -0.08$.
This result is consistent with values reported by \citet{liu:2005}, i.e.\ $-0.002 \pm 0.02$, and by \citet{salman:2020}, i.e.\ $-0.305 \pm 0.2$.
At 1~au, the CME mean speed from EUHFORIA ($354$~km~s$^{-1}$) is lower than observed at $Wind$ ($488$~km~s$^{-1}$) by $\sim 135$~km~s$^{-1}$, and the ratio of modelled and observed values is equal to $0.73$. Such underestimation of the CME impact speed at 1~AU can be explained by the slower-than-observed (higher-than-observed) speed (density) in the modelled solar wind (as discussed in Section~\ref{sec:results_sw}), which led to a higher drag in our simulation than occurred in reality. An additional possible concause to such a discrepancy is a slight underestimation in the CME initial speed based on coronal images.

\textbf{Density.}
The CME mean proton density in EUHFORIA simulations decreases with heliocentric distance. The fall off is well described by a power law with exponent $a_{n_{p}} = -2.38$, which is similar with the value of $-2.32 \pm 0.07$ reported by \citet{liu:2005} based on a statistical study. 
At 1~au, the CME mean proton density modelled with EUHFORIA ($2.37$~cm$^{-3}$) is similar to the one observed by $Wind$ ($1.90$~cm$^{-3}$), with the ratio of modelled and observed values equal to $1.25$.

\textbf{Magnetic field magnitude.}
The CME mean magnetic field magnitude in EUHFORIA simulations is found to decrease with heliocentric distance, 
with fall off of $a_{B} = -1.90$.
This value is well within the error bars of the values reported by \citet{gulisano:2010}, i.e.\
$-1.89 \pm 0.10$ for perturbed CMEs, 
$-1.85 \pm 0.11$ for non-perturbed CMEs, and
$-1.85 \pm 0.07$ for the full set.
On the other hand, our value is higher than those reported in earlier studies by \citet{liu:2005, wang:2005, leitner:2007}, 
who recovered exponents between $-1.30 \pm 0.09$ \citep{leitner:2007} and $-1.52$ \citep{wang:2005}.
As pointed out by \citet{gulisano:2010}, the discrepancy among different observational studies is most probably the result of different selection criteria used to identify the CME events, and of the different ranges of heliocentric distances considered in various studies.
More recently, \citet{winslow:2015} reported an exponent of $-1.95 \pm 0.19$ between Mercury and 1~au, while \citet{janvier:2019} and \citet{salman:2020} did not perform a fitting of the CME mean magnetic field at different heliocentric distances. However, as visible in panel~(c) of Figure~\ref{fig:spheromak_radial_boundaries_01}, their values at selected heliocentric distances appear overall consistent with our modeling results and with previous observational studies.
At 1~au, the mean CME magnetic field magnitude from EUHFORIA ($10.0$~nT) is underestimated with respect to the value observed at $Wind$ ($16.5$~nT), and the ratio of modelled and observed values is $\sim 0.61$.
In this respect, we note that the CME internal magnetic field has traditionally been the most difficult parameter to reproduce in numerical simulations, because of the severe observational limitations affecting our knowledge of the global magnetic structure of CMEs and ICMEs \citep[e.g.][]{owens:2016, wood:2017}.
The results shown in Figure~\ref{fig:euhforia_earth} are in line with this general trend, and additionally represent a significant improvement compared to other modelling attempts for the same CME event \citep[e.g.][]{shen:2014, singh:2020}. We refer the reader to Paper~1 for a more detailed discussion on the observational limitations and uncertainties affecting to the reproduction of the CME internal magnetic field.

\textbf{Temperature.}
The CME mean temperature in EUHFORIA simulations decreases with heliocentric distance with exponent $a_{T_p} = -1.19$, which is significantly faster than reported by \citet{liu:2005}, i.e.\ $-0.32 \pm 0.06$.
At 1~au, the CME mean proton temperature from EUHFORIA ($8.70 \times 10^{4}$~K) is higher than observed at $Wind$ ($5.44 \times 10^{4}$~K), and the ratio of modelled and observed values is $\sim 1.60$.

\textbf{Pressures and proton $\beta$.}
The CME mean proton thermal pressure in EUHFORIA simulations decreases with heliocentric distance with a slope of $a_{P_p} = -4.05$,
while CME mean magnetic pressure decreases with heliocentric distance with a slope of $a_{P_{mag}} = -3.96$.
As a result, the mean proton $\beta$ shows a mild increase with heliocentric distance, with a slope of $a_{\beta_p} = 0.11$ ($r_P = 0.62, \,\, r_S = 0.74$). 
We note that this behaviour might be a result of the particular threshold condition imposed to identify the magnetic ejecta in EUHFORIA time series, as also suggested by the histogram distributions in Figure~\ref{fig:spheromak_radial_boundaries_02}~(h) showing the median shifting towards lower values between 0.2 and 1.0~au, while the mean value remains approximately constant.
Because of the lack of statistical studies investigating the radial evolution of the pressure terms and of the proton and plasma $\beta$ in the case of real CMEs, we cannot directly compare the results from our simulations with observational values at other distances than 1~au. 
We note that 1~au, the mean pressure terms inside the magnetic ejecta in EUHFORIA simulations are underestimated compared to the observed values at $Wind$.
In particular, the ratio between modelled and observed values are:
$0.92$ for the mean proton thermal pressure, 
$0.34$ for the mean magnetic pressure, 
$0.36$ for the mean total pressure. 
The resulting mean $\beta_p$ ratio is $2.63$.

\subsection{Radial evolution of sheath and CME sizes}
\label{subsec:results_sizes}

Similarly to \citet{gulisano:2010}, we fit the temporal speed profile $v_\mathrm{CME}(t)$ in the magnetic ejecta in both EUHFORIA and $Wind$ time series 
using a least square fit with a linear function of time, i.e.\
\begin{equation}
    v_{\mathrm{CME}, fit} (t) = v_{ fit} \cdot t + v_{\mathrm{CME}, 0},
    \label{eqn:v_fit} 
\end{equation}
where $t$ is the time since the start of the magnetic ejecta, 
and $v_{fit}$ is the slope of the fitted linear function. 
The linear fit is used to define the following velocities: 
the velocity of the CME front $v_{start} = v_{\mathrm{CME}, fit}(t_{start})$,
the velocity of the CME back $v_{end} = v_{\mathrm{CME}, fit}(t_{end})$,
and the velocity of the CME centre $v_{c} = v_{\mathrm{CME}, fit}(t_{c})$ with $t_c = (t_{start}+t_{end})/2$.
As visible from Figure~\ref{fig:spheromak_zeta}, $v_{start}$, $v_{end}$, and $v_{c}$ are generally close to the non-fitted velocities 
$v_\mathrm{CME}(t_{start})$, $v_\mathrm{CME}(t_{end})$, and $v_\mathrm{CME}(t_{c})$, although the exact differences depend on the particular heliocentric distance considered.
We further define the full expansion velocity of the magnetic ejecta as
$\Delta v_\mathrm{CME} \simeq v_{\mathrm{CME}, fit}(t_{start}) - v_{\mathrm{CME}, fit}(t_{end}) $, 
and the magnetic ejecta duration as $\Delta t_\mathrm{CME} = t_{end} - t_{start}$.
We also compute an estimate of the CME size as $S_\mathrm{CME} = v_{c} \cdot \Delta t_\mathrm{CME}$ (we note that this is likely a slight overestimation of the CME size due to CME ageing, further discussed in Section~\ref{sec:results_interpretation}).

In the sheath region, we calculate the mean sheath velocity $\langle v \rangle _\mathrm{sh}$ and the duration of the sheath as $\Delta t_\mathrm{sh} = t_{start} - t_{shock}$.
These two quantities are used to compute an estimate of the sheath size, as $S_\mathrm{sh} = \langle v \rangle _\mathrm{sh} \cdot \Delta t_\mathrm{sh}$.

\begin{figure}
\centering
{ \includegraphics[width=0.95\hsize]{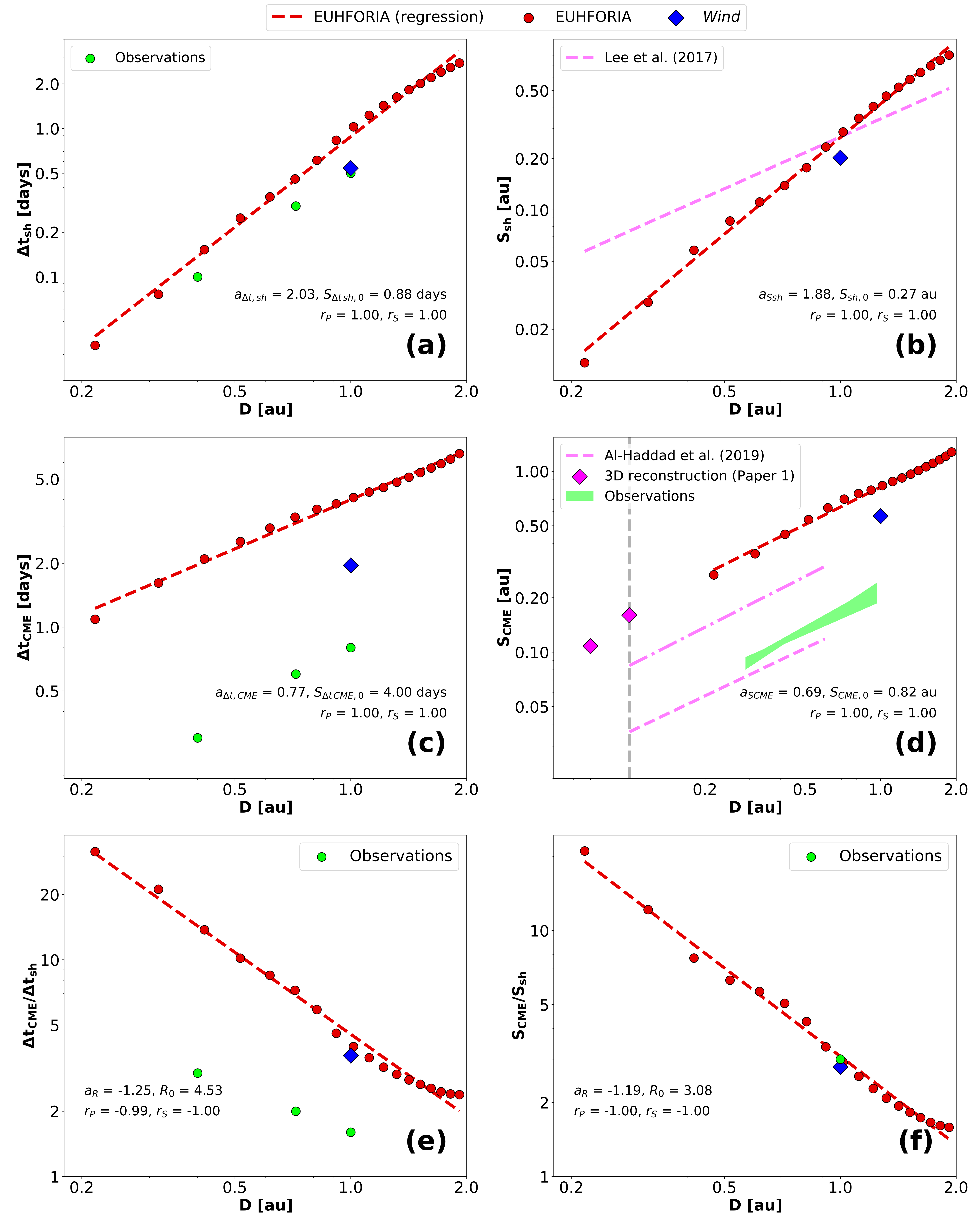}}
\caption{
Durations and radial sizes of the CME and its driven sheath as a function of $D$, based on EUHFORIA simulations.
(a): sheath duration.
(b): sheath size.
(c): ejecta duration determined starting from the condition $\beta_p \le 0.5$.
(d): ejecta size determined starting from the condition $\beta_p \le 0.5$.
(e): ejecta to sheath duration ratio.
(f): ejecta to sheath size ratio.
In panels (a), (c), (e) and (f), observational values at selected heliocentric distances are indicated as green dots, 
based on \citet{masias:2016} and \citet{janvier:2019}.
In panel (b), the sheath size expected from Equation~8 in \citet{lee:2017} is indicated as a magenta dashed line.
In panel (d), the range of values recovered from observation-based relations are indicated as a green shaded area.
The values recovered from \citet{alhaddad:2019} based on numerical simulations are indicated as dashed and dotted-dashed magenta lines.
Observation-based values in the solar corona reconstructed in Paper~1 are shown as magenta diamonds.
The position of the EUHFORIA inner boundary is indicated as a dashed grey line. 
Values observed at $Wind$ are shown as blue diamonds.
}
\label{fig:cme_size} 
\end{figure}

Figure \ref{fig:cme_size} shows the duration and size of the sheath and of the magnetic ejecta, as a function of the heliocentric distance $D$.
Table~\ref{tab:cme} reports the radial dependencies ($a_x$ parameter in Equations~\ref{eqn:power_law_lin} and~\ref{eqn:power_law_log}) derived from the fitting of the individual values at different heliocentric distances using Equation~\ref{eqn:power_law_lin} in log--log space. A comparison with theoretical and observational values (when available) is provided. 

\textbf{CME duration and size.} 
The linear fitting of the CME duration and size in log--log scale from EUHFORIA simulation provide slopes of 
$a_{\Delta t_\mathrm{CME}} = 0.77$ and $a_{S_\mathrm{CME}} = 0.69$.
For comparison, fitting the values reported by \citet{janvier:2019} at selected planetary locations, 
we recover a slope for the evolution of the CME duration with heliocentric distance equal to $1.08$;
slopes for the evolution of the CME size with heliocentric distance reported in previous studies range between $0.61$ \citep{wang:2005} and $0.92$ \citep{liu:2005}.
Distinguishing between perturbed and non-perturbed CMEs, 
\citet{gulisano:2010} reported a size slope of $0.89 \pm 0.15$ for non-perturbed magnetic ejectas, 
and a slope of $0.45 \pm 0.16$ for perturbed magnetic ejectas, with a value of $0.78 \pm 0.12$ for the full set of magnetic ejectas (perturbed and non-perturbed) considered (see Equation~3 therein). In this regard, EUHFORIA results fall between the two categories, i.e.\ the lower limit of non-perturbed ones, and the upper limit of perturbed ones.
The growth of the CME size in time in our simulation appears overall consistent with typical observations.

However, we note that in the EUHFORIA simulation, the magnetic ejecta results significantly more extended both in time and in the radial direction than typically reported from in situ observations and from self-consistent numerical simulations of flux-rope CMEs in the corona and heliosphere \citep{alhaddad:2019}. 
This can be seen by comparing the intercept of the linear fitting $S_{\mathrm{CME},0}$ with that reported by previous studies. 
In EUHFORIA, we obtain $S_{\mathrm{CME},0}=0.82$~au and a CME duration of $\Delta t_{\mathrm{CME},0}=4.00$~days.
For comparison, the values reported in literature range between $0.16$~au \citep[][perturbed CMEs]{gulisano:2010} and $0.32$~au \citep[][non-perturbed CMEs]{gulisano:2010}, i.e.\ the CME size in EUHFORIA is between 2.56 and 5.13 times larger than typically observed.
\citet{janvier:2019} reported a typical CME duration of $0.8$~days at 1~au, meaning that the duration of the CME in EUHFORIA is about 5 times longer than typically observed.
The overestimation in the CME size in EUHFORIA is also apparent when we compare the modelled values with in situ observations at $Wind$ (1~au) for the specific CME considered: in fact, the CME duration in EUHFORIA is 2.1 times longer than observed, while the CME size is about 1.5 times larger.

\textbf{Sheath duration and size.} 
The linear fitting of the sheath duration and size in log--log scale from EUHFORIA simulation provide slopes of 
$a_{\Delta t_\mathrm{sh}} = 2.03$ and $a_{S_\mathrm{sh}} = 1.88$.
For comparison, fitting the values provided by \citet{janvier:2019} at selected planetary locations, 
we recover a slope of $1.77$ for the growth of the CME-driven sheath duration.
We also calculate the upper limit of the expected sheath size as predicted by Equation~8 in \citet{lee:2017},
with $\theta_\mathrm{CME}=0^\circ$, $\gamma=1.5$, $R_\mathrm{CME} = D \, \tan(\omega_\mathrm{CME}/2)$ as in Paper~1, 
and the Alvf\'{e}n Mach number $M_{A}$ calculated similarly to \citet{scolini:2020b}. 
This relation predicted a slope of $0.99$. 
The sheath size in EUHFORIA simulations therefore grows faster than expected, although values near 1~au appear consistent with theoretical expectations.
On the other hand, the growth in the duration $\Delta t_\mathrm{sh}$ ($a_{\Delta t_\mathrm{sh}} = 2.03$) well matches the observational trend, making us confident of the performance of the model in reproducing this particular aspect of CME evolution (as the duration of a structure from in situ time series provides a more direct comparison to observations than its inferred size).

In the EUHFORIA simulation, the sheath also results significantly more extended in the radial direction than observed in situ.
This can be seen by comparing the intercept of the linear fitting $S_{\mathrm{sh},0}$ with that reported by previous studies. 
In EUHFORIA, we obtain $S_{\mathrm{sh},0}=0.27$~au.
The sheath duration is $\Delta t_{\mathrm{sh},0}=0.88$~days.
For comparison, \citet{janvier:2019} reported a typical sheath duration of $0.5$~days at 1~au, i.e.\ the duration of the sheath size in EUHFORIA is about 1.75 times larger than typically observed.
This result is also confirmed for the specific CME considered, with the sheath duration in EUHFORIA that is about 1.9 times longer than observed at $Wind$, while the sheath size is about 1.4 times larger.

\textbf{Duration and size ratio of CME substructures.} 
We find that the ejecta to sheath duration and size ratios both decrease with heliocentric distance 
(with fall offs of $a_{R_{\Delta t}} = -1.25 $ and $a_{R_S} = -1.19$), reflecting the different rate of duration/size increase in the sheath and magnetic ejecta, with the sheath growing faster than the ejecta. 
This result is consistent with those reported by \citet{janvier:2019} based on actual observations of interplanetary CMEs between Mercury and 1~au. 
The nature of the size increase is interpreted in different ways for the sheath and the magnetic ejecta.
While both structures expand in the solar wind while propagating away from the Sun, the faster increase in the sheath size is interpreted as the result of the additional solar wind piling up via a ``snow plow'' effect. The ultimate thickness of the sheath therefore depends on the amount of plasma and magnetic field that is accumulated at the front of the CME into the sheath, and on the amount of it that is able to escape towards the CME sides, in addition to the aforementioned intrinsic expansion \citep{siscoe:2008}. 
Additional magnetic reconnection phenomena leading to erosion/flux injection effects \citep{demoulin:2016} are also contributing to reducing/increasing the size of both the sheath and the magnetic ejecta, although a lesser extent.
Overall, it seems more plausible for the increase of the sheath duration compared with that of the magnetic ejecta to be due to a pile-up of material combined with sheath expansion, rather than to magnetic reconnection phenomena.

\subsection{Radial evolution of CME expansion}
\label{subsec:results_expansion}

\begin{figure}
\centering
{\includegraphics[width=.48\hsize]{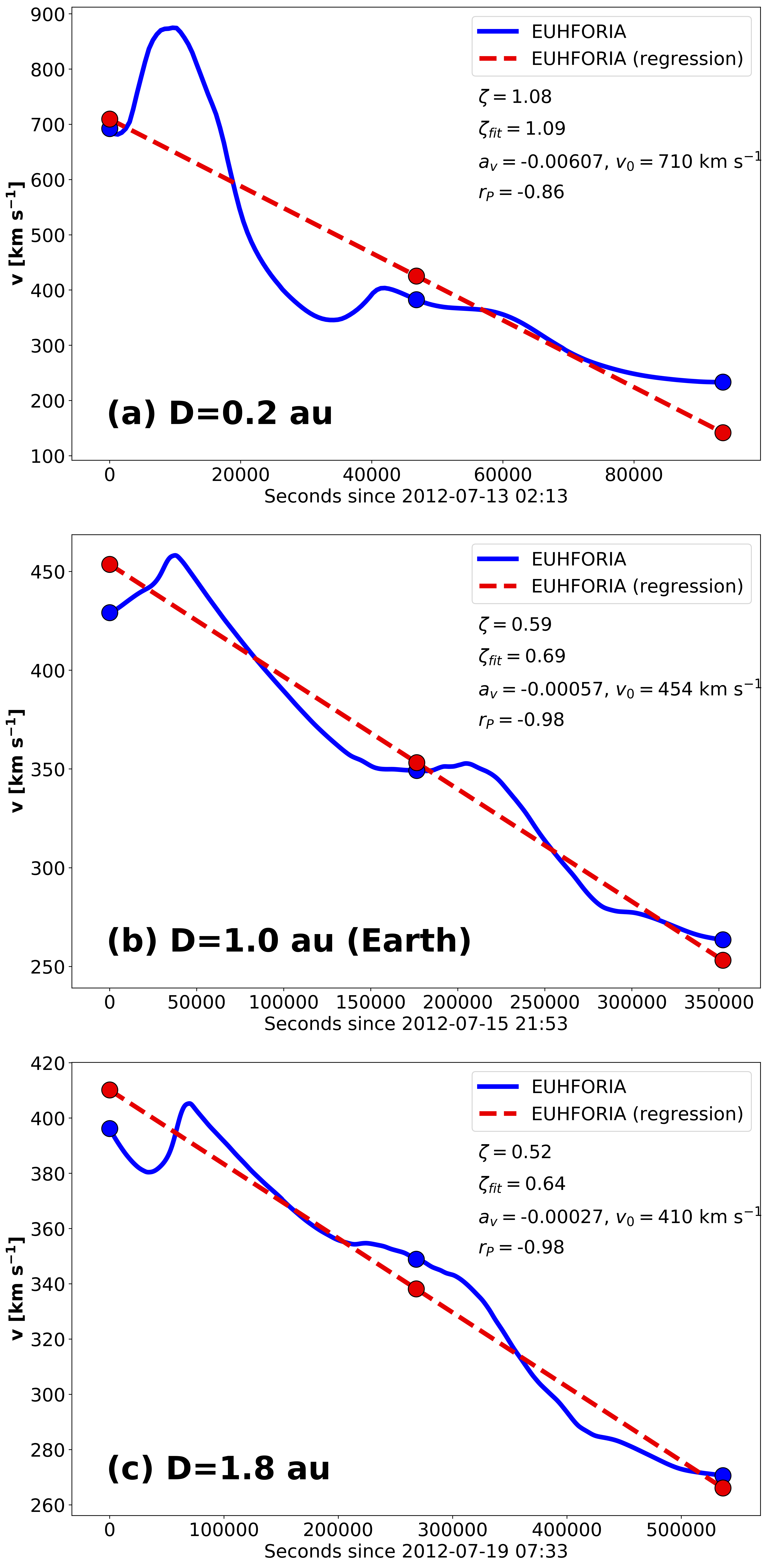}
\includegraphics[width=.483\hsize]{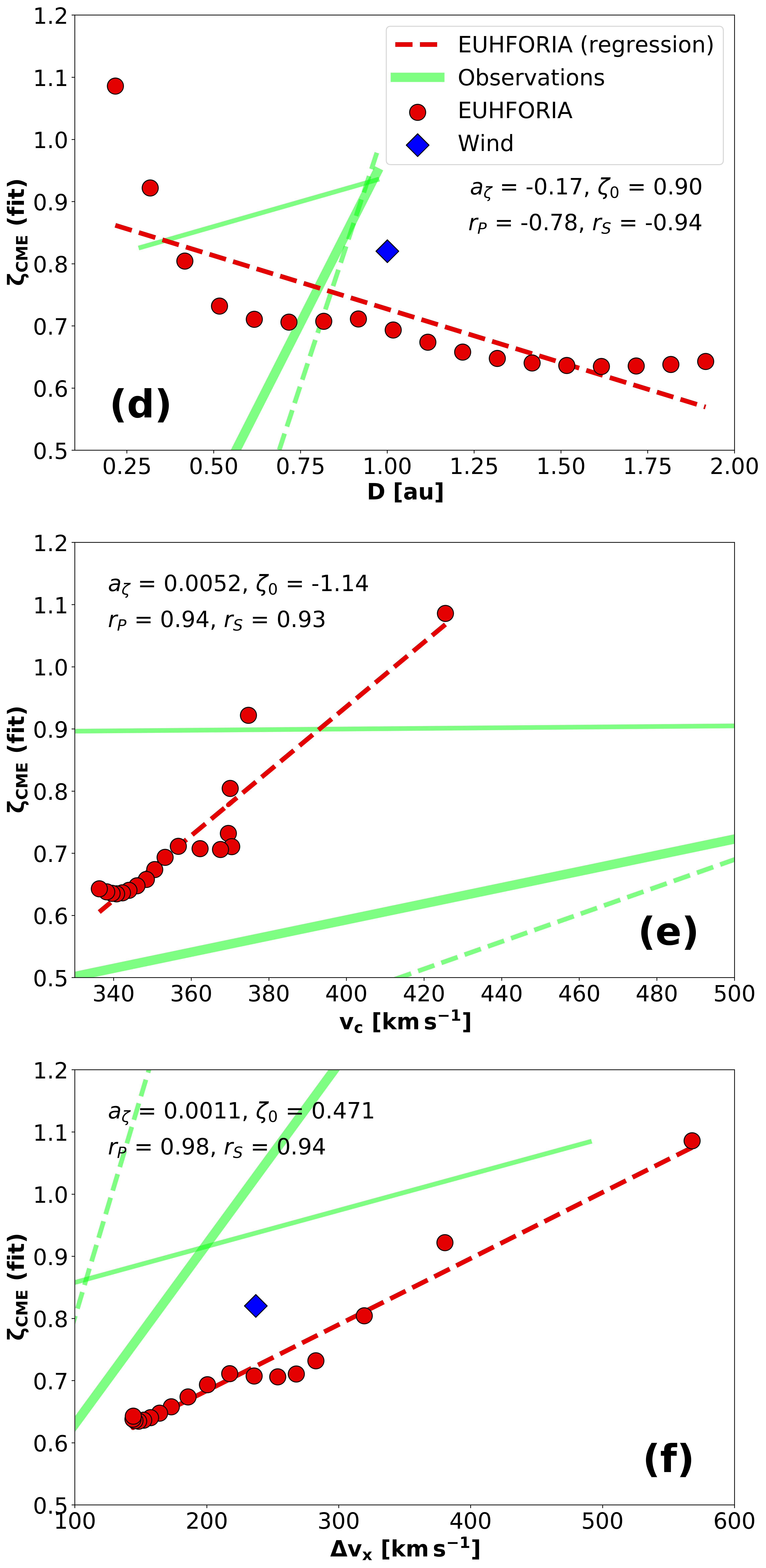}}
\caption{
Left column: 
derivation of the non-dimensional CME expansion rate $\zeta$ from the linear fitting of the CME speed profile in EUHFORIA simulations at $D=0.2$~au (a), $D=1.0$~au (b), and $D=1.8$~au (c).
Right column: 
non-dimensional CME expansion rate $\zeta_\mathrm{CME}$ as a function of $D$ (d), 
$v_\mathrm{c,CME}$ (e), 
and $\Delta v_\mathrm{x,CME}$ (f).
Values observed at $Wind$ are shown as blue diamonds.
The green lines represent the fit for perturbed CMEs (dashed line), non-perturbed CME (thin continuous line) and for both sets of CMEs (thick continuous line) as reported by \citet{gulisano:2010}.
}
\label{fig:spheromak_zeta} 
\end{figure}

One key parameter providing insights on the radial evolution of interplanetary CMEs is the non-dimensional CME expansion rate, 
defined by \citet{demoulin:2008} as
\begin{equation}
    \zeta_\mathrm{CME} = \frac{\Delta v_\mathrm{CME}}{\Delta t_\mathrm{CME}} \frac{D}{v_c^2}.
\end{equation}
We compute $\zeta_\mathrm{CME}$ at different heliocentric distances in EUHFORIA, and compare it with the value estimated from $Wind$ observations at 1~au.
This dimensionless parameter takes into consideration that faster (higher $v_c$) and longer (higher $\Delta t_\mathrm{CME}$) CMEs have higher expansion speeds (higher $\Delta v_\mathrm{CME}$), while the normalisation by the heliocentric distance considered ($D$) allows to compare observations at different distances from the Sun.
From observations, this parameter is found to range between $0.5$ and 1.5 for most non-perturbed CMEs
\citep[with a median value around 0.8; see][]{demoulin:2008, demoulin:2010, gulisano:2010}
and between $-1$ and $2$ for perturbed CMEs \citep{gulisano:2010}.
In EUHFORIA, we find that $\zeta_\mathrm{CME}$ varies between 0.63 and 1.09 depending on the heliocentric distance considered (Figure~\ref{fig:spheromak_zeta}~(d)). 
From $Wind$ observations, we find $\zeta_\mathrm{CME} = 0.82$ by fitting the observed speed profile with a linear fit throughout the whole magnetic ejecta period as indicated in Figure~\ref{fig:20120712_wind}.
Expected, observed and modelled values are overall consistent among each other. 
The implications of the recovered expansion rates for the global evolution of the CME under study are further discussed in Section~\ref{sec:results_interpretation}. 

\textbf{Correlations.}
Theoretically, $\zeta_\mathrm{CME}$ is expected to be independent from $\Delta t_\mathrm{CME}$, $v_c$ and $D$ in the case of non-perturbed CMEs.
In EUHFORIA simulations, we observe this is not the case (as reported in Figure \ref{fig:spheromak_zeta}).
In fact, the linear fitting of $\zeta_\mathrm{CME}$ as a function of the heliocentric distance provide correlation coefficients of $r_P=-0.78$ and $r_S=-0.94$, indicating the two are strongly (anti-)correlated. For comparison, \citet{gulisano:2010} reported a very weak correlation for non-perturbed magnetic ejectas, and a modest correlation ($r_P=0.49$) for perturbed ones. 
The fitting of $\zeta_\mathrm{CME}$ as a function of $v_c$ provides $r_P=0.94$ and $r_S=0.93$, indicating a strong correlation of the two parameters. On the other hand, \citet{gulisano:2010} reported a weak correlation ($r_P \le  0.3$) regardless of the magnetic ejecta type considered.
Finally, fitting $\zeta_\mathrm{CME}$ as a function of $\Delta v_\mathrm{CME}$ we report correlation coefficients of $r_P=0.98$ and $r_S=0.94$, 
meaning in EUHFORIA the two parameters are very strongly correlated. For comparison, \citet{gulisano:2010} reported a strong correlation coefficient ($r_P=0.79$) in the case of perturbed magnetic ejectas only.
This analysis suggests that the propagation behaviour of the CME in EUHFORIA simulations resembles that of perturbed magnetic ejectas.
This is also confirmed by the left column panels in Figure~\ref{fig:spheromak_zeta} (panels (a)--(c)), where we observe that in EUHFORIA, the CME speed profile is characterised by an irregular speed decrease that is significantly deviating from a monotonic decrease, as observed in the case of perturbed CMEs. 
This behaviour is observed at all the heliocentric distances considered, with more perturbed conditions closer to the Sun, where the fitting of the CME speed time profile $v_\mathrm{CME}(t)$ with a straight line gives a minimum Pearson correlation coefficient of $r_P = -0.86$. The physical origin of this perturbation in EUHFORIA simulations is investigated in the following paragraphs and Section~\ref{sec:results_interpretation}. 

\begin{figure}
\centering
{\includegraphics[width=0.6\hsize]{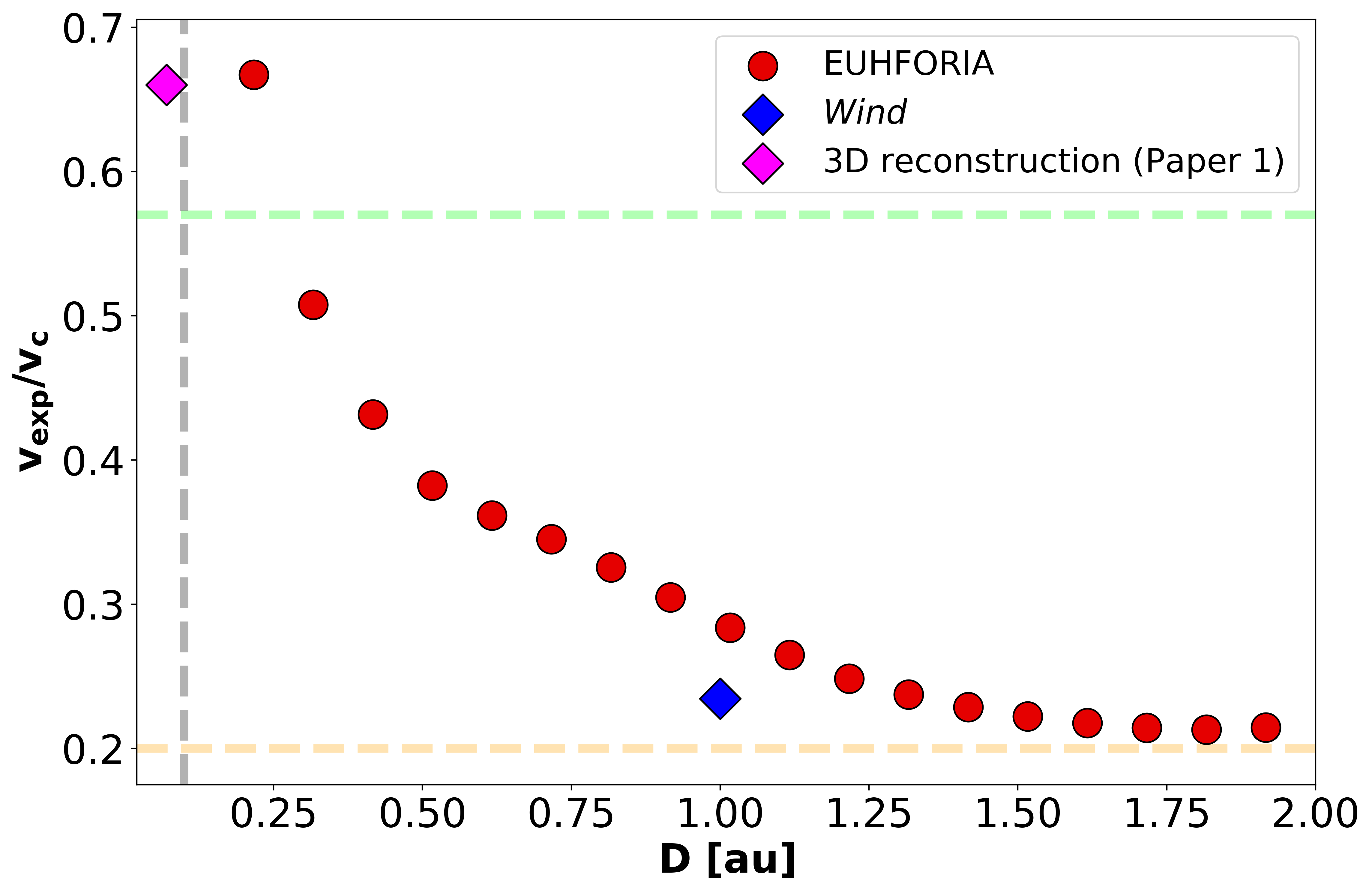}}
\caption{
Ratio between the expansion speed $v_{exp}$ and the translation speed $v_{trans}$ of the CME, as a function of the heliocentric distances.
The reconstructed value in the corona (at $14.9$~solar radii, corresponding to 0.07~au) from Paper~1 is indicated as a magenta diamond.
The observed value at $Wind$ is indicated as a blue diamond. 
Reference values measured in the corona \citep[$< 0.1$~au, from][]{gopalswamy:2009} and at 1~au \citep[from][]{lugaz:2020} are indicated as green and orange dashed lines, respectively. The location of the EUHFORIA heliospheric inner boundary is indicated by the dashed vertical grey line.
}
\label{fig:spheromak_vexp_vrad} 
\end{figure}

\textbf{Comparison of expansion and translation speeds.}
To further explore the nature of the CME, we compute at all heliocentric distances the relative importance of its expansion speed, 
calculated as $v_{exp} = \Delta v_\mathrm{CME} / 2$, and of its translation speed $v_{trans} \simeq v_c$.
Figure~\ref{fig:spheromak_vexp_vrad} summarises the results. We observe that closer to the Sun, the ratio of the expansion and translation speeds decreases faster than farther away.
In particular, close to the model inner boundary (i.e.\ at $D=0.2$~au), the expansion speed accounts for $67$\% of the translation speed, dropping by almost 50\% between 0.2 and 0.7~au, where it reaches a value of $35$\%. Between 0.7~au and $D=1.9$~au the decrease is more modest, reaching $20$\% at the outer distance considered. At 1~au the ratio drops to $28$\% (100~km~s$^{-1}$ vs 354~km~s$^{-1}$).
For comparison, the 3D reconstruction of the CME kinematics performed in Paper~1 estimated the expansion speed being $66$\% of the translation speed (503~km~s$^{-1}$ vs 763~km~s$^{-1}$) at a height of about 0.07~au ($\sim 14.9$~solar~radii), which is similar to the value extracted from EUHFORIA simulations at 0.2~au. At 1~au, $Wind$ data indicate the expansion speed accounted for $\sim 22$\% of the translation speed at 1~au (110~km~s$^{-1}$ vs 488~km~s$^{-1}$), i.e.\ a fraction of only $8$~percentage~points lower than estimated with EUHFORIA. 
These numbers are also consistent with the statistical observational picture, with previous studies reporting that the expansion speed is typically $\sim 57$\% of the translation speed of a CME in the corona \citep[below 0.1~au;][]{dallago:2003, schwenn:2005, gopalswamy:2009b}. At 1~au, the typical fraction drops below 20\% \citep{lugaz:2020}. Simulations and observations therefore agree in supporting a picture where the contribution of the expansion speed reduces as the CME approaches a more relaxed state toward the equilibrium with the surrounding solar wind, as further discussed in Section~\ref{subsec:results_magnetic_profile}.

\subsection{Radial evolution of CME magnetic field profile}
\label{subsec:results_magnetic_profile}

Recent works by \citet{masias:2016}, \citet{janvier:2019}, and \citet{demoulin:2020} highlighted the close relation between the asymmetry in the magnetic field time profile within a magnetic ejecta, and the spatial asymmetries of the magnetic structure possibly related to its level of relaxation. 
In this work, we investigate this relation in EUHFORIA simulations and we quantify the asymmetry in terms of two parameters: the time shift of the magnetic field peak compared to the centre of the magnetic ejecta ($\Delta t_{max}$), and the asymmetry parameter ($\Delta t_{asym}$), corresponding to the first moment of the magnetic field within the magnetic ejecta, introduced by \citet{janvier:2019}.

\bigskip
First, we define the time shift of the peak of the magnetic field with respect to the centre of the magnetic ejecta, as
\begin{equation}
    \Delta t_{max} = t_{max} - \frac{t_{start} + t_{end}}{2}.
\end{equation}
In order to account for the progressive expansion of the CME structure and to compare the results at different heliocentric distances, 
we further normalise $\Delta t_{max}$ with respect to the magnetic ejecta duration, as 
\begin{equation}
    \Delta t_{max,norm} = \frac{\Delta t_{max}}{\Delta t_\mathrm{CME}}.
    \label{eqn:magnetic_field_peak_shift_norm}
\end{equation}

\bigskip
Second, we compute the asymmetry parameter in the same way as \citet{janvier:2019} and \citet{lanabere:2020}, i.e.\ 
\begin{equation}
    \Delta t_{asym} = \frac{1}{\Delta t_\mathrm{CME} \, \langle B \rangle_\mathrm{CME}} 
    \int^{t_{end}}_{t_{start}} \,  \left (  t - \frac{t_{start} + t_{end}}{2} \right ) B(t) \, dt,
    \label{eqn:asymmetry_parameter}
\end{equation}
which we compute from discrete EUHFORIA time series as 
\begin{equation}
    \Delta t_{asym} \simeq \frac{1}{\Delta t_\mathrm{CME} \, \langle B \rangle_\mathrm{CME}} \sum^{t_{end}}_{t_{start}}\mathop{}_{\mkern-5mu t_i}
    \left (  t_i - \frac{t_{start} + t_{end}}{2} \right ) B_i \, (t_i-t_{i-1}).
    \label{eqn:asymmetry_parameter_discrete}
\end{equation}
In order to account for the progressive expansion of the CME structure during propagation, and to compare the results at different heliocentric distances, 
we further normalise $\Delta t_{asym}$ with respect to the magnetic ejecta duration, as 
\begin{equation}
    \Delta t_{asym,norm} = \frac{\Delta t_{asym}}{\Delta t_\mathrm{CME}}.
    \label{eqn:asymmetry_parameter_discrete_norm}
\end{equation}
We note that $\Delta t_{asym,norm}$ is a more robust indicator than $\Delta t_{max,norm}$ due to the fact that it considers the whole magnetic field profile (through the integral over time), while $\Delta t_{max,norm}$ only considers the single point associated with the peak of the magnetic field profile.

\begin{figure}
\centering
{\includegraphics[width=.8\hsize]{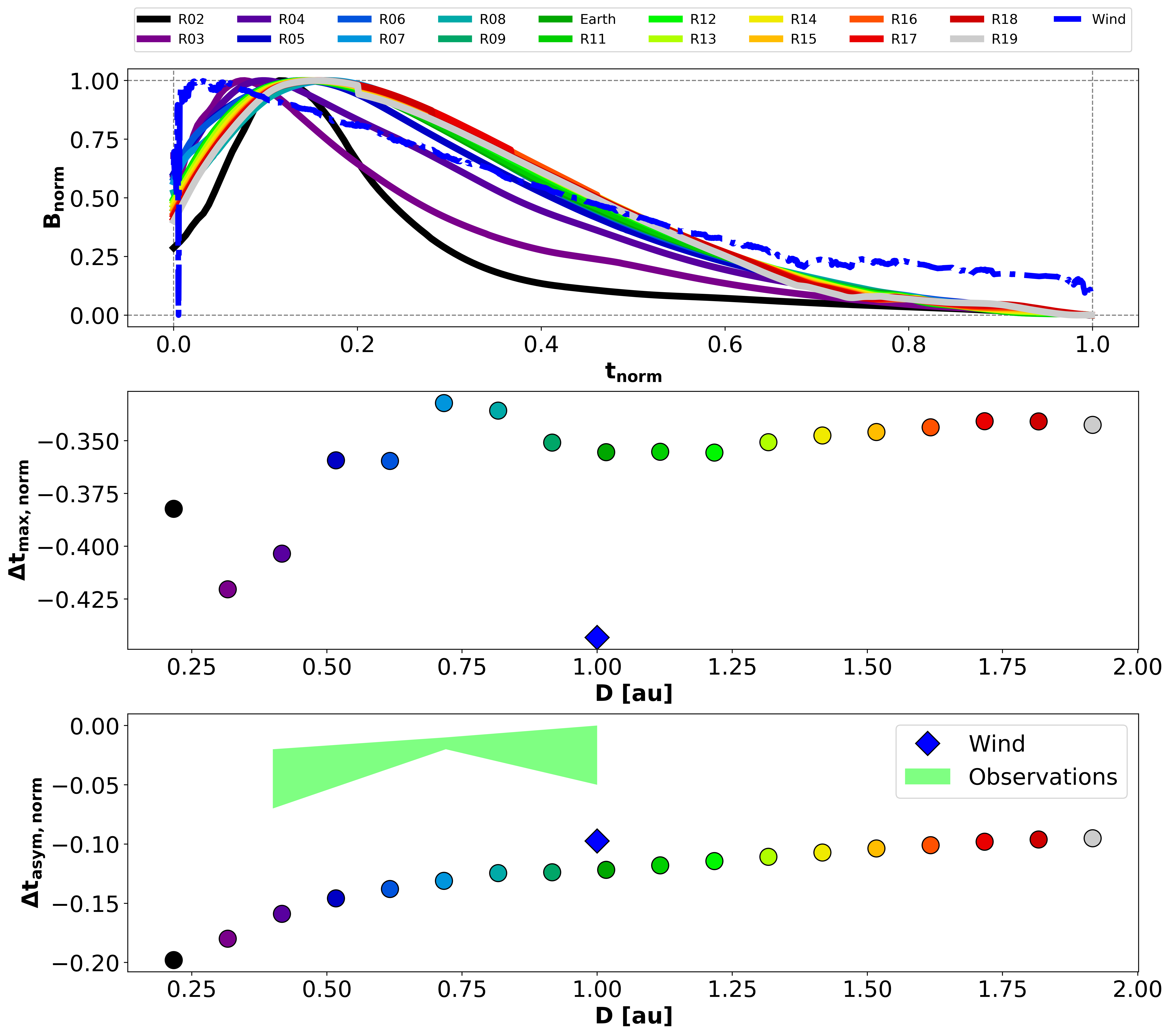}}
\caption{
Top panel: CME magnetic field profile in EUHFORIA at various heliocentric distances, 
expressed in terms of the normalised magnetic field $B_{norm}$ as a function of the normalised time $t_{norm}$.
The normalised magnetic field profile at $Wind$ is indicated by the dashed-dotted blue line.
Middle panel: normalised time shift between the magnetic field peak and the centre of the magnetic ejecta $\Delta t_{B_{peak, norm}}$, as a function of the heliocentric distances. The observed value at $Wind$ is indicated as a blue diamond.
Bottom panel: normalised asymmetry parameter $\Delta t_{asym, norm}$ for the magnetic field profile, as a function of the heliocentric distances.
The green shaded area indicates the range of asymmetry of the mean CME profiles at different heliocentric distances, based on the results by \citet{janvier:2019}. The observed value at $Wind$ is indicated as a blue diamond.
}
\label{fig:cme_magnetic_profile} 
\end{figure}

Figure~\ref{fig:cme_magnetic_profile} summarises the main characteristics of the CME magnetic field profile at different distances. 
To better compare between the time profiles at different heliocentric distances, 
in the top panel of Figure~\ref{fig:cme_magnetic_profile} we also plot the magnetic field profile normalised over the range of magnetic field values observed within the CME at different heliocentric distances, calculated as 
\begin{equation}
    B_{norm}(t) = \frac{B(t)-B_{min}}{B_{max} - B_{min}}.
\end{equation}
The temporal axis shows the time since the start of the magnetic ejecta, normalised over its duration at different heliocentric distances,
\begin{equation}
    t_{norm} = \frac{t-t_{start}}{t_{end} - t_{start}}.
\end{equation}
In the middle panel of Figure~\ref{fig:cme_magnetic_profile}, we plot the asymmetry in the magnetic field profile using as proxy the normalised shift of the magnetic field peak compared to the centre of the magnetic ejecta at different heliocentric distances. In the bottom panel, we plot the normalised asymmetry parameter in function of the heliocentric distance.

Results from both metrics show that the CME magnetic field profile is more asymmetric closer to the Sun than at higher heliocentric distances. 
Moreover, EUHFORIA simulations also show that the asymmetry indicated by the time shift of the magnetic field peak at $D=0.2$~au is initially lower, it increases up to $D=0.3$~au and then it progressively reduces. On the other hand, the magnitude of the asymmetry parameter follows a monotonically-decreasing trend with heliocentric distance. Beyond $D \simeq 0.7$~au, both metrics indicate that the magnetic field asymmetry changes less rapidly than closer to the Sun. 
Comparing EUHFORIA results at 1~au with observed values at $Wind$, we report a ratio of modelled ($-0.36$) and observed ($-0.44$) values equal to $\sim 0.82$ for the normalised time shift (corresponding to an underestimation of the asymmetry in EUHFORIA compared to observed values).
The ratio of modelled ($-0.12$) and observed ($-0.10$) values is $\sim 1.20$ for the normalised asymmetry parameter (corresponding to an overestimation of the asymmetry in EUHFORIA compared to observations). We note that the two metrics considered here provide two (complementary) ways of looking at the magnetic field time profile asymmetry of CMEs, and that the overestimation or underestimation provided should not be considered as particularly informative as it mostly relates to the specific definition of the parameter/metric considered. Rather, one should focus on the overall trend with heliocentric distance, which in both cases tends towards lower asymmetry values at high distances from the Sun. 
A decrease in the the asymmetry parameter with heliocentric distance has been previously observed in situ and interpreted as due to the relaxation of a CME magnetic structure tending towards a state of equilibrium with the surrounding solar wind \citep{janvier:2019}.
While a monotonic trend in the asymmetry parameter is typically observed, the non-monotonic behaviour of the time shift of the magnetic field peak as a function of the heliocentric distance recovered from EUHFORIA simulations is intriguing, and we further discuss its nature and origin in Section~\ref{sec:results_interpretation}.

\subsection{Radial evolution of CME wake}
\label{subsec:results_wake}

Previous studies assessed that perturbed conditions in the solar wind after the passage of a CME often have a long recovery period that can sometimes exceed the duration of the CME disturbance itself \citep{liu:2014, temmer:2017}. Observations at 1~au and at inner heliocentric distances highlighted that such a phenomenon, referred to as ``solar wind pre-conditioning'', can last up to 5~days after the end of the CME observed in situ, having important implications for CME propagation models and space weather forecasting algorithms, as the propagation of a CME precursor event might significantly affect the propagation of a following CME even when the two are launched several days apart. However, the actual duration of this pre-conditioning in EUHFORIA numerical simulations has never been assessed prior to this study.

In this work, we quantify the duration of the solar wind CME-induced perturbation after the passage of the magnetic ejecta, which we refer to as ``CME wake'', based on a 5\% threshold condition, i.e.\ by applying the following threshold condition to EUHFORIA time series at different heliocentric distances:
\begin{equation}
\left|X(t_i)-X_{sw}(t_i)\right| \ge 0.05 \left| X(t_i)-X_{sw}(t_i) \right|_{max}, \\
\label{eqn:wake_identification}   
\end{equation}
where $t_i$ is a generic time in the EUHFORIA time series,
$X$ and $X_{sw}$ are time profiles of a generic MHD variable obtained from simulations of the CME and ambient solar wind with EUHFORIA,  
and $\left|X(t_i)-X_{sw}(t_i)\right|_{max}$ corresponds to the maximum (peak) variation recorded in background-subtracted time series of that same variable.
At each spacecraft, the end of the CME wake relative to a given variable ($t_{X,wake}$) is marked based on the first time $t_i$ after the end of the magnetic ejecta ($t_{end}$) at which Equation~\ref{eqn:wake_identification} is not satisfied anymore in EUHFORIA speed, proton number density, magnetic field magnitude, and proton temperature time series. The duration of the CME wake after the end of the magnetic ejecta is further computed as $\Delta t_{X, wake} = t_{X, wake} - t_{end}$. Figure~\ref{fig:cme_wake} in Appendix~\ref{app:appendix} shows the result of this analysis at selected heliocentric distances.
We note that the search for periods of perturbed solar wind speed and magnetic field conditions after the passage of a magnetic ejecta as indicator of the CME wake is consistent with the typical wake conditions reported by \citet{temmer:2017} and \citet{janvier:2019}. On the other hand, to the best of our knowledge no previous study has ever quantified the effect of CME pre-conditioning to the solar wind with respect to density and temperature. 

From Figure~\ref{fig:cme_wake}, we note that all the variables except the temperature present a similar duration of the perturbation after the end of the CME passage.
The temperature, on the other hand, appears to be restored to pre-CME values already within the magnetic ejecta. At a first consideration, this might indicate that taking $\beta_p$ as only parameter to distinguish the CME duration in EUHFORIA time series might not be sufficient, and it may lead to a misinterpretation of the location of the CME rear edge that might also explain the longer ejecta duration and radial size reported in Figure~\ref{fig:cme_size}. If, in fact, the CME wake as determined from the temperature profile were to be considered as marking the real end of the magnetic ejecta, the estimated ejecta size would be reduced by a factor of $\sim 2$ compared to current estimated in Figure~\ref{fig:cme_size}, resulting in values more similar to observations at $Wind$.
%
While it should be noted that the identification of proxies to identify the exact duration of the CME wake is non-trivial, and it most probably requires further verification via an in-depth investigation of the 3D magnetic structure of the CME, there are two main reasons that make us consider the temperature proxy as not representative of the actual CME wake duration:
(1) in EUHFORIA, the choice of both the solar wind and CME initial temperatures at 0.1~au (i.e.\ the model inner boundary) relies on severe assumptions that are likely to significantly affect the modelling of the plasma temperature throughout the whole modelling domain; 
(2) the derivation of the temperatures of the various particle populations in the solar wind based on in situ measurements are subject to higher uncertainties than the other plasma and magnetic field parameters; as such, temperature estimations are not considered very reliable even when it comes to solar wind in situ measurements.
While leaving future investigations on this matter for future studies, here we point out that because of the time consistency between low $\beta_p$ signatures and other typical signatures indicating the passage of a magnetic ejecta (e.g.\ high speed, intense magnetic field, low densities) in in situ time series, and of the lack of any evidence supporting a more reliable modelling of the solar wind temperature compared to other parameters, we are inclined to believe that the $\beta_p$ condition do provide a reliable identification of the CME rear edge after all.

\begin{figure}
\centering
{ \includegraphics[width=0.7\hsize]{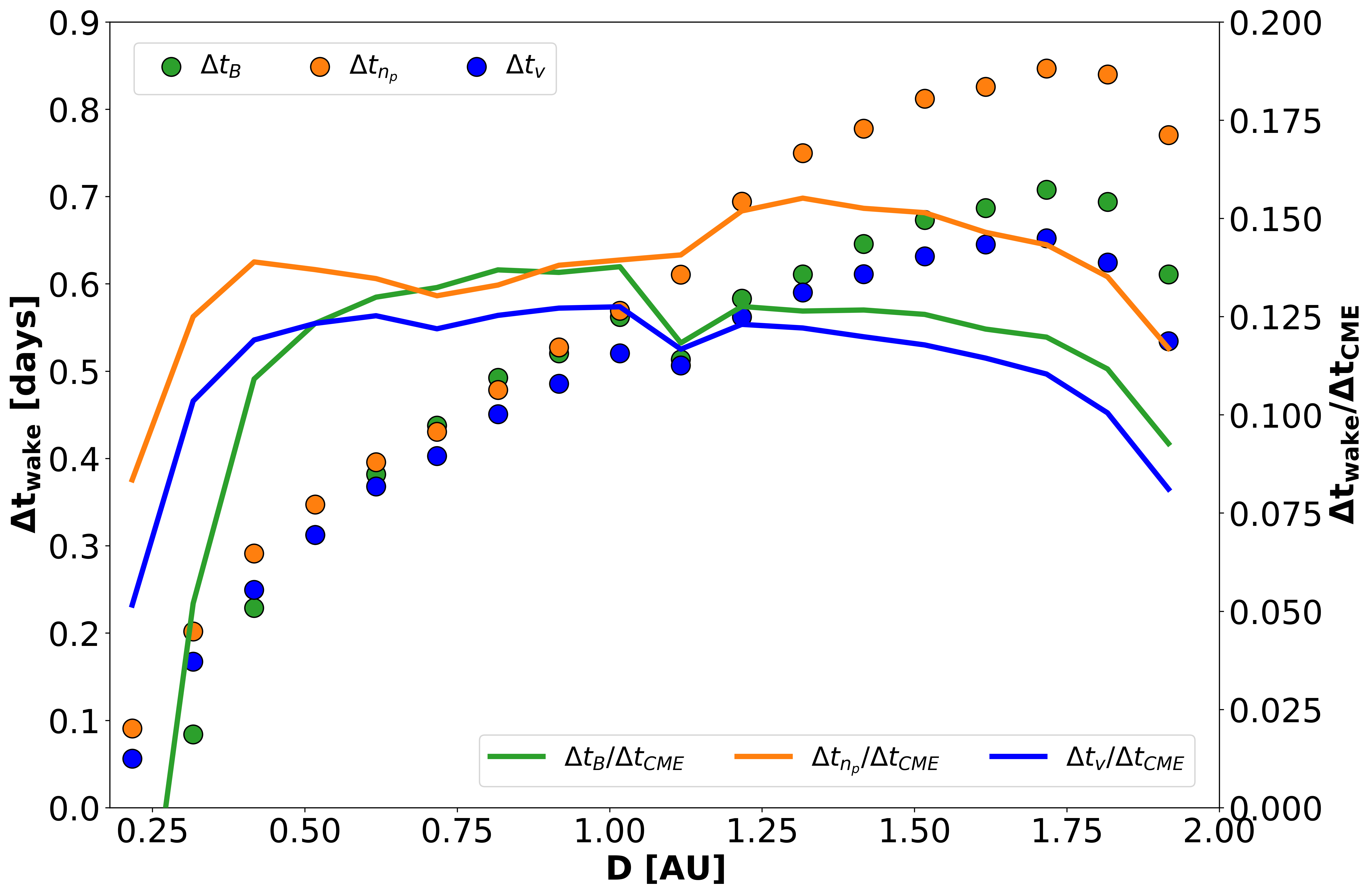} }
\caption{
Left: duration of the CME wake since the end of the magnetic ejecta for the magnetic field, density, and temperature, as a function of the heliocentric distance $D$ (indicated by the solid dots). Right: ratio between the duration of the CME wake and of the magnetic ejecta, as a function of the heliocentric distance $D$, for the same quantities (indicated by the solid lines).}
\label{fig:20120712_wake_distances} 
\end{figure}

Figure~\ref{fig:20120712_wake_distances} shows the duration of the CME wake and its ratio with respect to the duration of the magnetic ejecta, as a function of the heliocentric distance, for the different variables considered (except for the temperature).
We observe that the CME wake at all heliocentric distances and for all the variables considered never exceeds the duration of 1~day after $t_{end}$ (maximum of $\Delta t_{wake, n_p} = 0.85$ reached at 1.7~au).
This duration is significantly lower than reported by previous studies \citep[typically from 2 to 5 days, see e.g.][]{temmer:2017}, and it might be due to the specific characteristics of the CME considered. 
However, other possible explanations for the short duration of CME pre-conditioning detected in EUHFORIA simulations might be the non-ideal MHD nature of turbulence developing in the CME flank/wake region in the solar corona and interplanetary space as well as its short scales \citep[e.g.][]{fan:2018}, which require higher spatial and temporal resolutions than the ones used in our simulation to be resolved.
The analysis at different heliocentric distances also shows that the duration of the solar wind preconditioning grows as a function of $D$ up to about 1.7~au, when it reaches a peak before starting to decrease.
Modelling results also show that the duration of the CME wake remains significantly smaller than that of the ejecta, with their ratio never exceeding the value of 20\% of the duration of the magnetic ejecta (maximum of $\Delta t_{wake, n_p}/\Delta t_\mathrm{CME} = 16$\% reached at 1.3~au). Their ratio remains almost constant between 0.3 and 1.3~au, meaning at these range of distances the two structures grow with a similar behaviour. After 1.3~au, simulations show the ratio decreasing because of the stable duration of the wake with respect to a progressive growth of the CME radial size. Within 0.3~au, we observe a much shorter duration of the CME wake, most probably due to a combination of poor detection performances of our algorithm, and modelling artifacts due to the vicinity to the model inner boundary.

\section{Discussion and physical interpretation}
\label{sec:results_interpretation}

\subsection{Evolution of the CME expansion rate}
\label{subsec:discussion_expansion} 

Figure~\ref{fig:spheromak_zeta}~(d), showing higher-than-typical CME expansion rates close to the Sun, indicates that the magnetic ejecta undergoes a phase of rapid expansion in innermost region of the heliosphere (within 0.4~au from the Sun) that cannot be explained in terms of the sole balancing of the decreasing pressure in the surrounding solar wind. 
In particular, \citet{demoulin:2009} showed that for a CME isotropic expansion controlled by the equilibrium with the surrounding solar wind, the expansion rate $\zeta_\mathrm{CME}$ should be independent on the heliocentric distance and on the size of the ejecta.
For such an expansion and assuming conservation of the magnetic flux, the total pressure of the ambient solar wind scales as $\propto D^{-a_{P_{tot}}}$, where the exponent $a_{P_{tot}}$ relates to the CME expansion rate as $a_{P_{tot}} \approx - 4 \zeta_\mathrm{CME}$ \citep{demoulin:2009}. Under these conditions, the mean magnetic field inside the magnetic ejecta should fall off as $\propto D^{a_{P_{tot}}/2}$, while the ejecta size should scale proportionally to $\propto D^{-a_{P_{tot}}/4} $. 
From Figure~\ref{fig:sw_radial_02}~(g) we recover that $a_{P_{tot}} \sim -2.78$, 
and therefore theoretical arguments would expect the ejecta size to grow as $S_\mathrm{CME} \propto D^{0.70}$.
Overall, this trend is well reproduced in our simulations, from which we find the CME size growing as $S_\mathrm{CME} \propto D^{0.69}$ (Figure~\ref{fig:cme_size}~(d)) on average throughout the whole heliospheric domain. On the other hand, the modelled mean magnetic field inside the ejecta falls off slightly faster than expected, i.e.\ $\langle B \rangle_\mathrm{CME} \propto D^{-1.90}$ (from Figure~\ref{fig:spheromak_radial_boundaries_01}~(c)), while the expected trend is $\langle B \rangle_\mathrm{CME} \propto D^{-1.39}$. 
When fitting separately heliocentric distances where strong and weak $\zeta_\mathrm{CME}$ dependencies are observed,
we recover $a_{P_{tot}} \sim - 3.34$, $S_\mathrm{CME} \propto D^{0.78}$, and $\langle B \rangle_\mathrm{CME} \propto D^{-1.82}$ between 0.2 and 0.4~au; and $a_{P_{tot}} \sim - 2.50$, $S_\mathrm{CME} \propto D^{0.61}$, and $\langle B \rangle_\mathrm{CME} \propto D^{-1.91}$ from 0.5 to 1.9~au.
Within 0.4 au, theoretical arguments would expect the ejecta size to grow as $S_\mathrm{CME} \propto D^{0.83}$, while beyond 0.5 au the expected trend is $S_\mathrm{CME} \propto D^{0.63}$, i.e.\ both are very close to the modelled behaviour.
%
In both cases, the mean magnetic field inside the ejecta decays faster than expected, as theoretical expectations would suggest that $\langle B \rangle_\mathrm{CME} \propto D^{-1.67}$ and $\langle B \rangle_\mathrm{CME} \propto D^{-1.25}$, respectively.
We note, however, that such discrepancies for the scaling of the mean magnetic field are within observational uncertainties, i.e.\ they are within the observed ranges reported in previous studies \citep[e.g.][]{liu:2005, wang:2005, leitner:2007, gulisano:2010, winslow:2015}. Possible contributions to the rapid magnetic field decay with heliocentric distance (particularly between 0.5 and 1.9~au, where we have ruled out a significant contribution of magnetic erosion and over-expansion) include: a non-isotropic expansion of the CME, possibly related to a non-isotropic solar wind pressure, and numerical dissipation. Given our focus on the CME propagation along Sun-to-Earth radial direction only, we leave the investigation of the aforementioned phenomena for future studies.

Quantitatively, the ratio of the expansion and translation speeds of the CME appears consistent with observation-based estimates in the solar corona (in Paper~1) and at 1~au (in this work) carried out for the specific event considered, and it is also in line with results obtained in previous works \citep[e.g.][]{gopalswamy:2009, lugaz:2020}.

By filling the observational gaps between 0.1~au and 1~au for the specific event considered, EUHFORIA simulations allow us to identify two regimes of CME expansion: 
a first phase between 0.1 and 0.4~au (i.e.\ around the orbit of Mercury), characterised by a rapid expansion of the ejecta structure in response to its continuous tendency to establish a pressure balance with the surrounding, expanding solar wind (controlled primarily by the interplay between the magnetic pressure inside the ejecta, and the thermal pressure in the surrounding wind; see Paper~1 for a full discussion on this subject).
In a second phase of expansion, between 0.4~au and 1.9~au, the magnetic ejecta expands more gradually in response to the slowly-decaying pressure in the surrounding medium \citep[][]{demoulin:2009}.
Overall, the radial dependence of $\zeta_\mathrm{CME}$ in Figure~\ref{fig:spheromak_zeta}~(d) appears therefore primarily linked to a non-exponential decay of the solar wind pressure (i.e.\ the radial dependence of $a_{P_{tot}}$), rather than to sources of expansion internal to the CME \citep[consistent with what theorised by][]{demoulin:2009}.
This also implies that erosion/flux injection phenomena do not alter the expansion of the CME size in a significant way. As the efficiency of these phenomena in triggering magnetic reconnection at the solar wind--CME boundary primarily depends on the relative orientation of the magnetic fields therein, we note that this condition might be related to the magnetic configuration of the specific event considered. Further studies are therefore needed to generalise the picture to other magnetic configurations.

Additionally, by comparing the CME size extracted from EUHFORIA time series at 0.2~au with the reconstructed CME size in the solar corona at 0.07 and 0.1~au reported in Paper~1, we observe that the size growth between 0.1~au and 0.2~au is consistent with a growth behaviour as $S \propto D^{\zeta_\mathrm{CME}}$.
We conclude that the overestimation in the CME radial size at 1~au is primarily the result of an overestimation in the CME radial size already in the range 0.7 -- 0.1~au, and can therefore be traced back to the limited flexibility of the spheromak model in reproducing the plethora of evolutionary deformations observed to affect CME geometries in the corona and heliosphere. 
In particular, as a spherical structure the spheromak model in EUHFORIA relies on the assumption that during insertion in the heliospheric domain (i.e.\ at 0.1~au) the width of the CME is the same in all directions, (i.e.\ the structure is characterised by a circular cross section in all directions), therefore completely neglecting pancaking effects occurring before 0.1~au \citep[e.g.][]{savani:2011, isavnin:2016}.
As such, the modelled expansion of the CME between 0.1~au and 1.9~au is entirely physical and is not undermined by any modelling artifact related to the insertion of the CME structure at 0.1~au. 
Although in principle the exact identification the ejecta boundaries is a delicate task both in observations and modelling, the fact that the initial CME radial size at insertion in the simulation domain is in line with further in situ estimates at virtual spacecraft suggests this most likely did not significantly affect our estimation of the ejecta size in EUHFORIA simulations.

Overall, the dependence of the CME expansion rate on the heliocentric distance suggests that this was a perturbed CME event. However, this appears to be limited to distances $\le 0.4$~au, and relates to a rapid expansion of the CME structure close to the model inner boundary. In our simulations, this is not related to the presence of any high speed solar wind stream compressing the magnetic ejecta from the back, although a high speed stream was present in reality as visible from $Wind$ observations. This is a known limitation of the solar wind modelling in EUHFORIA that has been recently addressed by a number of publications and that is still currently under investigation \citep{asvestari:2019, hinterreiter:2019, samara:2020}.

\subsection{Evolution of the magnetic field profile} 
\label{subsec:discussion_profile} 

The magnetic field inside CMEs is one of the most important factors affecting both the large-scale evolution of CME structures in interplanetary space, and their potential as space weather drivers \citep[e.g.][]{gosling:1991, kilpua:2017}.
Understanding its radial evolution and the factors affecting its in situ observations is therefore extremely important to better assess the potential space weather impact of a CME at a target location.
At a given spacecraft, the asymmetry in the observed magnetic field profile within magnetic ejectas can be interpreted as the sum of two contributions: 
the expansion of the CME as it crosses the spacecraft \citep[the so-called ``ageing effect'';][]{demoulin:2008, demoulin:2018}, 
and the non-circular cross section of the magnetic ejecta itself \citep{demoulin:2009b}. 
Previous studies agree in considering the latter to be typically dominant at 1~au. 
However, the ageing effect has been found to be the main source of magnetic asymmetry for a significant minority of magnetic ejectas at 1~au, making the consideration and correction for its effects worthwhile, in particular, in the case of large events \citep{demoulin:2020}.
Furthermore, we note that the effect of ageing at different heliocentric distances than 1~au, particularly those closer to the Sun, cannot be discarded \textit{a priori} when interpreting the results obtained from EUHFORIA simulations.

The ageing effect is naturally linked to the $\zeta_\mathrm{CME}$ parameter, which provides an estimate for the expansion rate of a magnetic ejecta regardless of its size, speed, and distance from the Sun. A number of previous studies estimated that for $\zeta_\mathrm{CME} \in (0.6, 1.0)$, i.e.\ the typical values at 1~au, the effect of ageing alone could not explain the asymmetry observed, which would typically require a much higher expansion rates \citep[e.g.][]{demoulin:2008, demoulin:2018}. However, it should be stressed that more recent estimates considering magnetic ejectas with expansion rates $\zeta_\mathrm{CME} \in (0.8, 1.2)$ reported ageing effect as the main source of the observed magnetic asymmetry for more than one every four events \citep{demoulin:2020}.

From EUHFORIA simulations, we report values of $\zeta_\mathrm{CME}$ ranging between 1.09 and 0.63 at the different heliocentric distances considered (Subsection~\ref{subsec:results_magnetic_profile}). Such numbers are comparable with typical observational values at 1~au, which suggests the asymmetries in the CME magnetic field profile are primarily due to actual cross-section asymmetries, although ageing may still provide a significant contribution leading to a mix of the two effects \citep{demoulin:2020}. For the particular CME considered, and based on our simulations, we consider ageing may be most relevant for heliocentric distances closer than 0.5~au, where $\zeta_\mathrm{CME}$ was higher than 0.8, while at distances larger than 0.5~au the lower expansion rates ($\zeta_\mathrm{CME} < 0.8$) suggest cross-sectional asymmetries as the dominant source of magnetic asymmetry.
EUHFORIA simulations therefore confirm previous estimates obtained from CME observations between the orbit of Mercury and 1~au \citep[][]{janvier:2019}. Future investigations on the contribution of CME ageing to in situ observations within 0.5~au, including a comparison of simulation results with observations carried out by PSP in the range 0.1--0.3~au, are needed to corroborate EUHFORIA results at such small heliocentric distances.

The discussion above suggests that the asymmetry observed in situ is the manifestation of an actual asymmetry in the ejecta cross section. Overall, we find that the asymmetry is less pronounced at higher heliocentric distances than close to the Sun. 
Therefore, we conclude that the results in Figure~\ref{fig:cme_magnetic_profile}, qualitatively consistent with those presented in Figure~3 in \citet{janvier:2019}, should be interpreted as evidence of the progressive relaxation of the ejecta magnetic structure during propagation in interplanetary space.
The fact that the time shift of the magnetic field peak does not follow the expected monotonic trend close to the Sun (within 0.7~au) might be due to a modelling artifact caused by the fact that at the insertion in the heliospheric domain (at $D=0.1$~au), the spheromak magnetic structure is defined as symmetric between its front and its back and it also lacks a pre-existing shock and sheath ahead. 
Only at $D > 0.1$~au (i.e.\ after insertion in the heliospheric domain), its cross section starts being deformed as a consequence of the interaction with the surrounding solar wind, primarily in response to the formation of a sheath which develops conditions at the front and back of the ejecta that are different from one another. Results in Figure~\ref{fig:cme_magnetic_profile}~(b) suggest that this phase of rapid sheath and asymmetry formation is fully completed between 0.3 and 0.4~au, and that only at higher heliocentric distances the spheromak magnetic structure starts actually evolving towards a more relaxed configuration exhibiting a trend consistent with observations. The asymmetry parameter appears to be a more stable metric, not very sensitive to such a modelling artifact, and it is found to better describe the evolutionary relaxation of the ejecta at the various heliocentric distance \citep{janvier:2019}.

\section{Conclusions}
\label{sec:conclusions}

In this work, as a case study, we investigated the interplanetary propagation of a CME. We modelled its radial evolution analysing the evolution of its mean properties, magnetic field profile, and pre-conditioning of the ambient solar wind after its passage. Aiming to bridge the gap between observational and modelling studies, we focused on these specific properties as they are the most often analysed in statistical studies of interplanetary CMEs. The CME under study was a fast halo CME directed towards the Earth observed on 12 July 2012 and previously analysed by the authors in Paper~1. The double goal of this work was to validate numerical models of CME propagation, particularly the EUHFORIA model \citep{pomoell:2018} describing the CME internal magnetic field as a linear force-free spheromak configuration \citep{verbeke:2019b}; and to perform a comprehensive analysis of the radial evolution of interplanetary CMEs, filling observational gaps and providing new context for observational studies of CME propagation in the inner heliosphere.
The main results are as follows:
\begin{enumerate}
    \item We found that the radial dependence of the mean solar wind and CME properties modelled in EUHFORIA are both consistent with observational and theoretical expectations based on statistical sets of solar wind and CME data.
    \item The investigation of the CME expansion rate at different heliocentric distances indicated that the CME under study underwent a phase of rapid expansion up to $\sim 0.4$~au, while farther out the expansion rate was moderate and almost independent from the heliocentric distance. At all distances, the CME expansion was consistent with a trend controlled by the pressure balance with the surrounding solar wind \citep{demoulin:2009}. The ratio between the expansion and translation speeds extracted close to the model inner boundary and at 1~au was also consistent with observational values in the corona and at Earth's orbit.
    \item The early rapid expansion alone was not sufficient to explain the overestimation in the CME radial size in EUHFORIA simulations, suggesting that the overestimation of the CME radial size by the spheromak model is most probably caused by an intrinsic limitation of the specific CME model to account for deformations of the CME structure such as pancaking \citep{savani:2011, isavnin:2016} or other possible asymmetric shapes \citep{demoulin:2009b}.
    \item The analysis of the CME expansion rate extracted from EUHFORIA simulations also suggested that this was a perturbed CME event. However, the perturbation appears to be limited to distances $\le 0.4$~au and linked to a rapid expansion of the CME structure close to the model inner boundary. It is important to mention that EUHFORIA simulations exhibited no evidence of a solar wind high speed stream compressing the magnetic ejecta from the back, although a high speed stream was present in $Wind$ observations. The poor reproduction of solar wind high speed streams in EUHFORIA is a known limitation that has been recently investigated by numerous authors \citep{asvestari:2019, hinterreiter:2019, samara:2020}.
    \item The asymmetry in the magnetic field profile indicated a progressive relaxation of the magnetic ejecta structure during propagation. The analysis of the time shift of the magnetic peak with respect to the centre of the magnetic ejecta allowed to estimate the effects of the insertion through the inner boundary to last up to $0.4$~au, i.e.\ slightly longer than estimated by \citet{scolini:2020} for simulations employing the cone CME model.
    \item Moreover, the expansion rate $\zeta_\mathrm{CME} \le 1.1$ suggested that CME ageing is most likely not a significant contribution to the asymmetry in the magnetic field profile at any heliocentric distance sampled; this may be particularly beyond 0.5~au, where the expansion rate was found to be lower than 0.8 \citep{demoulin:2020}.
    \item We report that the modelled duration of the CME wake is underestimated compared to typical observations \citep{temmer:2017}, and that this may be partially due to limitations of the ideal MHD and large-scale approach used in treating the formation of the CME wake. We argue that a better description of waves and turbulence might be required to realistically reproduce this feature in numerical simulations.
\end{enumerate}

We conclude that EUHFORIA combined with the spheromak model is able to provide a physically-consistent description of the radial evolution of the solar wind and CMEs throughout the inner heliosphere, at least in proximity of the ecliptic plane and inside the CME and in its nearby environment. However, our study also highlights some intrinsic limitations of the spheromak model in the reproduction of the CME radial size, which will have to be improved in a refined version of EUHFORIA in order to achieve more accurate descriptions of the interplanetary propagation of CMEs. In this respect, it is also important to mention that modelling results depend on the initial configuration and exact model used for the CME initiation as well as for the ambient solar wind \citep[e.g.][]{alhaddad:2019}. 

The future comparison with in situ observations of CMEs in the range 0.1--0.3~au carried out by PSP is needed in order to ultimately validate modelling results close to the corona--heliosphere boundary, i.e.\ in the proximity of the Alfv\'{e}n surface. As PSP orbits will access the LASCO/C3 coronagraph field of view (within 30 solar radii) in the near future, we will have the unprecedented opportunity to actually measure the CME characteristics as close as 15--30 solar radii, and to directly compare them with results obtained from EUHFORIA simulations of the newly-observed CME events.
Results from ESA's Solar Orbiter \citep[][]{muller:2020} mission will also help extending current observational data sets to heliocentric distances as close as than 0.28~au and at heliocentric latitudes as high as $25^\circ$, calling for further studies specifically assessing the performances of numerical models outside of the ecliptic plane.

\begin{acknowledgements}
C.S. was supported by the Research Foundation -- Flanders (FWO, strategic base PhD fellowship No. 1S42817N). 
S.D. acknowledges partial support from the Argentinian grants UBACyT (UBA) and PIP-CONICET-11220130100439CO. 
S.D. is member of the Carrera del Investigador Científico, CONICET.
L.R. and A.N.Z. thank the European Space Agency (ESA) and the Belgian Federal Science Policy Office (BELSPO) for their support in the framework of the PRODEX Programme.
This project also received funding from the European Union’s Horizon 2020 research and innovation programme under grant agreement No 870405 (EUHFORIA~2.0; \url{https://euhforia.com/euhforia-2-0/}). These results were also obtained in the framework of the projects C14/19/089 (C1 project Internal Funds KU Leuven), G.0D07.19N (FWO-Vlaanderen), and C~90347 (ESA Prodex).

$Wind$ data can be accessed via the OMNIWeb Plus data portal (\url{https://omniweb.gsfc.nasa.gov/ftpbrowser/wind_min_merge.html}).
This paper uses data from the Heliospheric Shock Database (\url{http://ipshocks.fi}), generated and maintained at the University of Helsinki.

EUHFORIA is developed as a joint effort between the University of Helsinki and KU Leuven. 
The full validation of solar wind and CME modelling is being performed within the BRAIN-be CCSOM project (Constraining CMEs and Shocks by Observations and Modeling throughout the inner heliosphere; \url{http://www.sidc.be/ccsom/}) and BRAIN-be SWiM project (Solar WInd Modeling with EUHFORIA for the new heliospheric missions).
The simulations were carried out at the VSC -- Flemish Supercomputer Centre, funded by the Hercules foundation and the Flemish Government -- Department EWI. 

C.S. thanks Jens Pomoell for useful discussions.
\end{acknowledgements}

\bibliographystyle{aa} 

\begin{thebibliography}{126}
\expandafter\ifx\csname natexlab\endcsname\relax\def\natexlab#1{#1}\fi

\bibitem[{{Al-Haddad} {et~al.}(2019){Al-Haddad}, {Lugaz}, {Poedts}, {Farrugia},
  {Nieves-Chinchilla}, \& {Roussev}}]{alhaddad:2019}
{Al-Haddad}, N., {Lugaz}, N., {Poedts}, S., {et~al.} 2019, \apj, 884, 179

\bibitem[{{Al-Haddad} {et~al.}(2013){Al-Haddad}, {Nieves-Chinchilla}, {Savani},
  {M{\"o}stl}, {Marubashi}, {Hidalgo}, {Roussev}, {Poedts}, \&
  {Farrugia}}]{alhaddad:2013}
{Al-Haddad}, N., {Nieves-Chinchilla}, T., {Savani}, N.~P., {et~al.} 2013,
  \solphys, 284, 129

\bibitem[{{Arge} {et~al.}(2004){Arge}, {Luhmann}, {Odstrcil}, {Schrijver}, \&
  {Li}}]{arge:2004}
{Arge}, C.~N., {Luhmann}, J.~G., {Odstrcil}, D., {Schrijver}, C.~J., \& {Li},
  Y. 2004, J. Atmospheric Sol.-Terr. Phys., 66, 1295

\bibitem[{{Asvestari} {et~al.}(2019){Asvestari}, {Heinemann}, {Temmer},
  {Pomoell}, {Kilpua}, {Magdalenic}, \& {Poedts}}]{asvestari:2019}
{Asvestari}, E., {Heinemann}, S.~G., {Temmer}, M., {et~al.} 2019, J. Geophys.
  Res. (Space Phys.), 124, 8280

\bibitem[{{Bothmer} \& {Schwenn}(1996)}]{bothmer:1996}
{Bothmer}, V. \& {Schwenn}, R. 1996, Adv. Space Res., 17, 319

\bibitem[{{Bougeret} {et~al.}(1984){Bougeret}, {King}, \&
  {Schwenn}}]{bougeret:1984}
{Bougeret}, J.~L., {King}, J.~H., \& {Schwenn}, R. 1984, \solphys, 90, 401

\bibitem[{{Brueckner} {et~al.}(1995){Brueckner}, {Howard}, {Koomen},
  {Korendyke}, {Michels}, {Moses}, {Socker}, {Dere}, {Lamy}, {Llebaria},
  {Bout}, {Schwenn}, {Simnett}, {Bedford}, \& {Eyles}}]{brueckner:1995}
{Brueckner}, G.~E., {Howard}, R.~A., {Koomen}, M.~J., {et~al.} 1995, \solphys,
  162, 357

\bibitem[{{Burlaga} {et~al.}(1981){Burlaga}, {Sittler}, {Mariani}, \&
  {Schwenn}}]{burlaga:1981}
{Burlaga}, L., {Sittler}, E., {Mariani}, F., \& {Schwenn}, R. 1981, \jgr, 86,
  6673

\bibitem[{Cane \& Richardson(2003)}]{cane:2003}
Cane, H.~V. \& Richardson, I.~G. 2003, J. Geophys. Res. (Space Phys.), 108

\bibitem[{{Cargill}(2004)}]{cargill:2004}
{Cargill}, P.~J. 2004, \solphys, 221, 135

\bibitem[{{Cranmer} {et~al.}(2017){Cranmer}, {Gibson}, \&
  {Riley}}]{cranmer:2017}
{Cranmer}, S.~R., {Gibson}, S.~E., \& {Riley}, P. 2017, \ssr, 212, 1345

\bibitem[{{Dal Lago} {et~al.}(2003){Dal Lago}, {Schwenn}, \&
  {Gonzalez}}]{dallago:2003}
{Dal Lago}, A., {Schwenn}, R., \& {Gonzalez}, W.~D. 2003, Adv. Space Res., 32,
  2637

\bibitem[{{Dasso} {et~al.}(2006){Dasso}, {Mandrini}, {D{\'e}moulin}, \&
  {Luoni}}]{dasso:2006}
{Dasso}, S., {Mandrini}, C.~H., {D{\'e}moulin}, P., \& {Luoni}, M.~L. 2006,
  \aap, 455, 349

\bibitem[{{Dasso} {et~al.}(2009){Dasso}, {Mandrini}, {Schmieder}, {Cremades},
  {Cid}, {Cerrato}, {Saiz}, {D{\'e}moulin}, {Zhukov}, {Rodriguez}, {Aran},
  {Menvielle}, \& {Poedts}}]{dasso:2009}
{Dasso}, S., {Mandrini}, C.~H., {Schmieder}, B., {et~al.} 2009, J. Geophys.
  Res. (Space Phys.), 114, A02109

\bibitem[{{Dasso} {et~al.}(2007){Dasso}, {Nakwacki}, {D{\'e}moulin}, \& {Mand
  rini}}]{dasso:2007}
{Dasso}, S., {Nakwacki}, M.~S., {D{\'e}moulin}, P., \& {Mand rini}, C.~H. 2007,
  Sol. Phys., 244, 115

\bibitem[{{D{\'e}moulin}(2009)}]{demoulin:2009c}
{D{\'e}moulin}, P. 2009, \solphys, 257, 169

\bibitem[{{D{\'e}moulin}(2010)}]{demoulin:2010}
{D{\'e}moulin}, P. 2010, in AIP Conference Series, Vol. 1216, 12th
  International Solar Wind Conference, ed. M.~{Maksimovic}, K.~{Issautier},
  N.~{Meyer-Vernet}, M.~{Moncuquet}, \& F.~{Pantellini}, 329--334

\bibitem[{{D{\'e}moulin} \& {Dasso}(2009{\natexlab{a}})}]{demoulin:2009}
{D{\'e}moulin}, P. \& {Dasso}, S. 2009{\natexlab{a}}, \aap, 498, 551

\bibitem[{{D{\'e}moulin} \& {Dasso}(2009{\natexlab{b}})}]{demoulin:2009b}
{D{\'e}moulin}, P. \& {Dasso}, S. 2009{\natexlab{b}}, \aap, 507, 969

\bibitem[{{D{\'e}moulin} {et~al.}(2018){D{\'e}moulin}, {Dasso}, \&
  {Janvier}}]{demoulin:2018}
{D{\'e}moulin}, P., {Dasso}, S., \& {Janvier}, M. 2018, \aap, 619, A139

\bibitem[{{D{\'e}moulin} {et~al.}(2020){D{\'e}moulin}, {Dasso}, {Lanabere}, \&
  {Janvier}}]{demoulin:2020}
{D{\'e}moulin}, P., {Dasso}, S., {Lanabere}, V., \& {Janvier}, M. 2020, \aap,
  639, A6

\bibitem[{{D{\'e}moulin} {et~al.}(2016){D{\'e}moulin}, {Janvier},
  {Mas{\'\i}as-Meza}, \& {Dasso}}]{demoulin:2016}
{D{\'e}moulin}, P., {Janvier}, M., {Mas{\'\i}as-Meza}, J.~J., \& {Dasso}, S.
  2016, \aap, 595, A19

\bibitem[{{D{\'e}moulin} {et~al.}(2008){D{\'e}moulin}, {Nakwacki}, {Dasso}, \&
  {Mandrini}}]{demoulin:2008}
{D{\'e}moulin}, P., {Nakwacki}, M.~S., {Dasso}, S., \& {Mandrini}, C.~H. 2008,
  \solphys, 250, 347

\bibitem[{{Domingo} {et~al.}(1995){Domingo}, {Fleck}, \&
  {Poland}}]{domingo:1995}
{Domingo}, V., {Fleck}, B., \& {Poland}, A.~I. 1995, \solphys, 162, 1

\bibitem[{{Fan} {et~al.}(2018){Fan}, {He}, {Yan}, {Tomczyk}, {Tian}, {Song},
  {Wang}, \& {Zhang}}]{fan:2018}
{Fan}, S., {He}, J., {Yan}, L., {et~al.} 2018, \solphys, 293, 6

\bibitem[{{Fox} {et~al.}(2016){Fox}, {Velli}, {Bale}, {Decker}, {Driesman},
  {Howard}, {Kasper}, {Kinnison}, {Kusterer}, {Lario}, {Lockwood}, {McComas},
  {Raouafi}, \& {Szabo}}]{fox:2016}
{Fox}, N.~J., {Velli}, M.~C., {Bale}, S.~D., {et~al.} 2016, \ssr, 204, 7

\bibitem[{{Goldstein} {et~al.}(1998){Goldstein}, {Neugebauer}, \&
  {Clay}}]{goldstein:1998}
{Goldstein}, R., {Neugebauer}, M., \& {Clay}, D. 1998, \jgr, 103, 4761

\bibitem[{{Good} {et~al.}(2018){Good}, {Forsyth}, {Eastwood}, \&
  {M{\"o}stl}}]{good:2018}
{Good}, S.~W., {Forsyth}, R.~J., {Eastwood}, J.~P., \& {M{\"o}stl}, C. 2018,
  \solphys, 293, 52

\bibitem[{{Good} {et~al.}(2015){Good}, {Forsyth}, {Raines}, {Gershman},
  {Slavin}, \& {Zurbuchen}}]{good:2015}
{Good}, S.~W., {Forsyth}, R.~J., {Raines}, J.~M., {et~al.} 2015, \apj, 807, 177

\bibitem[{{Gopalswamy} {et~al.}(2009{\natexlab{a}}){Gopalswamy}, {Dal Lago},
  {Yashiro}, \& {Akiyama}}]{gopalswamy:2009b}
{Gopalswamy}, N., {Dal Lago}, A., {Yashiro}, S., \& {Akiyama}, S.
  2009{\natexlab{a}}, Central European Astrophysical Bulletin, 33, 115

\bibitem[{{Gopalswamy} {et~al.}(2009{\natexlab{b}}){Gopalswamy},
  {M{\"a}kel{\"a}}, {Xie}, {Akiyama}, \& {Yashiro}}]{gopalswamy:2009}
{Gopalswamy}, N., {M{\"a}kel{\"a}}, P., {Xie}, H., {Akiyama}, S., \& {Yashiro},
  S. 2009{\natexlab{b}}, J. Geophys. Res. (Space Phys.), 114, A00A22

\bibitem[{{Gosling} {et~al.}(1991){Gosling}, {McComas}, {Phillips}, \&
  {Bame}}]{gosling:1991}
{Gosling}, J.~T., {McComas}, D.~J., {Phillips}, J.~L., \& {Bame}, S.~J. 1991,
  \jgr, 96, 7831

\bibitem[{{Gulisano} {et~al.}(2012){Gulisano}, {D{\'e}moulin}, {Dasso}, \&
  {Rodriguez}}]{gulisano:2012}
{Gulisano}, A.~M., {D{\'e}moulin}, P., {Dasso}, S., \& {Rodriguez}, L. 2012,
  \aap, 543, A107

\bibitem[{{Gulisano} {et~al.}(2010){Gulisano}, {D{\'e}moulin}, {Dasso}, {Ruiz},
  \& {Marsch}}]{gulisano:2010}
{Gulisano}, A.~M., {D{\'e}moulin}, P., {Dasso}, S., {Ruiz}, M.~E., \& {Marsch},
  E. 2010, \aap, 509, A39

\bibitem[{{Heber} {et~al.}(2009){Heber}, {Wieler}, {Baur}, {Olinger},
  {Friedmann}, \& {Burnett}}]{heber:2009}
{Heber}, V.~S., {Wieler}, R., {Baur}, H., {et~al.} 2009, \gca, 73, 7414

\bibitem[{{Hellinger} {et~al.}(2011){Hellinger}, {Matteini},
  {{\v{S}}tver{\'a}k}, {Tr{\'a}vn{\'\i}{\v{c}}ek}, \&
  {Marsch}}]{hellinger:2011}
{Hellinger}, P., {Matteini}, L., {{\v{S}}tver{\'a}k}, {\v{S}}.,
  {Tr{\'a}vn{\'\i}{\v{c}}ek}, P.~M., \& {Marsch}, E. 2011, J. Geophys. Res.
  (Space Phys.), 116, A09105

\bibitem[{{Hellinger} {et~al.}(2013){Hellinger}, {Tr{\'a}vn{\'\i}{\v{c}}ek},
  {{\v{S}}tver{\'a}k}, {Matteini}, \& {Velli}}]{hellinger:2013}
{Hellinger}, P., {Tr{\'a}vn{\'\i}{\v{c}}ek}, P.~M., {{\v{S}}tver{\'a}k},
  {\v{S}}., {Matteini}, L., \& {Velli}, M. 2013, J. Geophys. Res. (Space
  Phys.), 118, 1351

\bibitem[{{Hinterreiter} {et~al.}(2019){Hinterreiter}, {Magdalenic}, {Temmer},
  {Verbeke}, {Jebaraj}, {Samara}, {Asvestari}, {Poedts}, {Pomoell}, {Kilpua},
  {Rodriguez}, {Scolini}, \& {Isavnin}}]{hinterreiter:2019}
{Hinterreiter}, J., {Magdalenic}, J., {Temmer}, M., {et~al.} 2019, \solphys,
  294, 170

\bibitem[{{Illing} \& {Hundhausen}(1985)}]{illing:1985}
{Illing}, R.~M.~E. \& {Hundhausen}, A.~J. 1985, \jgr, 90, 275

\bibitem[{{Isavnin}(2016)}]{isavnin:2016}
{Isavnin}, A. 2016, \apj, 833, 267

\bibitem[{{Janvier} {et~al.}(2014){Janvier}, {D{\'e}moulin}, \&
  {Dasso}}]{janvier:2014}
{Janvier}, M., {D{\'e}moulin}, P., \& {Dasso}, S. 2014, \aap, 565, A99

\bibitem[{{Janvier} {et~al.}(2019){Janvier}, {Winslow}, {Good}, {Bonhomme},
  {D{\'e}moulin}, {Dasso}, {M{\"o}stl}, {Lugaz}, {Amerstorfer}, {Soubri{\'e}},
  \& {Boakes}}]{janvier:2019}
{Janvier}, M., {Winslow}, R.~M., {Good}, S., {et~al.} 2019, J. Geophys. Res.
  (Space Phys.), 124, 812

\bibitem[{{Kasper} {et~al.}(2012){Kasper}, {Stevens}, {Korreck}, {Maruca},
  {Kiefer}, {Schwadron}, \& {Lepri}}]{kasper:2012}
{Kasper}, J.~C., {Stevens}, M.~L., {Korreck}, K.~E., {et~al.} 2012, \apj, 745,
  162

\bibitem[{{Kilpua} {et~al.}(2017){Kilpua}, {Koskinen}, \&
  {Pulkkinen}}]{kilpua:2017}
{Kilpua}, E., {Koskinen}, H. E.~J., \& {Pulkkinen}, T.~I. 2017, Liv. Rev. Sol.
  Phys., 14, 5

\bibitem[{{Kilpua} {et~al.}(2015){Kilpua}, {Lumme}, {Andreeova}, {Isavnin}, \&
  {Koskinen}}]{kilpua:2015}
{Kilpua}, E.~K.~J., {Lumme}, E., {Andreeova}, K., {Isavnin}, A., \& {Koskinen},
  H.~E.~J. 2015, J. Geophys. Res. (Space Phys.), 120, 4112

\bibitem[{{Klein} \& {Burlaga}(1982)}]{klein:1982}
{Klein}, L.~W. \& {Burlaga}, L.~F. 1982, \jgr, 87, 613

\bibitem[{{Korreck} {et~al.}(2020){Korreck}, {Szabo}, {Nieves Chinchilla},
  {Lavraud}, {Luhmann}, {Niembro}, {Higginson}, {Alzate}, {Wallace}, {Paulson},
  {Rouillard}, {Kouloumvakos}, {Poirier}, {Kasper}, {Case}, {Stevens}, {Bale},
  {Pulupa}, {Whittlesey}, {Livi}, {Goetz}, {Larson}, {Malaspina}, {Morgan},
  {Narock}, {Schwadron}, {Bonnell}, {Harvey}, \& {Wygant}}]{korreck:2020}
{Korreck}, K.~E., {Szabo}, A., {Nieves Chinchilla}, T., {et~al.} 2020, \apjs,
  246, 69

\bibitem[{{Lanabere} {et~al.}(2020){Lanabere}, {Dasso}, {D{\'e}moulin},
  {Janvier}, {Rodriguez}, \& {Mas{\'\i}as-Meza}}]{lanabere:2020}
{Lanabere}, V., {Dasso}, S., {D{\'e}moulin}, P., {et~al.} 2020, \aap, 635, A85

\bibitem[{{Lavraud} {et~al.}(2014){Lavraud}, {Ruffenach}, {Rouillard},
  {Kajdic}, {Manchester}, \& {Lugaz}}]{lavraud:2014}
{Lavraud}, B., {Ruffenach}, A., {Rouillard}, A.~P., {et~al.} 2014, J. Geophys.
  Res. (Space Phys.), 119, 26

\bibitem[{{Lee} {et~al.}(2017){Lee}, {Moon}, {Lee}, {Kim}, \& {Cho}}]{lee:2017}
{Lee}, J.-O., {Moon}, Y.~J., {Lee}, J.-Y., {Kim}, R.~S., \& {Cho}, K.~S. 2017,
  \apj, 838, 70

\bibitem[{{Leitner} {et~al.}(2007){Leitner}, {Farrugia}, {M{\"o}stl},
  {Ogilvie}, {Galvin}, {Schwenn}, \& {Biernat}}]{leitner:2007}
{Leitner}, M., {Farrugia}, C.~J., {M{\"o}stl}, C., {et~al.} 2007, J. Geophys.
  Res. (Space Phys.), 112, A06113

\bibitem[{{Lepping} {et~al.}(2005){Lepping}, {Wu}, \&
  {Berdichevsky}}]{lepping:2005}
{Lepping}, R.~P., {Wu}, C.~C., \& {Berdichevsky}, D.~B. 2005, Ann. Geophys.,
  23, 2687

\bibitem[{{Liu} {et~al.}(2005){Liu}, {Richardson}, \& {Belcher}}]{liu:2005}
{Liu}, Y., {Richardson}, J.~D., \& {Belcher}, J.~W. 2005, \planss, 53, 3

\bibitem[{{Liu} {et~al.}(2014){Liu}, {Luhmann}, {Kajdi{\v{c}}}, {Kilpua},
  {Lugaz}, {Nitta}, {M{\"o}stl}, {Lavraud}, {Bale}, {Farrugia}, \&
  {Galvin}}]{liu:2014}
{Liu}, Y.~D., {Luhmann}, J.~G., {Kajdi{\v{c}}}, P., {et~al.} 2014, Nature
  Comm., 5, 3481

\bibitem[{{Lopez}(1987)}]{lopez:1987}
{Lopez}, R.~E. 1987, \jgr, 92, 11189

\bibitem[{{Lopez} \& {Freeman}(1986)}]{lopez:1986}
{Lopez}, R.~E. \& {Freeman}, J.~W. 1986, \jgr, 91, 1701

\bibitem[{{Lugaz} {et~al.}(2017){Lugaz}, {Temmer}, {Wang}, \&
  {Farrugia}}]{lugaz:2017}
{Lugaz}, N., {Temmer}, M., {Wang}, Y., \& {Farrugia}, C.~J. 2017, \solphys,
  292, 64

\bibitem[{{Lugaz} {et~al.}(2020){Lugaz}, {Winslow}, \& {Farrugia}}]{lugaz:2020}
{Lugaz}, N., {Winslow}, R.~M., \& {Farrugia}, C.~J. 2020, J. Geophys. Res.
  (Space Phys.), 125, e27213

\bibitem[{{Manchester} {et~al.}(2017){Manchester}, {Kilpua}, {Liu}, {Lugaz},
  {Riley}, {T{\"o}r{\"o}k}, \& {Vr{\v{s}}nak}}]{manchester:2017}
{Manchester}, W., {Kilpua}, E. K.~J., {Liu}, Y.~D., {et~al.} 2017, \ssr, 212,
  1159

\bibitem[{{Manchester} {et~al.}(2014){Manchester}, {Kozyra}, {Lepri}, \&
  {Lavraud}}]{manchester:2014}
{Manchester}, W.~B., {Kozyra}, J.~U., {Lepri}, S.~T., \& {Lavraud}, B. 2014, J.
  Geophys. Res. (Space Phys.), 119, 5449

\bibitem[{Mariani \& Neubauer(1990)}]{mariani:1990}
Mariani, F. \& Neubauer, F.~M. 1990, The Interplanetary Magnetic Field, ed.
  R.~Schwenn \& E.~Marsch (Berlin, Heidelberg: Springer), 183--206

\bibitem[{{Marsch}(2006)}]{marsch:2006}
{Marsch}, E. 2006, Liv. Rev. Sol. Phys., 3, 1

\bibitem[{{Marsch}(2012)}]{marsch:2012}
{Marsch}, E. 2012, \ssr, 172, 23

\bibitem[{{Mas{\'\i}as-Meza} {et~al.}(2016){Mas{\'\i}as-Meza}, {Dasso},
  {D{\'e}moulin}, {Rodriguez}, \& {Janvier}}]{masias:2016}
{Mas{\'\i}as-Meza}, J.~J., {Dasso}, S., {D{\'e}moulin}, P., {Rodriguez}, L., \&
  {Janvier}, M. 2016, \aap, 592, A118

\bibitem[{{M{\"o}stl} {et~al.}(2020){M{\"o}stl}, {Weiss}, {Bailey}, {Reiss},
  {Amerstorfer}, {Hinterreiter}, {Bauer}, {McIntosh}, {Lugaz}, \&
  {Stansby}}]{moestl:2020}
{M{\"o}stl}, C., {Weiss}, A.~J., {Bailey}, R.~L., {et~al.} 2020, \apj, 903, 92

\bibitem[{{M{"u}ller} {et~al.}(2020){M{"u}ller}, {St. Cyr}, {Zouganelis},
  {Gilbert}, {Marsden}, {Nieves-Chinchilla}, {Antonucci}, {Auch{\`e}re},
  {Berghmans}, {Horbury}, {Howard}, {Krucker}, {Maksimovic}, {Owen}, {Rochus},
  {Rodriguez-Pacheco}, {Romoli}, {Solanki}, {Bruno}, {Carlsson}, {Fludra},
  {Harra}, {Hassler}, {Livi}, {Louarn}, {Peter}, {Sch{"u}hle}, {Teriaca}, {del
  Toro Iniesta}, {Wimmer-Schweingruber}, {Marsch}, {Velli}, {De Groof},
  {Walsh}, \& {Williams}}]{muller:2020}
{M{"u}ller}, D., {St. Cyr}, O.~C., {Zouganelis}, I., {et~al.} 2020, \aap, 642,
  A1

\bibitem[{{Nakwacki} {et~al.}(2011){Nakwacki}, {Dasso}, {D{\'e}moulin}, {Mand
  rini}, \& {Gulisano}}]{nakwacki:2011}
{Nakwacki}, M.~S., {Dasso}, S., {D{\'e}moulin}, P., {Mand rini}, C.~H., \&
  {Gulisano}, A.~M. 2011, \aap, 535, A52

\bibitem[{{Nieves-Chinchilla} {et~al.}(2020){Nieves-Chinchilla}, {Szabo},
  {Korreck}, {Alzate}, {Balmaceda}, {Lavraud}, {Paulson}, {Narock}, {Wallace},
  {Jian}, {Luhmann}, {Morgan}, {Higginson}, {Arge}, {Bale}, {Case}, {Dudfok de
  Wit}, {Giacalone}, {Goetz}, {Harvey}, {Jones-Melosky}, {Kasper}, {Larson},
  {Livi}, {McComas}, {MacDowall}, {Malaspina}, {Pulupa}, {Raouafi},
  {Schwadron}, {Stevens}, \& {Whittlesey}}]{nieves:2020}
{Nieves-Chinchilla}, T., {Szabo}, A., {Korreck}, K.~E., {et~al.} 2020, \apjs,
  246, 63

\bibitem[{{Odstrcil} {et~al.}(2020){Odstrcil}, {Mays}, {Hess}, {Jones},
  {Henney}, \& {Arge}}]{odstrcil:2020}
{Odstrcil}, D., {Mays}, M.~L., {Hess}, P., {et~al.} 2020, \apjs, 246, 73

\bibitem[{{Ogilvie} {et~al.}(1995){Ogilvie}, {Chornay}, {Fritzenreiter},
  {Hunsaker}, {Keller}, {Lobell}, {Miller}, {Scudder}, {Sittler}, {Torbert},
  {Bodet}, {Needell}, {Lazarus}, {Steinberg}, {Tappan}, {Mavretic}, \&
  {Gergin}}]{ogilvie:1995}
{Ogilvie}, K.~W., {Chornay}, D.~J., {Fritzenreiter}, R.~J., {et~al.} 1995,
  \ssr, 71, 55

\bibitem[{{Owens}(2016)}]{owens:2016}
{Owens}, M.~J. 2016, \apj, 818, 197

\bibitem[{{Pal} {et~al.}(2020){Pal}, {Dash}, \& {Nandy}}]{pal:2020}
{Pal}, S., {Dash}, S., \& {Nandy}, D. 2020, \grl, 47, e86372

\bibitem[{{Palmerio} {et~al.}(2017){Palmerio}, {Kilpua}, {James}, {Green},
  {Pomoell}, {Isavnin}, \& {Valori}}]{palmerio:2017}
{Palmerio}, E., {Kilpua}, E.~K.~J., {James}, A.~W., {et~al.} 2017, \solphys,
  292, 39

\bibitem[{{Parker}(1958)}]{parker:1958}
{Parker}, E.~N. 1958, \apj, 128, 664

\bibitem[{{Parker}(1965)}]{parker:1965}
{Parker}, E.~N. 1965, \ssr, 4, 666

\bibitem[{{Paularena} {et~al.}(2001){Paularena}, {Wang}, {von Steiger}, \&
  {Heber}}]{paularena:2001}
{Paularena}, K.~I., {Wang}, C., {von Steiger}, R., \& {Heber}, B. 2001, \grl,
  28, 2755

\bibitem[{{Perrone} {et~al.}(2019{\natexlab{a}}){Perrone}, {Stansby},
  {Horbury}, \& {Matteini}}]{perrone:2019a}
{Perrone}, D., {Stansby}, D., {Horbury}, T.~S., \& {Matteini}, L.
  2019{\natexlab{a}}, \mnras, 483, 3730

\bibitem[{{Perrone} {et~al.}(2019{\natexlab{b}}){Perrone}, {Stansby},
  {Horbury}, \& {Matteini}}]{perrone:2019b}
{Perrone}, D., {Stansby}, D., {Horbury}, T.~S., \& {Matteini}, L.
  2019{\natexlab{b}}, \mnras, 488, 2380

\bibitem[{{Pomoell} \& {Poedts}(2018)}]{pomoell:2018}
{Pomoell}, J. \& {Poedts}, S. 2018, J. Space Weather Space Clim., 8, A35

\bibitem[{{Reisenfeld} {et~al.}(2007){Reisenfeld}, {Burnett}, {Becker},
  {Grimberg}, {Heber}, {Hohenberg}, {Jurewicz}, {Meshik}, {Pepin}, {Raines},
  {Schlutter}, {Wieler}, {Wiens}, \& {Zurbuchen}}]{reisenfeld:2007}
{Reisenfeld}, D.~B., {Burnett}, D.~S., {Becker}, R.~H., {et~al.} 2007, \ssr,
  130, 79

\bibitem[{{Richardson} \& {Cane}(1995)}]{richardson:1995}
{Richardson}, I.~G. \& {Cane}, H.~V. 1995, \jgr, 100, 23397

\bibitem[{{Richardson} \& {Cane}(2010)}]{richardson:2010}
{Richardson}, I.~G. \& {Cane}, H.~V. 2010, \solphys, 264, 189

\bibitem[{{Richardson} {et~al.}(2006){Richardson}, {Liu}, {Wang}, \&
  {Burlaga}}]{richardson:j:2006}
{Richardson}, J.~D., {Liu}, Y., {Wang}, C., \& {Burlaga}, L.~F. 2006, Adv.
  Space Res., 38, 528

\bibitem[{{Rodriguez} {et~al.}(2016){Rodriguez}, {Mas{\'\i}as-Meza}, {Dasso},
  {D{\'e}moulin}, {Zhukov}, {Gulisano}, {Mierla}, {Kilpua}, {West}, {Lacatus},
  {Paraschiv}, \& {Janvier}}]{rodriguez:2016}
{Rodriguez}, L., {Mas{\'\i}as-Meza}, J.~J., {Dasso}, S., {et~al.} 2016,
  \solphys, 291, 2145

\bibitem[{{Rodriguez} {et~al.}(2008){Rodriguez}, {Zhukov}, {Dasso}, {Mand
  rini}, {Cremades}, {Cid}, {Cerrato}, {Saiz}, {Aran}, {Menvielle}, {Poedts},
  \& {Schmieder}}]{rodriguez:2008}
{Rodriguez}, L., {Zhukov}, A.~N., {Dasso}, S., {et~al.} 2008, Annales
  Geophysicae, 26, 213

\bibitem[{{Ruffenach} {et~al.}(2015){Ruffenach}, {Lavraud}, {Farrugia},
  {D{\'e}moulin}, {Dasso}, {Owens}, {Sauvaud}, {Rouillard}, {Lynnyk},
  {Foullon}, {Savani}, {Luhmann}, \& {Galvin}}]{ruffenach:2015}
{Ruffenach}, A., {Lavraud}, B., {Farrugia}, C.~J., {et~al.} 2015, J. Geophys.
  Res. (Space Phys.), 120, 43

\bibitem[{{Ruffenach} {et~al.}(2012){Ruffenach}, {Lavraud}, {Owens}, {Sauvaud},
  {Savani}, {Rouillard}, {D{\'e}moulin}, {Foullon}, {Opitz}, {Fedorov},
  {Jacquey}, {G{\'e}not}, {Louarn}, {Luhmann}, {Russell}, {Farrugia}, \&
  {Galvin}}]{ruffenach:2012}
{Ruffenach}, A., {Lavraud}, B., {Owens}, M.~J., {et~al.} 2012, J. Geophys. Res.
  (Space Phys.), 117, A09101

\bibitem[{{Salman} {et~al.}(2020){Salman}, {Winslow}, \& {Lugaz}}]{salman:2020}
{Salman}, T.~M., {Winslow}, R.~M., \& {Lugaz}, N. 2020, J. Geophys. Res. (Space
  Phys.), 125, e27084

\bibitem[{{Samara} {et~al.}(2021){Samara}, {Pinto}, {Magdaleni\'{c}},
  {Jer\v{c}i\'{c}}, {Scolini}, {Wijsen}, {Jebaraj}, {Rodriguez}, \&
  {Poedts}}]{samara:2020}
{Samara}, E., {Pinto}, R.~F., {Magdaleni\'{c}}, J., {et~al.} 2021, A\&A (in
  press), arXiv:2102.06617

\bibitem[{{Savani} {et~al.}(2011){Savani}, {Owens}, {Rouillard}, {Forsyth},
  {Kusano}, {Shiota}, \& {Kataoka}}]{savani:2011}
{Savani}, N.~P., {Owens}, M.~J., {Rouillard}, A.~P., {et~al.} 2011, \apj, 731,
  109

\bibitem[{{Schwenn}(1983)}]{schwenn:1983}
{Schwenn}, R. 1983, in NASA Conference Publication, Vol. 228, NASA Conference
  Publication, 0.489

\bibitem[{{Schwenn}(1990)}]{schwenn:1990b}
{Schwenn}, R. 1990, {Large-Scale Structure of the Interplanetary Medium}, ed.
  R.~{Schwenn} \& E.~{Marsch}, 99

\bibitem[{{Schwenn} {et~al.}(2005){Schwenn}, {dal Lago}, {Huttunen}, \&
  {Gonzalez}}]{schwenn:2005}
{Schwenn}, R., {dal Lago}, A., {Huttunen}, E., \& {Gonzalez}, W.~D. 2005, Ann.
  Geophys., 23, 1033

\bibitem[{{Schwenn} \& {Marsch}(1990)}]{schwenn:1990}
{Schwenn}, R. \& {Marsch}, E. 1990, Physics and Chemistry in Space, 20

\bibitem[{{Schwenn} {et~al.}(1981){Schwenn}, {Mohlhauser}, {Marsch}, \&
  {Rosenbauer}}]{schwenn:1981}
{Schwenn}, R., {Mohlhauser}, K.~H., {Marsch}, E., \& {Rosenbauer}, H. 1981, in
  Solar Wind 4, 126

\bibitem[{{Scolini} {et~al.}(2020{\natexlab{a}}){Scolini}, {Chan{\'e}},
  {Pomoell}, {Rodriguez}, \& {Poedts}}]{scolini:2020}
{Scolini}, C., {Chan{\'e}}, E., {Pomoell}, J., {Rodriguez}, L., \& {Poedts}, S.
  2020{\natexlab{a}}, Space Weather, 18, e02246

\bibitem[{{Scolini} {et~al.}(2020{\natexlab{b}}){Scolini}, {Chan{\'e}},
  {Temmer}, {Kilpua}, {Dissauer}, {Veronig}, {Palmerio}, {Pomoell},
  {Dumbovi{\'c}}, {Guo}, {Rodriguez}, \& {Poedts}}]{scolini:2020b}
{Scolini}, C., {Chan{\'e}}, E., {Temmer}, M., {et~al.} 2020{\natexlab{b}},
  \apjs, 247, 21

\bibitem[{{Scolini} {et~al.}(2019){Scolini}, {Rodriguez}, {Mierla}, {Pomoell},
  \& {Poedts}}]{scolini:2019}
{Scolini}, C., {Rodriguez}, L., {Mierla}, M., {Pomoell}, J., \& {Poedts}, S.
  2019, \aap, 626, A122

\bibitem[{{Scolini} {et~al.}(2018){Scolini}, {Verbeke}, {Poedts}, {Chan{\'e}},
  {Pomoell}, \& {Zuccarello}}]{scolini:2018b}
{Scolini}, C., {Verbeke}, C., {Poedts}, S., {et~al.} 2018, Space Weather, 16,
  754

\bibitem[{{Shen} {et~al.}(2014){Shen}, {Shen}, {Zhang}, {Hess}, {Wang}, {Feng},
  {Cheng}, \& {Yang}}]{shen:2014}
{Shen}, F., {Shen}, C., {Zhang}, J., {et~al.} 2014, Journal of Geophysical
  Research (Space Physics), 119, 7128

\bibitem[{{Shen} {et~al.}(2017){Shen}, {Wang}, {Shen}, \& {Feng}}]{shen:2017}
{Shen}, F., {Wang}, Y., {Shen}, C., \& {Feng}, X. 2017, \solphys, 292, 104

\bibitem[{{Singh} {et~al.}(2020){Singh}, {Kim}, {Pogorelov}, \&
  {Arge}}]{singh:2020}
{Singh}, T., {Kim}, T.~K., {Pogorelov}, N.~V., \& {Arge}, C.~N. 2020, Space
  Weather, 18, e02405

\bibitem[{{Siscoe} \& {Odstrcil}(2008)}]{siscoe:2008}
{Siscoe}, G. \& {Odstrcil}, D. 2008, J. Geophys. Res. (Space Phys.), 113,
  A00B07

\bibitem[{{Temmer} {et~al.}(2021){Temmer}, {Holzknecht}, {Dumbovic}, {Vrsnak},
  {Sachdeva}, {Heinemann}, {Dissauer}, {Scolini}, {Asvestari}, {Veronig}, \&
  {Hofmeister}}]{temmer:2021}
{Temmer}, M., {Holzknecht}, L., {Dumbovic}, M., {et~al.} 2021, J. Geophys. Res.
  (Space Phys.), 126

\bibitem[{{Temmer} {et~al.}(2017){Temmer}, {Reiss}, {Nikolic}, {Hofmeister}, \&
  {Veronig}}]{temmer:2017}
{Temmer}, M., {Reiss}, M.~A., {Nikolic}, L., {Hofmeister}, S.~J., \& {Veronig},
  A.~M. 2017, \apj, 835, 141

\bibitem[{{Thernisien}(2011)}]{thernisien:2011}
{Thernisien}, A. 2011, \apjs, 194, 33

\bibitem[{{Thernisien} {et~al.}(2009){Thernisien}, {Vourlidas}, \&
  {Howard}}]{thernisien:2009}
{Thernisien}, A., {Vourlidas}, A., \& {Howard}, R.~A. 2009, \solphys, 256, 111

\bibitem[{{Totten} {et~al.}(1995){Totten}, {Freeman}, \& {Arya}}]{totten:1995}
{Totten}, T.~L., {Freeman}, J.~W., \& {Arya}, S. 1995, \jgr, 100, 13

\bibitem[{{Venzmer} \& {Bothmer}(2018)}]{venzmer:2018}
{Venzmer}, M.~S. \& {Bothmer}, V. 2018, \aap, 611, A36

\bibitem[{{Verbeke} {et~al.}(2019){Verbeke}, {Pomoell}, \&
  {Poedts}}]{verbeke:2019b}
{Verbeke}, C., {Pomoell}, J., \& {Poedts}, S. 2019, \aap, 627, A111

\bibitem[{{Verscharen} {et~al.}(2019){Verscharen}, {Klein}, \&
  {Maruca}}]{verscharen:2019}
{Verscharen}, D., {Klein}, K.~G., \& {Maruca}, B.~A. 2019, Liv. Rev. Sol.
  Phys., 16, 5

\bibitem[{{von Steiger} {et~al.}(2000){von Steiger}, {Schwadron}, {Fisk},
  {Geiss}, {Gloeckler}, {Hefti}, {Wilken}, {Wimmer-Schweingruber}, \&
  {Zurbuchen}}]{vonsteiger:2000}
{von Steiger}, R., {Schwadron}, N.~A., {Fisk}, L.~A., {et~al.} 2000, \jgr, 105,
  27217

\bibitem[{{Vourlidas} {et~al.}(2013){Vourlidas}, {Lynch}, {Howard}, \&
  {Li}}]{vourlidas:2013}
{Vourlidas}, A., {Lynch}, B.~J., {Howard}, R.~A., \& {Li}, Y. 2013, \solphys,
  284, 179

\bibitem[{{Vr{\v{s}}nak} {et~al.}(2010){Vr{\v{s}}nak}, {{\v{Z}}ic},
  {Falkenberg}, {M{\"o}stl}, {Vennerstrom}, \& {Vrbanec}}]{vrsnak:2010}
{Vr{\v{s}}nak}, B., {{\v{Z}}ic}, T., {Falkenberg}, T.~V., {et~al.} 2010, \aap,
  512, A43

\bibitem[{{Wang} {et~al.}(2005){Wang}, {Du}, \& {Richardson}}]{wang:2005}
{Wang}, C., {Du}, D., \& {Richardson}, J.~D. 2005, J. Geophys. Res. (Space
  Phys.), 110, A10107

\bibitem[{{Webb} \& {Howard}(2012)}]{webb:2012}
{Webb}, D.~F. \& {Howard}, T.~A. 2012, Liv. Rev. Sol. Phys., 9, 3

\bibitem[{{Weiss} {et~al.}(2021){Weiss}, {M{\"o}stl}, {Amerstorfer}, {Bailey},
  {Reiss}, {Hinterreiter}, {Amerstorfer}, \& {Bauer}}]{weiss:2021}
{Weiss}, A.~J., {M{\"o}stl}, C., {Amerstorfer}, T., {et~al.} 2021, \apjs, 252,
  9

\bibitem[{{Wimmer-Schweingruber} {et~al.}(2006){Wimmer-Schweingruber},
  {Crooker}, {Balogh}, {Bothmer}, {Forsyth}, {Gazis}, {Gosling}, {Horbury},
  {Kilchenmann}, {Richardson}, {Richardson}, {Riley}, {Rodriguez}, {von
  Steiger}, {Wurz}, \& {Zurbuchen}}]{wimmer:2006}
{Wimmer-Schweingruber}, R.~F., {Crooker}, N.~U., {Balogh}, A., {et~al.} 2006,
  \ssr, 123, 177

\bibitem[{{Winslow} {et~al.}(2015){Winslow}, {Lugaz}, {Philpott}, {Schwadron},
  {Farrugia}, {Anderson}, \& {Smith}}]{winslow:2015}
{Winslow}, R.~M., {Lugaz}, N., {Philpott}, L.~C., {et~al.} 2015, J. Geophys.
  Res. (Space Phys.), 120, 6101

\bibitem[{{Winslow} {et~al.}(2016){Winslow}, {Lugaz}, {Schwadron}, {Farrugia},
  {Yu}, {Raines}, {Mays}, {Galvin}, \& {Zurbuchen}}]{winslow:2016}
{Winslow}, R.~M., {Lugaz}, N., {Schwadron}, N.~A., {et~al.} 2016, J. Geophys.
  Res. (Space Phys.), 121, 6092

\bibitem[{{Winslow} {et~al.}(2018){Winslow}, {Schwadron}, {Lugaz}, {Guo},
  {Joyce}, {Jordan}, {Wilson}, {Spence}, {Lawrence}, {Wimmer-Schweingruber}, \&
  {Mays}}]{winslow:2018}
{Winslow}, R.~M., {Schwadron}, N.~A., {Lugaz}, N., {et~al.} 2018, \apj, 856,
  139

\bibitem[{{Winslow} {et~al.}(2021){Winslow}, {Scolini}, {Lugaz}, \&
  {Galvin}}]{winslow:2021}
{Winslow}, R.~M., {Scolini}, C., {Lugaz}, N., \& {Galvin}, A.~B. 2021, ApJ
  (submitted)

\bibitem[{{Wood} {et~al.}(2017){Wood}, {Wu}, {Lepping}, {Nieves-Chinchilla},
  {Howard}, {Linton}, \& {Socker}}]{wood:2017}
{Wood}, B.~E., {Wu}, C.-C., {Lepping}, R.~P., {et~al.} 2017, \apjs, 229, 29

\bibitem[{{Wood} {et~al.}(2012){Wood}, {Wu}, {Rouillard}, {Howard}, \&
  {Socker}}]{wood:2012}
{Wood}, B.~E., {Wu}, C.~C., {Rouillard}, A.~P., {Howard}, R.~A., \& {Socker},
  D.~G. 2012, \apj, 755, 43

\bibitem[{{Zurbuchen} {et~al.}(2002){Zurbuchen}, {Fisk}, {Gloeckler}, \& {von
  Steiger}}]{zurbuchen:2002}
{Zurbuchen}, T.~H., {Fisk}, L.~A., {Gloeckler}, G., \& {von Steiger}, R. 2002,
  \grl, 29, 1352

\bibitem[{{Zurbuchen} \& {Richardson}(2006)}]{zurbuchen:2006}
{Zurbuchen}, T.~H. \& {Richardson}, I.~G. 2006, \ssr, 123, 31

\end{thebibliography}



\begin{appendix}

\section{Radial evolution of CME time profile}
\label{app:appendix}

\begin{figure}
\centering
\includegraphics[width=\hsize]{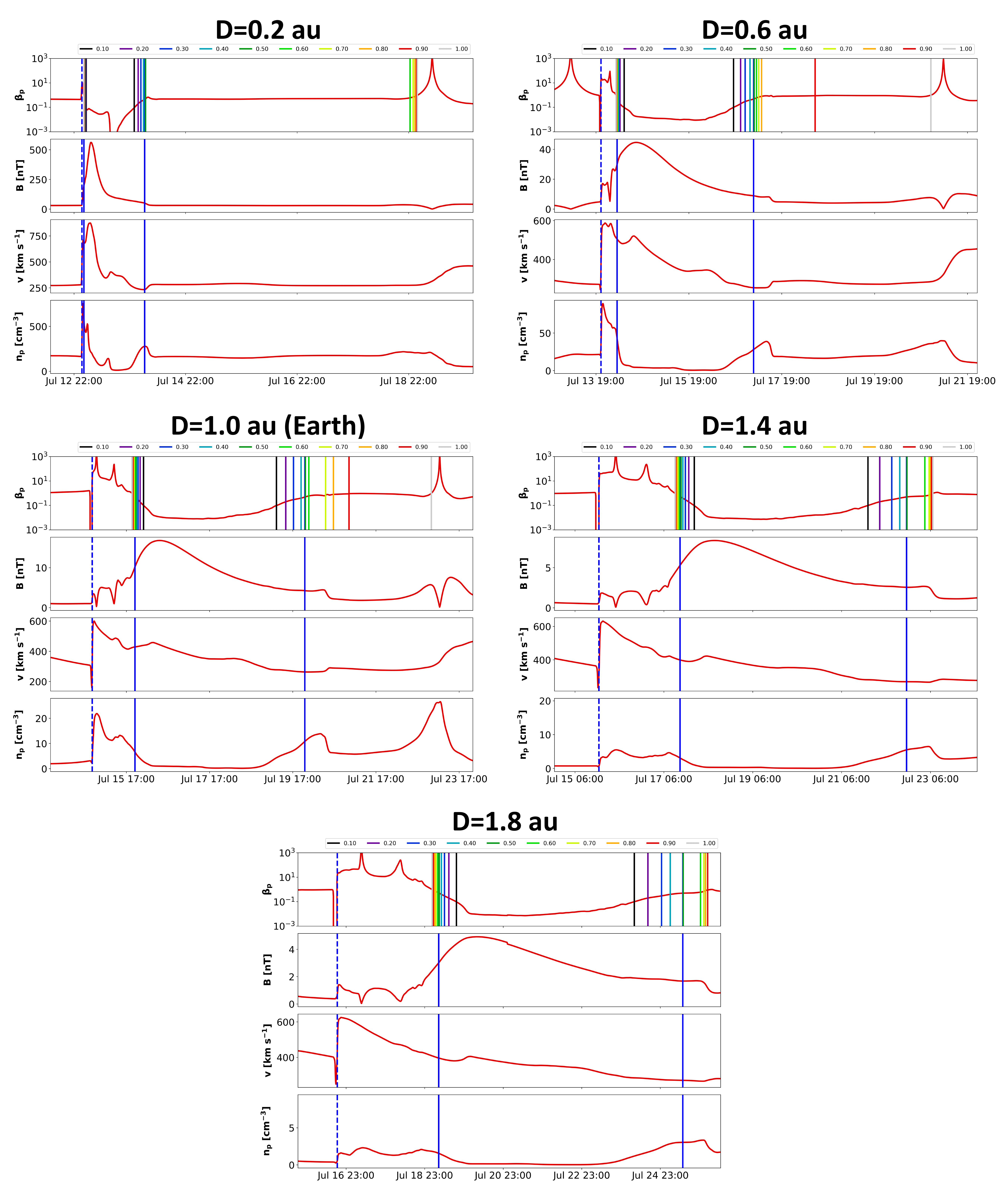}
\caption{Identification of the CME boundaries in EUHFORIA time series at various heliocentric distances.
At each heliocentric distance, the proton $\beta_p$, magnetic field intensity $B$, speed $v$ and proton number density $n_p$ are provided. 
In all panels, the arrival time of the CME-driven shock is marked by the dashed blue line.
In the top panel, the boundaries of the magnetic ejecta identified by imposing different $\beta_p$ thresholds are marked by coloured vertical lines, while the ultimate boundaries of the magnetic ejecta obtained by applying the threshold condition $\beta_p \le 0.5$ are marked as continuous blue lines in the second, third, and fourth panels.
}
\label{fig:spheromak_beta_thresholds}
\end{figure}

\begin{figure}
\centering
\includegraphics[width=\hsize]{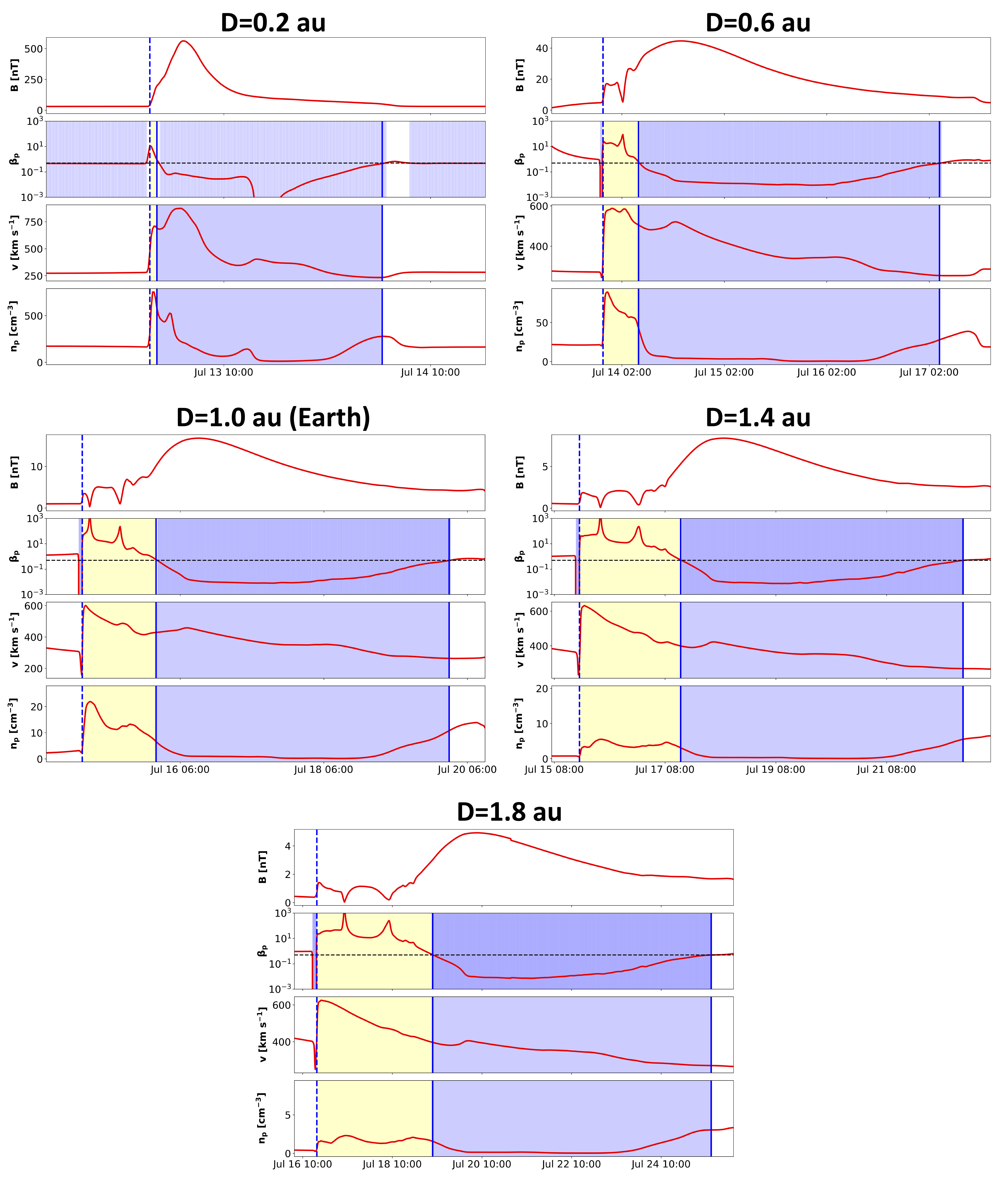}
\caption{Identification of the sheath and CME boundaries in EUHFORIA time series (in red) at various heliocentric distances.
At each heliocentric distance, the magnetic field intensity $B$, proton $\beta_p$, speed $v$ and proton number density $n_p$ are provided.
The arrival time of the CME-driven shock is marked by the dashed blue line. The identification of the magnetic ejecta boundaries (marked as continuous blue lines, same as in Figure~\ref{fig:spheromak_beta_thresholds}) has been obtained by applying the threshold condition $\beta_p \le 0.5$ (indicated as blue shaded regions). The duration of the CME sheath, based on the identification of $t_{shock}$ and $t_{start}$, is marked in yellow. In the proton $\beta$ panels, the $\beta_p=0.5$ threshold is indicated as an horizontal dashed black line.
}
\label{fig:spheromak_boundaries}
\end{figure}

\begin{figure}
\centering
{\includegraphics[width=\hsize]{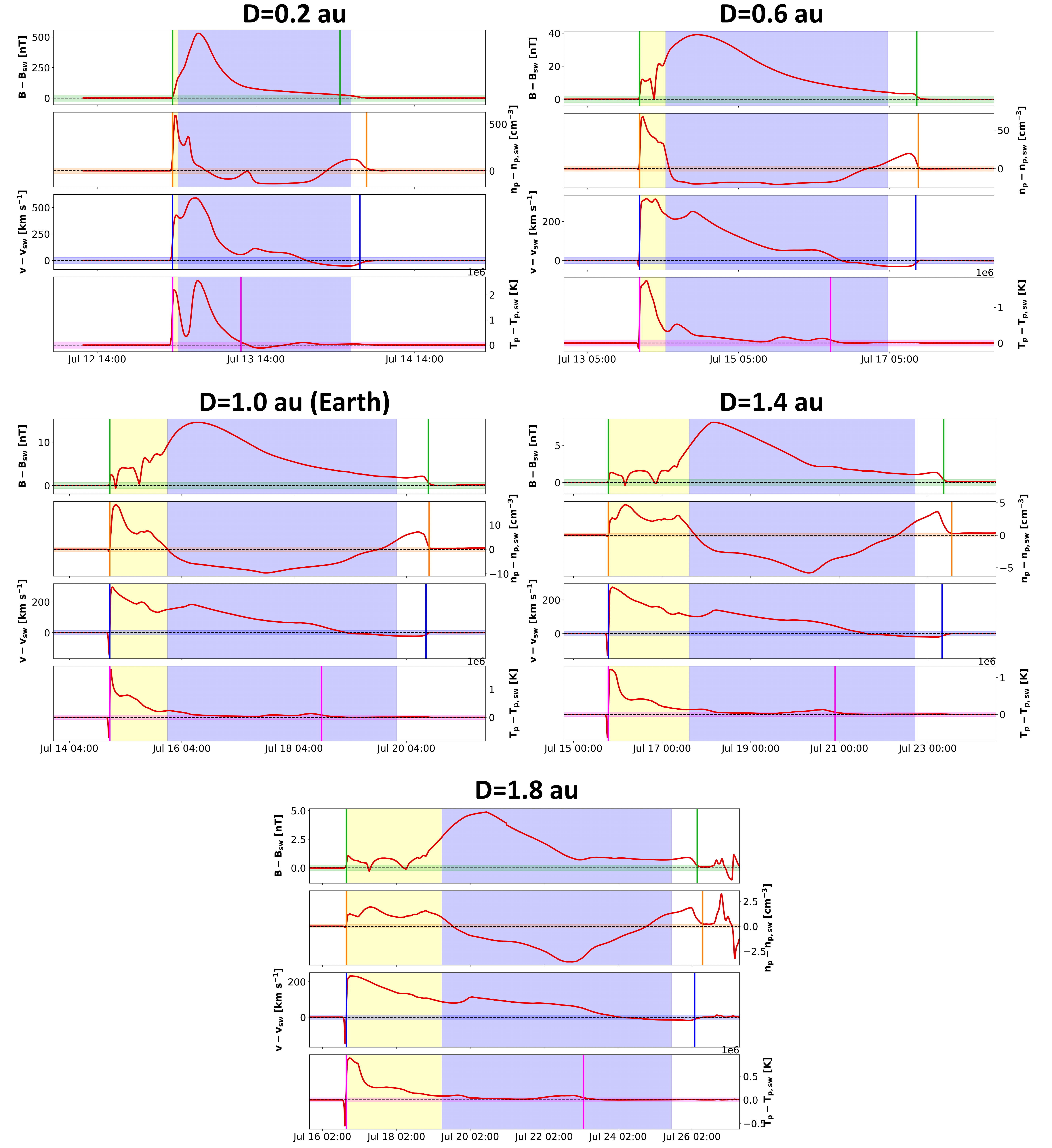}}
\caption{Identification of the start and end of the CME perturbation to the background solar wind
based on the 5\% threshold (marked by the coloured horizontal rectangles in each panel) in EUHFORIA time series at various heliocentric distances.
The CME wake for each variable is marked by the coloured vertical lines.
The period of the sheath and CME passage (same as in Figure~\ref{fig:spheromak_boundaries}) are marked as yellow and blue shaded areas, respectively.}
\label{fig:cme_wake} 
\end{figure}

This appendix provides a summary of the CME time series extracted at selected virtual spacecraft 
and the further analysis performed to identify the shock, magnetic ejecta boundaries, and CME wake. 
Figure~\ref{fig:spheromak_beta_thresholds} shows the performance of different $\beta_p$ thresholds in the identification of the CME/magnetic ejecta boundaries in EUHFORIA time series at various heliocentric distances.
Figure~\ref{fig:spheromak_boundaries} shows the identification of the sheath and CME/magnetic ejecta boundaries in EUHFORIA time series (in red) at various heliocentric distances based on Equation~\ref{eqn:shock_identification} and on the $\beta_p=0.5$ threshold.
Figure~\ref{fig:cme_wake} shows the identification of the start and end of the CME pre-conditioning of the background solar wind (CME wake) based on the 5\% threshold in EUHFORIA time series at various heliocentric distances.

\end{appendix}

\end{document}